\documentclass[a4paper,11pt]{article}
\pdfoutput=1 
\usepackage{jcappub} 
\usepackage{aas_macros}
\usepackage[T1]{fontenc} 
\usepackage[utf8]{inputenc}
\usepackage[dvipsnames]{xcolor}
\usepackage{mwe,bm,verbatim,cprotect,xcolor,xspace}
\usepackage{orcidlink}
\usepackage{subfig,csquotes,float, ulem}
\usepackage{bbold}
\usepackage[export]{adjustbox}
\graphicspath{{figs/}}

\def\be{\begin{equation}}
\def\ee{\end{equation}}

\def\ba#1\ea{\begin{align*}#1\end{align*}}
\def\bg#1\eg{\begin{gather}#1\end{gather}}

\newcommand\numberthis{\addtocounter{equation}{1}\tag{\theequation}}

\def\switch#1#2{#2}  
\newlength{\diagramwidth}
\switch{
\setlength{\diagramwidth}{3.0em}
}{
\setlength{\diagramwidth}{1cm}
}

\newcommand{\refeq}[1]{Eq.~(\ref{eqn:#1})}          
\newcommand{\refeqs}[2]{Eqs.~(\ref{eqn:#1})--(\ref{eqn:#2})}          
          
\newcommand{\reffig}[1]{Fig.~\ref{fig:#1}}  
\newcommand{\reffigs}[2]{Fig.~\ref{fig:#1}--\ref{fig:#2}}  
          
\newcommand{\reftab}[1]{Tab.~\ref{tab:#1}}          
\newcommand{\refsec}[1]{Sec.~\ref{sec:#1}}
\newcommand{\refapp}[1]{App.~\ref{app:#1}}

\newcommand{\codefont}[1]{{\texttt{#1}}}

\def\freeIC{\textsc{freeIC}}
\def\fixedIC{\textsc{fixedIC}}

\def\Plin{P_{\rm L}}

\renewcommand{\emph}[1]{\textit{#1}}

\newcommand{\vx}{\bm{x}}

\newcommand{\vk}{\bm{k}}
\newcommand{\vq}{\bm{q}}
\newcommand{\vp}{\bm{p}}

\newcommand{\<}{\langle}
\renewcommand{\>}{\rangle}

\renewcommand{\d}{\delta}

\newcommand{\dLambda}{\d^{(1)}_\Lambda}
\newcommand{\dLambdazero}{\d^{(1)}_{\Lambda_0}}
\newcommand{\ddet}{\delta_{\mathrm{det}}}
\newcommand{\ddetLambda}{\delta_{\mathrm{det},\Lambda}}

\newcommand{\eps}{\epsilon}

\newcommand{\hubble}{\mathcal{H}}

\def\Mpch{\,h^{-1}\text{Mpc}}
\def\iMpch{\,h\,\text{Mpc}^{-1}}
\def\Mpchcubed{\, h^{-3} \, \text{Mpc}^{3}}

\def\Plin{P_\text{L}}

\def\shat{\hat{s}}

\def\kmax{k_{\rm max}}

\def\M{\mathcal{M}}

\def\G{\mathcal{G}}
\def\J{\mathcal{J}}

\newcommand{\resp}[2]{R^{#1}_{\,\,\,#2}}
\newcommand{\resptwo}[3]{{R_2}^{#1}_{\,\,\,#2,#3}}
\newcommand{\dd}{\mathrm{d}}

\newcommand{\UnitM}{\mathbb{1}}

\newcommand{\bdelta}{b_{\delta}}
\newcommand{\bdeltatwo}{b_{\delta^2}}
\newcommand{\blapldelta}{b_{\nabla^2\delta}}
\newcommand{\btrMoneMone}{b_{\tr [M^{(1)}M^{(1)}]}}
\newcommand{\bsigmasigma}{b_{\sigma^2}}

\newcommand{\perm}[1]{ \expandafter\ifstrempty\expandafter{#1} {\mbox{perm.}} {\mbox{$#1$ perm.}} }

\DeclareMathOperator{\tr}{tr}

\let\Re\relax
\DeclareMathOperator{\Re}{Re}
\let\Im\relax
\DeclareMathOperator{\Im}{Im}
\def\lapl{\nabla^2}
\newcommand{\levicivita}{}
\def\levicivita#1#{\tensor#1{\epsilon}}

\definecolor{RoyalBlue}{rgb}{0.25,.41,.88}
\definecolor{ForestGreen}{rgb}{.13,.54,.13}
\definecolor{DeepPurple}{rgb}{.45,.1,.75}

\makeatletter
\renewcommand*{\thesection}{\arabic{section}}

\renewcommand*{\p@subsection}{}
\makeatother

\newcommand{\figurespath}{figs}

\title{Consistency tests of field level inference with the EFT likelihood}

\author[a]{Andrija Kosti\'{c}\orcidlink{0000-0002-8219-0025},}
\author[b,c]{Nhat-Minh Nguyen\orcidlink{0000-0002-2542-7233},}
\author[a]{Fabian Schmidt\orcidlink{0000-0002-6807-7464},}
\author[a]{Martin~Reinecke}

\affiliation[a]{Max Planck Institute for Astrophysics, Karl-Schwarzschild-Stra{\ss}e 1, 85748 Garching, Germany}
 \affiliation[b]{Leinweber Center for Theoretical Physics, University of Michigan, 450 Church St, Ann Arbor, MI 48109-1040, USA}
\affiliation[c]{Department of Physics, University of Michigan, 450 Church Street, Ann
Arbor, MI 48109, US}

\emailAdd{akostic@mpa-garching.mpg.de}
\emailAdd{nguyenmn@umich.edu}
\emailAdd{fabians@mpa-garching.mpg.de}
\emailAdd{martin@mpa-garching.mpg.de}


\abstract{Analyzing the clustering of galaxies at the field level in principle promises access to all the cosmological information available. Given this incentive, in this paper we investigate the performance of field-based forward modeling approach to galaxy clustering using the effective field theory (EFT) framework of large-scale structure (LSS). We do so by applying this formalism to a set of consistency and convergence tests on synthetic datasets. We explore the high-dimensional joint posterior of LSS initial conditions by combining Hamiltonian Monte Carlo sampling for the field of initial conditions, and slice sampling for cosmology and model parameters. We adopt the Lagrangian perturbation theory forward model from \cite{paper_nLPT}, up to second order, for the forward model of biased tracers. We specifically include model mis-specifications in our synthetic datasets within the EFT framework.
We achieve this by generating synthetic data at a higher cutoff scale $\Lambda_0$, which controls which Fourier modes enter the EFT likelihood evaluation, than the cutoff $\Lambda$ used in the inference. In the presence of model mis-specifications, we find that the EFT framework still allows for robust, unbiased joint inference of a) cosmological parameters --- specifically, the scaling amplitude of the initial conditions --- b) the initial conditions themselves, and c) the bias and noise parameters. In addition, we show that in the purely linear case, where the posterior is analytically tractable, our samplers fully explore the posterior surface. We also demonstrate convergence in the cases of nonlinear forward models.
Our findings serve as a confirmation of the EFT field-based forward model framework developed in \cite{paperI,paperII,paperIIb,cabass/schmidt,cabass/schmidt:2020,paper_realspace}, and as another step towards field-level cosmological analyses of real galaxy surveys.}

\begin{document}
\date{\today}

\maketitle
\flushbottom

\clearpage

\section{Introduction}
\label{sec:intro} 
Current and future galaxy surveys such as DESI \citep{aghamousa2016desi}, Euclid \cite{amendola2018cosmology}, PFS \cite{PFS}, and the Vera Rubin Observatory \citep{abell2009lsst} offer a wealth of modes for probing the physics of structure formation.
The traditional approach to cosmology inference from galaxy clustering is to compress the galaxy density field into summary statistics, such as two-point (see \cite{eBOSS:2021} and references therein), three-point \cite{2015MNRAS.451..539G,2017MNRAS.465.1757G,philcox2022boss,damico/etal:2022}, and four-point functions \cite{2022JCAP...09..050G, philcox2022probing, hou2022measurement}.

An alternative approach, and the one we follow in this paper, attempts to extract information at the field level, by explicitly forward modeling the entire observed galaxy density field including all the relevant physics and observational effects. This physical Bayesian forward modeling approach {\citep{jasche2013bayesian, wang2014ELUCID, jasche2019physical, lavaux2019BOSS, kitaura2020cosmic, bosetal2019bayesiancosmicdensityrsd} (see also references therein)} in principle allows for exploiting information on cosmology beyond $n$-point functions via explicit marginalization over the initial conditions.
While the amount of cosmological information available beyond the low-order $n$-point functions is still unclear, this approach, at the very least, allows for a consistent treatment of Baryon Acoustic Oscillation reconstruction \cite{paperII,babic/etal}, and thus is well motivated. Observational systematic effects can be explicitly encoded into the forward model (e.g. \cite{porqueres2019explicit,lavaux2019BOSS}), and might be easier to disentangle from cosmological signals at the field level as compared to summary statistics.

So far, the field-level inference approaches have typically used an empirical galaxy bias model and a simplified likelihood to infer from galaxy clustering data.
Using dark matter halos in N-body simulations as the reference data, \cite{nguyen2021} demonstrated that most empirical bias models and likelihoods that have been widely adopted by this approach thus far (e.g. \cite{jasche2019physical, lavaux2019BOSS}) can significantly bias the inferred cosmological fields.
Therefore, the key issue for this approach currently lies in a rigorous physical model, or alternatively, a sufficiently flexible effective model (e.g. \cite{KodiRamanah2019PaintingHF, Charnock2020NeuralPE, Modi2018CosmologicalRF}) to connect the matter and tracer fields.

The effective field theory of large-scale structure (EFTofLSS) \cite{baumann/etal,carroll2014consistent,carassco2012eftoflss} provides a systematic way of incorporating the complex nonlinear physics of galaxy formation on small scales, by using the fact that the galaxy formation is spatially localized and that galaxies and matter comove on large scales (the latter is ensured by the equivalence principle).
In particular, the EFTofLSS provides a parametric model for the matter-tracer relation up to the given order in matter density perturbations for any tracer field of interest (see \cite{biasreview} for a review). This, in turn, allows for robust extraction of cosmological information from the tracer data up to quasilinear scales \cite{philcox2022cosmology, damico/etal:2022}, i.e. wavenumbers smaller than the nonlinear scale wavenumber $k_{\rm NL}$ ($k_{\rm NL} \sim 0.2 \iMpch$ at $z=0$).

Until recently, the EFT predictions were restricted to summary statistics, but Refs.~\cite{paperI, cabass/schmidt, paperIIb} (see \cite{schmittfull2019modeling} for related work) presented a derivation of a field-level, EFT-based forward model and likelihood. 
This paper is the next in a series of papers developing this approach \cite{paperI, paperII, paperIIb, paper_nLPT, paper_realspace, stadler2023leftyrsd}, with the crucial addition of the \textit{marginalization over the initial conditions field by explicitly sampling it}. 
Specifically, we test the consistency of the EFT likelihood -- previously demonstrated for fixed initial conditions -- on a set of synthetic tracer data and a range of forward models including model mis-specification. The tests of the EFT likelihood approach presented here serve as a crucial stepping stone towards applying the method to more realistic tracers such as dark matter halo or simulated galaxy catalogs.
Throughout this paper, we use the following fiducial cosmology: $\Omega_m=0.3, \Omega_\Lambda=0.7, h=0.7, n_s=0.967, \sigma_8=0.85$ and a box size of $L=2000 \Mpch$. We keep cosmological parameters fixed to the fiducial $\Lambda$CDM cosmology but vary a scaling parameter $\alpha$ multiplying the initial conditions, which corresponds to varying $\sigma_8$ while keeping all other cosmological parameters fixed.

The structure of the paper is as follows.
In \refsec{forward_data_models}, we elaborate on the specific forward models we use for consistency tests of the EFT likelihood. We first describe the simple \textsc{linear} toy models and then physical models based on Lagrangian perturbation theory (LPT), {the \textsc{1lpt} and \textsc{2lpt}} models in \refsec{linear_fwd_models} and \refsec{onelpt_and_2lpt_fwd_models}, respectively. {In \refsec{likelihood}, we give the description of the EFT likelihoods} and the expression for the full field-level posterior being sampled in \refsec{full_posterior}.
We outline the sampling methods and the code implementation in \refsec{sampling_methods} and \refsec{code_implementation}, respectively. The synthetic datasets are described in \refsec{synthetic_datasets} and the consistency test results are presented in \refsec{analysis}. We conclude and discuss future outlook in \refsec{conclusions_and_summary}. In the appendices, additional details complementing the main results in \refsec{analysis} are presented and discussed.

\section{Forward models}
\label{sec:forward_data_models}

{In this section, we present all forward models used throughout the paper (\refsec{linear_fwd_models} and \refsec{onelpt_and_2lpt_fwd_models}) and we describe the EFT likelihoods, the last piece of our inference framework in \refsec{likelihood}}.

{Our forward models aim at modeling the following quantity}
\be
\d_d(\vx) \equiv \frac{n_d(\vx)}{\bar n_d} - 1 = \ddet(\vx) + \epsilon(\vx),
\label{eqn:generic_fwd_model_form}
\ee
with $n_d$ representing the number density field of synthetic tracers, $\bar n_d$ its spatial mean, and $\d_d$ the fractional overdensity.
Throughout this paper, we will work with tracers defined at a fixed time $\tau$, corresponding to today's epoch in the fiducial cosmology. Hence, we will drop the time argument for clarity.

We effectively marginalize over $\bar n_d$ by working with $\d_d$ and excluding the $\vec k=0$ mode from the analysis. Thus, the field-level forward models consist of two parts: the mean-field prediction $\ddet(\vx)$, which is a deterministic function of the initial conditions, and the likelihood, constructed from the assumptions about the underlying noise field $\epsilon(\vx)$.
We describe the different deterministic forward models used in this paper next, before turning to the likelihoods which couple to all of these deterministic forward models. 

The general form of the deterministic contribution is
\be
\ddet(\vx) = \sum_O b_O O(\vx),
\label{eqn:generic_ddet}
\ee
where $O$ denote the bias operators and $b_O$ the corresponding coefficients. Depending on the gravity model and the bias model we use, \refeq{generic_ddet} takes on different forms. Since we utilize an EFT approach, our forward models are defined for a specific cutoff $\Lambda$. The motivation for this cutoff is twofold:
\begin{itemize}
	\item[1.] The EFT model developed applies only to large scales. Hence, we want to restrict the likelihood evaluation only to the modes below the cutoff $k < \Lambda$. This in turn means that we need to apply a sharp-$k$ cut to the field operators $O \rightarrow O_{\Lambda}$.
	\item[2.] As first pointed out by \cite{carroll2014consistent}, and then shown in detail in \citep{paperIIb}, it is also necessary to perform a cutoff at the level of initial conditions $\Lambda_{\rm in} \equiv \Lambda$. This allows for proper renormalization of the dynamical evolution of the large-scale modes we want to model.
\end{itemize}
Throughout, we generate the synthetic data at a higher or equal cutoff $\Lambda_0$ than the value $\Lambda$ used in the inference, motivated by the fact that real-world tracers resemble data with $\Lambda_0 \gg \Lambda$. In fact, real data effectively has no cutoff, i.e. $\Lambda_0 \to \infty$. Below, we describe the specifics of the forward models employed throughout the paper. {We focus first on the simplest limiting cases, which involve only linear density fields, and then explain how we build up the \textsc{1LPT} and \textsc{2LPT} forward models.}

\subsection{\textsc{linear} forward models}
\label{sec:linear_fwd_models}

In \textsc{linear} forward models, the gravity model is restricted to linear evolution, which is incorporated by applying the appropriate transfer function to the initial conditions. On top of this, we also include tracer bias.
We consider two different bias expansions.
The first one involves only the linear bias $\bdelta$ and can be expressed as follows
\ba
\ddetLambda^{\textsc{linear}_1}(\vk) 
&= \bdelta \d^{(1)}_{\Lambda}(\vk),
\numberthis
\label{eqn:trivial_fwd_model}\\
\d_{\Lambda}^{(1)}(\vk) 
&\equiv W_{\Lambda}(k)\, \d^{(1)}(\vk) \\
&= W_{\Lambda}(k)\, \alpha\, T(k) \shat(\vk),
\numberthis
\label{eqn:deltaLambda}
\ea
where $\shat(\vk)$ denotes a unit Gaussian field which describes the initial conditions (see also discussion in \refsec{full_posterior}),  while $T(k)$ denotes the transfer function (recall that we keep the time fixed and implicit throughout). Since we keep the cosmological parameters fixed, we do not write them explicitly here. The scaling parameter $\alpha$ is defined such that, for $\alpha=1$, $\d^{(1)}(\vk)$ corresponds to a realization of the linear density field in the fiducial cosmology. $W_{\Lambda}(k)$ denotes the isotropic sharp-$k$ filter\footnote{See \refapp{fourier_hartley} for more details and our Fourier space convention.} at $k=\Lambda$. For this forward model (\refeq{trivial_fwd_model}), it is possible to analytically derive the posterior for the initial conditions $\shat$ (see \refapp{shat_posterior} for detailed calculation), and hence to test whether our inference approach fully explores the posterior in this case.

The simplest extension of the above {forward model} is to include also the quadratic field with the corresponding $\bdeltatwo$ bias parameter:
\ba
\ddetLambda^{\textsc{linear}_2}(\vk) &= \bdelta \d^{(1)}_{\Lambda}(\vk) + \bdeltatwo W_{\Lambda}(k) \int_{\vk'} \d^{(1)}_{\Lambda}(\vk - \vk')
\d^{(1)}_{\Lambda}(\vk'), 
\numberthis
\label{eqn:trivial_fwd_model_2nd_order_bias}
\ea
with $\d^{(1)}_{\Lambda}$ given by \refeq{trivial_fwd_model}. The bias expansion here is not complete in the EFT sense; we employ this forward model merely as the simplest possible generalization from an entirely linear forward model. 
The complete second-order bias expansion is considered in the next section.
Nevertheless, due to the nonlinearity induced by the quadratic bias, the $\shat$ posterior is non-Gaussian and it is non-trivial to make exact statements about its statistical moments (see \refapp{b2_running} for further discussion).
The models given by \refeqs{trivial_fwd_model}{trivial_fwd_model_2nd_order_bias} thus serve as toy models for which there exists full or approximate analytical expression of the $\shat$ posterior. They both assume all the relevant information is contained within the linear density field and do not involve nontrivial gravitational displacements, which are however essential when attempting field-level inference on real data.

\subsection{\textsc{1LPT} and \textsc{2LPT} forward models}
\label{sec:onelpt_and_2lpt_fwd_models}

The forward models in \refeqs{trivial_fwd_model}{trivial_fwd_model_2nd_order_bias} are valid only for describing tracers at linear order.
In order to obtain more accurate descriptions at higher orders -- which properly account for gravitational evolution -- we turn to Lagrangian perturbation theory.
The Lagrangian formulation of structure formation captures the effect of bulk flows non-perturbatively. This is especially useful for forward modeling.
To be more precise, we consider a first- and second-order LPT, labeling them with \textsc{1LPT} and \textsc{2LPT}, respectively. 

We begin by writing the Eulerian position along the fluid line at conformal time $\tau$ as 
\be
\vx_{\rm fl}(\vq, \tau) = \vq + \bm{\psi}(\vq, \tau),
\ee
with $\vq$ being the Lagrangian coordinate, $\bm{\psi}$ the displacement field and $\lim_{\tau\to 0} \bm{\psi}(\vq, \tau) = 0$. {We also re-instate the explicit $\tau$ dependence throughout this overview for clarity}.
Combining the mass conservation condition and the Poisson equation for the non-relativistic (cold) dark matter fluid yields the following equation of motion for $\bm{\psi}$ \cite{Rampf:2012,zheligovsky/frisch,matsubara:2015,paper_nLPT}
\be
\tr
\left[
(\mathbb{1} + \mathbf{M})^{-1}
(\mathbf{M}'' + \hubble \, \mathbf{M}')
\right]
=
-\frac{3}{2}\Omega_m\hubble^2
\left[
\lvert 
\mathbb{1} + \mathbf{M}
\rvert^{-1}
-
1
\right],
\label{eqn:displ_diff_eq}
\ee
where $M_{ij}(\vq, \tau) \equiv \partial_{(q_i} \psi_{j)}(\vq,\tau)$ is the symmetric part of the Lagrangian distortion tensor, the primes denote derivatives with respect to $\tau$, $\Omega_m$ is the corresponding matter density parameter and $\hubble$ the conformal Hubble rate. We restrict to the symmetric part of $\partial_{q_i} \psi_j$ throughout, since the antisymmetric part, corresponding to the curl of $\bm{\psi}$, is only nonzero at third order in perturbations {and} here we restrict ourselves to second order. Thus, $\bm{\psi}$ is a longitudinal vector and can be written as
\be
\bm{\psi}(\vq, \tau) = \frac{\mathbf{\nabla}}{\nabla^2} \sigma(\vq, \tau); \quad
\sigma(\vq,\tau) \equiv \tr{ \mathbf{M}}(\vq,\tau).
\label{eqn:s_field_eq}
\ee
Lagrangian perturbation theory then proceeds by expanding \cite{bouchet1994perturbative}
\be
\mathbf{M} = \mathbf{M}^{(1)} + \mathbf{M}^{(2)} + \ldots,
\label{eqn:Mexp}
\ee
and analogously for $\bm{\psi}$ and $\sigma$, where $\mathbf{M}^{(n)}$ involves exactly $n$ powers of the linear density field $\delta^{(1)}$. In fact, $\sigma^{(1)}(\vq,\tau) = -\delta^{(1)}(\vq,\tau)$.
For a given expansion history, \refeq{displ_diff_eq} can be formally integrated to yield recurrence relations relating $\mathbf{M}^{(n)}$ to the lower-order contributions \cite{Rampf:2012,zheligovsky/frisch,matsubara:2015}.

The perturbative contributions to $\mathbf{M}$ in \refeq{Mexp} can serve as building blocks for a general bias expansion of the form given in \refeq{generic_ddet}. The reason is that the Lagrangian distortion tensor along the fluid trajectory captures all leading gravitational observables for a comoving observer (see section 2.5 in \cite{biasreview} and \cite{MSZ}).
Specifically, one needs to construct all scalar contractions of the $\mathbf{M}^{(i)}$, {$i\leq n$}, that are relevant at the given order $n$. At second order, this yields
\be
O_L(\vq, \tau) \in 
\left\{
\tr
\left[
\mathbf{M}^{(1)}
\right](\vq, \tau),
\left( 
\tr
\left[
\mathbf{M}^{(1)}
\right](\vq, \tau)
\right)^2,
\tr
\left[
\mathbf{M}^{(1)}\mathbf{M}^{(1)}
\right](\vq, \tau)
\right \},
\label{eqn:bias_operators_nLPT}
\ee
where we emphasize that $O_L\equiv O(\vq, \tau)$ is in Lagrangian coordinates.
Thus, up to second order, we require three distinct bias operators and the associated bias coefficients.

In order to obtain the corresponding Eulerian fields we use a weighted particle approach \cite{schmittfull2019modeling,paper_nLPT}. We consider $(N_{g}^{\rm{Eul}})^3$ {pseudo} particles on a regular grid in $\vq$, and assign each of them 3 weights corresponding to the three bias operators. Then, each of these {pseudo} particles is displaced from $\vq$ to $\vx = \vq+\bm{\psi}(\vq,\tau)$, and the masses are deposited to the grid using a mass-conserving assignment scheme (we choose cloud-in-cell assignment here). This yields the three Eulerian operators corresponding to the Lagrangian operators listed above.  
In fact, we replace the weight field $\tr[\mathbf{M}^{(1)}]$ with a unit weight field, so that the resulting Eulerian field is the LPT matter density field $\d_{n\rm LPT}$ ($n=1,2$), and use this linear bias term in Eulerian frame. 

Finally, we also include the leading-order higher-derivative bias contribution (see sec 2.6 in \cite{biasreview}), $\lapl\d_{n\rm LPT}$, computed in the Eulerian frame, in order to capture the finite spatial size of the regions forming the tracer of interest. Therefore, the full list of bias parameters and corresponding operators is
\ba
\ddetLambda^{n\rm LPT}(\vx, \tau) 
=\:& b_\d {\d_{n\rm LPT,\Lambda}}(\vx,\tau) \\
&+ 
\bsigmasigma [\sigma^{(1)}_{\Lambda}]^2(\vx,\tau)
+ 
\btrMoneMone \tr{\left[ M_{\Lambda}^{(1)}M_{\Lambda}^{(1)} \right]}(\vx,\tau) \\
&+ 
\blapldelta \lapl {\d_{n\rm LPT,\Lambda}}(\vx,\tau),
\numberthis
\label{eqn:fwd_model_2lpt2d}
\ea
where we emphasize the presence of the cutoff $\Lambda$. The linear displacement tensor $M_{\Lambda}^{(1)}$ is related to the linear density perturbation via
\ba
M_{ij, \Lambda}^{(1)}(\vq) 
&= -\frac{\partial_{q_i}\partial_{q_j}}{\nabla_q^2} \d_{\Lambda}^{(1)}(\vq),
\ea
where $\d^{(1)}_{\Lambda}$ is defined in \refeq{deltaLambda}.

\subsection{Field-level likelihood}
\label{sec:likelihood}

Instead of modeling directly the tracer number counts, as done for example in \cite{jasche2019physical,jasche2013bayesian}, the EFT likelihood aims to describe the tracer density field $\delta_d$.
It is obtained by integrating out the modes with $k>\Lambda$ in the initial conditions \cite{cabass/schmidt}.
These small-scale modes also produce a stochastic contribution to the predicted galaxy density field $\ddetLambda$. This effect is encoded by the noise field, $\epsilon$. Since this effective noise field arises from the superposition of many independent modes, it is Gaussian to leading order.
Moreover, because the tracer formation is spatially local, the power spectrum of the noise is constant to leading order, with corrections scaling as $k^2$ \cite{biasreview}. In other words,
\be
\< \eps(\vk) \eps(\vk') \> = (2\pi)^3\d_D(\vk+\vk')\,P_{\eps}\,(1 + \sigma_{\eps,2}k^2 + \cdots),
\label{eqn:cov_Eps}
\ee
where $\d_D$ denotes the Dirac delta function, and
we have denoted with $P_\eps \sim \bar{n}_d^{-1}$ the leading order contribution to the noise covariance. Note that the scale-dependent correction, $\sigma_{\eps,2}$, is written here as fractional correction by convention. In general, the gravitational evolution of small-scale modes under the influence of large-scale modes generates density-dependent noise terms, which cause the noise covariance to be non-diagonal in Fourier space.
We do not consider these contributions here, since all synthetic datasets used here do not contain such density-dependent noise. {The exception is the \textsc{2LPT} synthetic dataset (see \reftab{synthetic_datasets}). In this case, the second-order bias coupled with the presence of a cutoff mismatch between synthetic data and inference forward models can generate density-dependent noise}. We also set the subleading noise contribution $\sigma_{\eps,2}$ to zero throughout this paper. More detailed investigations of the impact of the noise model are relegated to future work.

{Our data model is then given by}
\ba
\d_{d,\Lambda}(\vx) &= \ddetLambda(\vx) + \eps(\vx),\\
\eps(\vx) &\hookleftarrow \G(\eps; 0, \sigma_{\eps}^2 \mathbb{1})),
\numberthis
\label{eqn:data_model}
\ea
where $\ddetLambda$ is given by one of the forward models we consider here, namely \refeqs{trivial_fwd_model}{trivial_fwd_model_2nd_order_bias} or \refeq{fwd_model_2lpt2d}. In the second line of \ref{eqn:data_model} we emphasize that the noise field is assumed to be Gaussian distributed with covariance structure given by the leading term from \refeq{cov_Eps}. Specifically, the noise variance $\sigma_\eps^2$ on the real-space grid of size $N_g^{\Lambda}$ {and box size $L$} is defined through the following relation 
\be
  P_\eps = \sigma_{\epsilon}^2 \frac{L^3}{(N_g^{\Lambda})^3}.
  \label{eqn:sigma_eps}
\ee
In this paper, we consider two forms of EFT likelihoods: one whose arguments explicitly include bias parameters, hence labeled \emph{unmarginalized likelihood}, and the other one where bias parameters are analytically marginalized out, the \emph{marginalized likelihood}. Below we detail the two likelihoods in that order. The notation closely follows that in \cite{paperII}.
{We also} switch to the discrete Fourier space representation of the fields (see \refapp{fourier_hartley} for details on our Fourier space convention).

{From our assumption of Gaussian noise it follows directly that the likelihood is likewise Gaussian. Namely,}
\ba
\ln\mathcal{L}\left(\d_{d, \Lambda} \Big| \alpha, \shat,\{b_O\}, \sigma_{\eps}\right) \,
=&
-\frac{1}{2}
\sum_{\vk \neq 0}^{\kmax}
\left[
\ln{2\pi\sigma^2_{\eps}}
+
\frac{1}{\sigma^2_{\eps}}
\Big\lvert
\d_{d, \Lambda}(\vk) - \d_{\mathrm{det},\Lambda}[\alpha, \shat, \{b_O\}](\vk)
\Big\rvert^2
\right]\\
&+ \text{const.} \,\, ,
\numberthis
\label{eqn:unmarg_like}
\ea
where we have explicitly stated the dependence of $\ddetLambda$ on bias parameters, $\{b_O\}$, the scaling parameter $\alpha$ and the initial conditions $\shat$. Note that $\ddetLambda$ is a deterministic function of these parameters and that the likelihood is evaluated only up to $k_{\mathrm{max}}$, strictly allowing only for modes below the cutoff. In order to maximize the information gain, we choose $k_{\mathrm{max}} = \Lambda$. We accumulate all the terms which depend neither on $\alpha$, $\{b_O\}$, $\shat$ nor $\sigma_{\eps}$ in $\text{const.}$, since they represent only irrelevant normalizing factors.

The marginalized likelihood is obtained by analytically marginalizing over all bias parameters in \refeq{unmarg_like}. Given that the likelihood depends quadratically on any bias coefficient (in case priors on the bias parameters are Gaussian or uniform on $(-\infty,\infty)$), this is a straightforward calculation (see section 2.2 in \cite{paperII}).
The expression is
\ba
-\ln\mathcal{L}\left(\d_{d, \Lambda} \Big| \alpha,\shat, \sigma_{\eps} \right) \,
&=
\frac{1}{2}
\tr{\ln{A_{OO'}}}
+\frac{1}{2}
\tr{\ln{C_{\rm{prior}}}}
\begin{aligned}[t]
&+\frac{1}{2}
\sum_{\vk \neq 0}^{\kmax}
\left(
\ln \sigma_{\eps}^2
+
\frac1{\sigma_{\eps}^{2}} \lvert\d_{d,\Lambda}(\vk)\rvert^2
\right)\\ 
&-
\frac{1}{2}
\sum_{\{O, O'\}}
B_O (A^{-1})_{OO'}B_{O'} + \text{const.} \,\, ,
\end{aligned} \\
B_O\equiv B_O[\shat,\alpha] &= 
\sum_{\vk \neq 0}^{\kmax}
\frac{\d^{*}_{d,\Lambda}(\vk)O[\shat,\alpha](\vk)}{\sigma_{\eps}^2} + \sum_{O'}(C^{-1}_{\rm{prior}})_{OO'}\mu_{b_{O'}}, \\
A_{OO'} \equiv A_{OO'}[\shat,\alpha] &= 
\sum_{\vk \neq 0}^{\kmax}
\frac{O^{*}[\shat,\alpha](\vk)O'[\shat,\alpha](\vk')}{\sigma_{\eps}^2} + \left(C^{-1}_{\rm{prior}}\right)_{OO'},
\numberthis
\label{eqn:marg_like}
\ea
where, once again, $\text{const.}$ encapsulates the terms independent of the parameters of interest. The marginalization was performed under the assumption of a Gaussian prior on bias parameters with covariance $C_{\rm{prior}}$ and mean $\mu_{b_{O}}$. We choose a fairly uninformative prior as given in \refeq{priors_marg}. {As indicated,} the information on $\alpha$ and the initial conditions $\shat$ is propagated through the $B_O$ and $A_{OO'}$ operators.

\section{Sampling the full posterior}
\label{sec:sampling_full_posterior}

Here, we provide the final expression for the posterior being sampled and elaborate more on some specific choices of our sampling scheme, as well as some additional details of our code implementation.

\subsection{Full posterior}
\label{sec:full_posterior}

The results of the previous section now allow us to write the full posterior which we aim to sample from. The expression can be obtained readily from
\ba
\mathcal{P}\left(\alpha, \shat, \{b_O\}, \sigma_{\eps} \Big| \d_{d, \Lambda}\right)
&=
\frac{
\mathcal{L}(\d_{d,\Lambda} | \alpha, \shat, \{b_O\},\sigma_{\eps})\mathcal{P}(\alpha, \shat, \{b_O\},\sigma_{\eps})}
{\mathcal{P}(\d_{d,\Lambda})},
\numberthis
\label{eqn:full_posterior}
\ea
where $\mathcal{P}(\alpha, \shat, \{b_O\}, \sigma_{\eps} | \d_{d, \Lambda})$ represents the posterior probability of the parameters of interest given the synthetic data, $\mathcal{L}(\d_{d,\Lambda} | \alpha, \shat, \{b_O\},\sigma_{\eps})$ the corresponding likelihood (see \refeqs{unmarg_like}{marg_like}), while $\mathcal{P}(\alpha, \shat, \{b_O\},\sigma_{\eps})$ represents the associated prior. {The $\mathcal{P}(\d_{d,\Lambda})$ represents the evidence, which does not play any role in our inference framework.} As noted in \refsec{intro}, we keep all other cosmological parameters fixed to the fiducial values listed there.

We assume minimal prior knowledge on $\sigma_{\eps}$ and the $\{ b_O \}$. Moreover, as physical parameters describing the properties of the tracers, they are a priori independent of the initial conditions $\shat$. Therefore, the joint prior structure is entirely factorized. We choose the following prior configuration throughout the paper for the unmarginalized likelihood:
\ba
\mathcal{P}(\sigma_{\eps}) &= \mathcal{U}(0.05, 100.), \\
\mathcal{P}(\alpha) &= \mathcal{U}(0.5,1.5), \\
\mathcal{P}(\bdelta) = \mathcal{U}(0.01,10), 
&\quad 
\mathcal{P}(\bdeltatwo) = \mathcal{U}(-10,10), \\
\mathcal{P}(\bsigmasigma) =
\mathcal{P}(\btrMoneMone) &=
\mathcal{P}(\blapldelta) = \mathcal{U}(-25,25),
\numberthis
\label{eqn:priors}
\ea
where $\mathcal{U}(l,r)$ denotes the uniform distribution on interval $[l,r]$. Note that here we choose to keep the prior on $\bdelta$ strictly positive, as we do not consider the modeling of negative bias tracers, such as voids, within this paper. We chose priors of the higher-order bias coefficients to be symmetric around zero, as a priori these bias coefficients could take on either sign.

For the inference with the marginalized likelihood, the only difference is in the priors for the bias parameters, which are taken to be of the following form
\ba
\mathcal{P}(\bdelta) = \mathcal{G}(0.01,10); 
&\quad 
\mathcal{P}(\bdeltatwo) = \mathcal{G}(0,10) \\
\mathcal{P}(\bsigmasigma) =
\mathcal{P}(\btrMoneMone) &=
\mathcal{P}(\blapldelta) = \mathcal{G}(0,25).
\numberthis
\label{eqn:priors_marg}
\ea
Given the signal-to-noise level of our synthetic datasets, these priors are essentially uninformative, and we expect an entirely negligible difference in parameter inferences between marginalized and unmarginalized likelihoods.

Finally, for the prior $\mathcal{P}(\shat)$ on the initial conditions, we consider the following two choices:
\be
\mathcal{P}(\shat) =
\begin{cases}
	\d_{D}(\shat - \shat_{\mathrm{true}}) & \text{for {\fixedIC} case} \\[3pt]
	\mathcal{G}(\shat; 0, {S}) & \text{for {\freeIC} case},
\end{cases}
\label{eqn:shat_prior}
\ee
where we have separated the cases with fixed initial conditions to the ground-truth (\fixedIC) and with initial conditions explicitly sampled (\freeIC).
In the latter case, the prior covariance structure of the discretized $\shat$ field is given by ${S^{\vx_i}}_{\vx_j} \equiv \< \shat(\vx_i)\shat(\vx_j) \>_{\mathcal{P}(\shat)} = \d_D^{i,j}$, while in Fourier space it becomes ${S^{\vk}}_{\vk'} \equiv \< \shat(\vk)\shat(\vk') \>_{\mathcal{P}(\shat)} = (N_g^{\Lambda})^3 \d_D^{\vk,\vk'}$, with $N_g^\Lambda$ being the grid size corresponding to the cutoff $\Lambda$, and $\delta_D$ denoting the Kronecker delta (see \refapp{fourier_hartley} for details on our Fourier convention).

The {\fixedIC} case corresponds to that considered in the application to dark matter halo catalogs in previous papers of this series \cite{paperI, paperII, paperIIb, paper_nLPT, paper_realspace}, where {the $\alpha$ parameter} was shown. Moreover, the same setup was used to measure bias parameters of simulated halos and galaxies in \cite{2021JCAP...08..029B, 2021JCAP...10..063L}. 
For real-data applications, as we have no knowledge of the true initial conditions a priori, it is crucial for our method to properly marginalize over all plausible realizations of the initial conditions.
{Thus, the parameter posterior obtained on our synthetic datasets with the {\fixedIC} prior serves as a good reference point for the {\freeIC} {case}. Specifically, we expect consistency between the {\fixedIC} and {\freeIC} {posterior parameter means}, which we demonstrate in \refsec{analysis}.}

\subsection{Sampling methods}
\label{sec:sampling_methods}

In order to explore the posterior surface of the parameters of our model, we utilize a combination of slice sampling (see Sec. 29.7 in \cite{mackay2003information} and \cite{neal2003slice}) and Hamiltonian Monte Carlo (HMC) sampling techniques (see \cite{betancourt2017conceptual, neal2011mcmc} and Sec. 30.1 in \cite{mackay2003information}). Below, we motivate this choice of sampling techniques, which was inspired by the findings made in the development of the \textsc{borg} code (see, e.g. \cite{jasche2019physical}).

The slice sampling technique is used for sampling $\alpha$, bias parameters $\{b_O\}$, and the noise parameter $\sigma_{\eps}$.
This means that we actually sample the 1D probability density function of these parameters conditioned on the current realization of initial conditions $\shat$. We adopt sequential univariate slice sampling, i.e. we sequentially sample individual parameters. While multivariate slice sampling methods do exist, they require additional tuning to be more efficient than our approach (see the discussion in section 5 of \cite{neal2003slice}). 

When it comes to sampling the posterior of the initial conditions $\shat$, we use the HMC sampling technique. The reason behind this choice is that the number of Monte Carlo samples needed for generating an independent sample scales more efficiently with the problem dimensionality than for standard Monte Carlo methods. This scaling goes as $\sim N_\text{dim}$ in the case of a standard random-walk algorithm, while it goes as $\sim N_\text{dim}^{1/4}$ for the HMC method (see, for example, Sec. 4.4 of \cite{neal2011mcmc}). For the problem we consider here, typically $N_\text{dim} \sim 10^5 - 10^6$. Therefore, HMC currently appears to be the most (if not only) practical sampling method to tackle such a problem.

In order to utilize HMC, one needs a fully differentiable forward model with respect to the initial conditions. This requirement is necessary for the crucial step of generating a new proposal of initial conditions, consistent with the data likelihood. This new proposal is generated by numerically integrating along the Hamiltonian flow defined by the likelihood and prior gradients with respect to the initial conditions $\shat$. For this, we choose the second-order leapfrog integrator, although see, e.g. \cite{mannseth2016application,2021MNRAS.502.3976H}, for the applicability of higher-order integration schemes.

Given the structure of our code, \codefont{LEFTfield}\footnote{\emph{Lagrangian, Effective-Field-Theory-based forward model of cosmological density fields.}}, described in \refsec{code_implementation}, the {analytical} derivative of the full forward model can be readily obtained through successive applications of the chain rule.

\begin{figure}[t!]
\centering
\includegraphics[width=\linewidth]{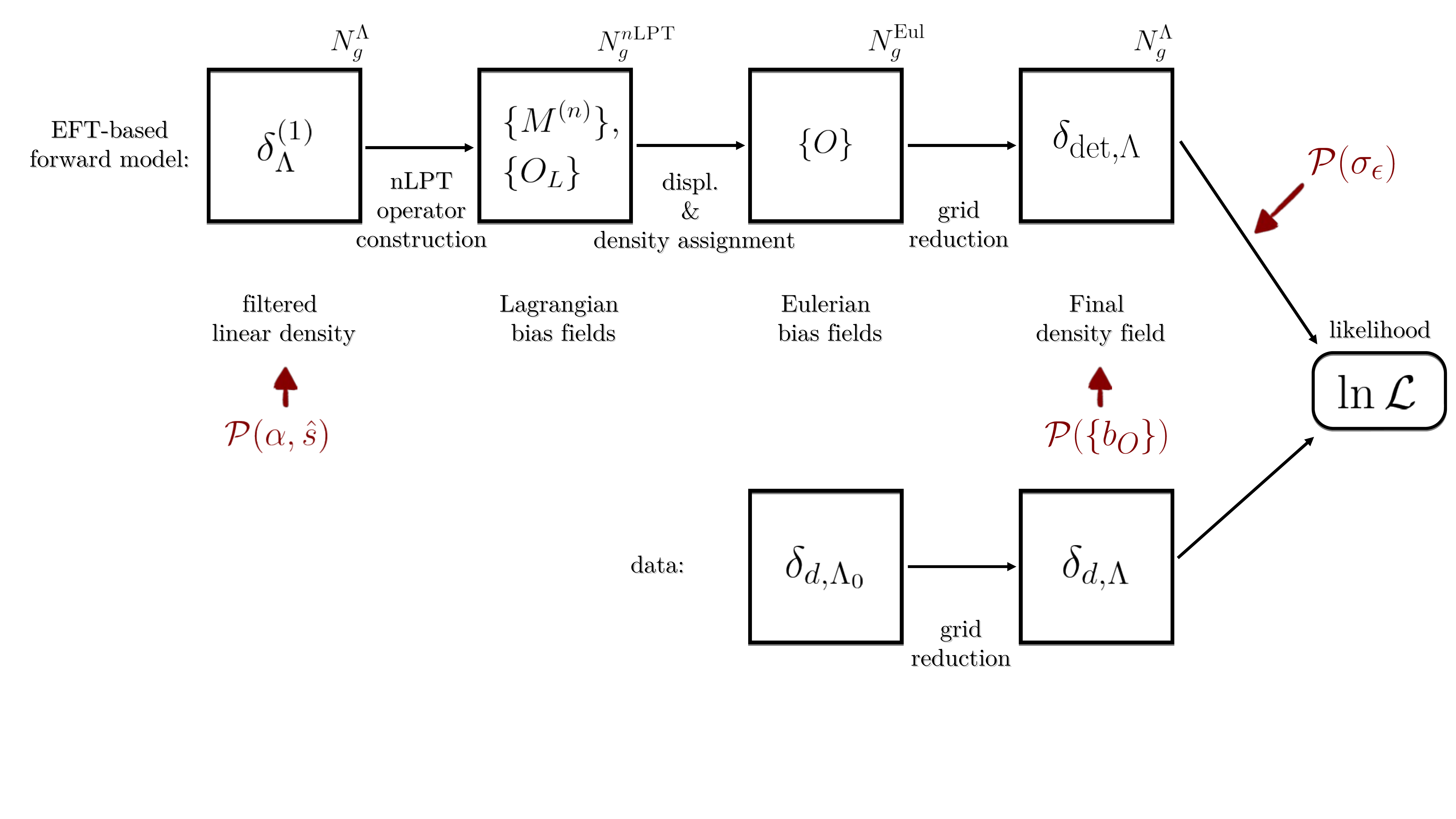}
\caption{A depiction of our consistency test architecture, which consists of two branches. The top one describes all the relevant elements of the forward models we consider. All forward model elements are described in \refsec{forward_data_models} and \refsec{code_implementation}. The bottom row schematically shows the synthetic data generation process. Note that we use the same forward models from the top row to generate the dataset in the bottom row but at a different cutoff value $\Lambda_0$. Our ensemble of synthetic datasets is described in \refsec{synthetic_datasets}.}
\label{fig:fwd_model_flowchart}
\end{figure}
Finally, readers may ask why not try to combine the two sampling approaches such that all the parameters are sampled within the HMC scheme at the same time. This however is unfeasible, due to the large difference between the derivative norms of the variation with respect to $\shat(\vk)$ and the variation of the rest of the model parameters. Variation of $\alpha$ or bias parameters affects all the modes of the forward model, while the derivative with respect to a given $\shat(\vk)$ only has to vary a single mode at a time. 
This in turn requires the HMC trajectories to be integrated with small step sizes, in order to reach a reasonable acceptance rate, and {ultimately} results in a very slow exploration of the (joint) posterior surface.

Therefore, decoupling the sampling of $\alpha$, bias, and noise parameters from the sampling of initial conditions seems to provide the fastest exploration time. Note that, in such a block-sampling scheme one could still use the HMC method to sample cosmological and bias parameters, conditioned on a given realization of the initial conditions. One could separate the former and latter into two separate HMC sampling blocks with two different equations of motion (to be integrated). However, since the dimension of the cosmological and bias parameter space is currently negligible, we opt for the robustness of slice sampling.

\subsection{Code implementation}
\label{sec:code_implementation}

The forward models and likelihoods described in \refsec{forward_data_models} are implemented in the differentiable code \codefont{LEFTfield}. \codefont{LEFTfield} adopts the C++17 standard and represents a substantially extended and more efficient version of the code presented in \cite{paper_nLPT}. Most importantly, \codefont{LEFTfield} implements the gradient of the likelihood with respect to the initial conditions $\shat$, which as noted is crucial for the HMC sampling approach. A public release of the code, subject to additional tests and tidying up, is planned for the future.
  
The code is structured in a modular way, breaking down the forward model into a series of simple steps, called ``forward elements'', which are templatized on generic input and output types and implement the general behavior of composed operator chains; specifically, the respective input and output types need to match for all composed forward elements.
The sequence of high-level operations contained in the $n\rm{LPT}$ forward models are represented in the top row of \reffig{fwd_model_flowchart} (note that each of the blocks, in general, consists of multiple forward elements). The forward model starts from a sample of initial conditions from which one obtains the initial density field $\d^{(1)}_\Lambda$, shown in top left of \reffig{fwd_model_flowchart}. The size of this grid is chosen based on the cutoff $\Lambda$. This is indicated with $N_g^\Lambda$. The next element is the bias operator construction. For this, we use the set of bias operators appearing in \refeq{bias_operators_nLPT} in the case of \textsc{1LPT} and \textsc{2LPT} models. At this step, care is also taken for representing all the physical modes of the forward model, by choosing appropriate grid sizes, indicated with $N_g^{n\textsc{LPT}}$. Afterward, these fields are displaced utilizing a weighted particle scheme to the final Eulerian positions. This results in a mapping $O_L \to O$, where now the $O$ operators are assigned onto a grid of size $N_g^{\text{Eul}}$, chosen in advance, with $N_g^{\rm{Eul}} \ge N_g^{n\textsc{LPT}}$ in order to keep all physical modes of the forward model represented on the Eulerian grid. In the end, the set of the displaced bias operators $\{O\}$ is resized in Fourier space to a smaller and final grid corresponding to $N_g^\Lambda$, using the sharp-$k$ cutoff.  Finally, the deterministic prediction $\ddetLambda$ from \refeq{fwd_model_2lpt2d} is constructed in the last piece of the top row. This also involves drawing a new set of relevant bias parameters from the corresponding prior. The grid reduction and grid padding are both performed in Fourier space. Several options for mass assignment schemes are implemented, including nearest-grid-point (NGP, which is not differentiable however), cloud-in-cell (CIC), and triangular-shaped cloud (TSC). In this paper, we use the CIC scheme throughout.
The very last piece of the forward model is the evaluation of the likelihood, given by either \refeq{unmarg_like} or \refeq{marg_like}.
Note that the forward models of \refsec{linear_fwd_models} are much simpler, but have the same overall structure.

In the case of the HMC sampling block, the full gradient of the likelihood with respect to the initial conditions, $\shat$, needs to be evaluated. In order to do so, we utilize the chain rule, collecting every $\shat$-dependent term of each element in the forward model, from right to left in the flowchart. In addition, for every sample of $\shat$, the slice sampler generates a new sample of the other parameters of interest. This process is repeated until {the desired number of samples} is achieved (see \refsec{analysis} and \refapp{results_convergence_tests} for more details on our convergence requirements and verification).

{The bulk of the computing time is spent in the HMC sampling of the initial conditions.}
For reference, we provide some benchmark computing times here. For the \textsc{2LPT} forward model with $64^3$ grid size, \codefont{LEFTfield} generates $\sim 200$ samples per CPU hour, roughly corresponding to $\sim 1$ effective sample per CPU hour, running on a single Intel(R) Xeon(R) Gold 6138 CPU @ 2.00GHz with 20 cores and using \codefont{OpenMP} parallelization.

\section{Synthetic datasets}
\label{sec:synthetic_datasets}

In this section, we describe how precisely we generate the synthetic data sets (the first element in the bottom row of \reffig{fwd_model_flowchart}). In general, a dataset is generated from each of the aforementioned forward models. We introduce model mis-specification through the mismatch between the cutoff $\Lambda_0$ in the synthetic data and a varying cutoff $\Lambda$ in our forward models, as indicated in the two elements in the bottom row of \reffig{fwd_model_flowchart}. We always fix $\Lambda_0=0.14 \iMpch$. Throughout, we label the specific realization of initial conditions used for synthetic data generation by $\shat_{\rm{true}}$. All parameters of the synthetic datasets are summarized in \reftab{synthetic_datasets}. 

\subsection{\textsc{linear} model synthetic data}
\label{sec:trivial_model_data}

\begin{table}[t!]
    \centering
    \begin{tabular}{ | c |  c  c  c  c  c  c  c  c | } 
     \hline
     Dataset \textbackslash \ Parameter & $\alpha$ & $\bdelta$ & $\bdeltatwo$ & $\bsigmasigma$ & $\btrMoneMone$ & $\blapldelta$ & $\sigma_{\eps}$ & $\tilde{\sigma}_{\eps}$ \\ 
     \hline 
     $\mathcal{D}^{\textsc{linear}}_1$  & $1.0$ & $1.0$ & $0$  & $0$ & $0$ & $0$ & $0.5$ & $0.586$ \\
     $\mathcal{D}^{\textsc{linear}}_2$  & $1.0$ & $1.0$ & $1.0$ & $0$ & $0$ & $0$ & $0.5$ & $0.586$ \\ 
     $\mathcal{D}^{\textsc{linear} - \text{informative}}_2$  & $1.0$ & $1.0$ & $1.0$ & $0$ & $0$ & $0$ & $0.001$ & $0.002$ \\
     $\mathcal{D}^{\textsc{2LPT}}_1$	& $1.0$ & $1.0$ & $0$ & $0$ & $0$ & $0$ & $0.5$ & $0.586$ \\
     $\mathcal{D}^{\textsc{2LPT}}_2$	& $1.0$ & $0.87$ & $0$ & $-0.2$ & $-0.2$ & $0.2$ & $0.4$ & $0.469$ \\
     \hline
    \end{tabular}
    \caption{Parameters used for generating different synthetic datasets. The noise levels in these datasets correspond to $P_\eps = 2743 \Mpchcubed$, except for the dataset from third row, which has $P_\eps = 1.1\times 10^{-2} \Mpchcubed$. {All datasets listed in this table are obtained using a cutoff of $\Lambda_0 = 0.14 \iMpch$.} The parameters listed in this table are described in \refsec{forward_data_models}, except $\tilde{\sigma}_{\eps}$, which is defined in \refeq{sigma_tilde}.}
    \label{tab:synthetic_datasets}
\end{table}

For the models described in \refeqs{trivial_fwd_model}{trivial_fwd_model_2nd_order_bias} we generate two sets of synthetic data: one with $\bdeltatwo=0$ and the other with $\bdeltatwo\neq0$. Below, we specify the parameter values adopted and explain our choice.

The first case of synthetic data, $\mathcal{D}^{\textsc{linear}}_1$, is obtained with the parameters listed in the first row of \reftab{synthetic_datasets}. Note that the cutoff set by $\Lambda_0$ determines the grid size, which in this case is $N_g^{{\Lambda_0}} = 90$. {This grid size most closely corresponds} to the Nyquist frequency for the cutoff $\Lambda_0=0.14 \iMpch$ and a box size $L = 2000 \Mpch$. The parameter $\sigma_{\eps}=0.5$ is the square root of the noise variance on this grid, which corresponds to a Poisson shot-noise for tracers with comoving number density of $\bar{n} \approx 3.645 \times 10^{-4} (\iMpch)^3$. 
Since the noise power spectrum $P_\eps$ is a physical quantity (in particular, independent of the grid size), it follows from \refeq{sigma_eps} that the combination $\sigma_{\epsilon} (N_g^{\Lambda})^{-3/2}$ must be independent of the grid size corresponding to the cutoff $\Lambda$.
{This implies that $\sigma_{\epsilon}$ itself depends on the grid size $N_g^\Lambda$}. Therefore, instead of working with $\sigma_{\epsilon}$, we define the following quantity
\be
\tilde{\sigma}_{\eps} \equiv 10^3 \frac{\sigma_{\epsilon}}{(N_g^{{\Lambda}})^{3/2}}
= 10^3 \left(\frac{P_\eps}{L^3}\right)^{1/2},
\label{eqn:sigma_tilde}
\ee
which is grid-size-independent by construction.
The prefactor $10^3$ is introduced for numerical convenience. We will mainly quote $\tilde{\sigma}_{\eps}$ instead of $\sigma_{\eps}$ in our posterior analyses in \refsec{analysis}.
As can be seen from \reftab{synthetic_datasets}, we adopt a comparable noise level for all synthetic datasets except for $\mathcal{D}_2^{\textsc{linear}-\text{informative}}$, which we describe below.

The second synthetic dataset, $\mathcal{D}^{\textsc{linear}}_2$ comes in two variants, listed in the second and third row of \reftab{synthetic_datasets}. We always choose $|\bdeltatwo|=|\bdelta|$ in order to introduce a non-negligible mode coupling through the quadratic term in \refeq{trivial_fwd_model_2nd_order_bias}. In addition, we consider $\mathcal{D}^{\textsc{linear} - \text{informative}}_2$ (third row of \reftab{synthetic_datasets}), using a very low noise level ($\tilde{\sigma}_{\eps} = 0.002$) and hence representing a very informative dataset. This dataset was included to further investigate the dependence of the inferred $\bdeltatwo$ as a function of the cutoff $\Lambda$ {(see \refsec{linear_b1_b2_sigma_tests})}.

Apart from the case of $\mathcal{D}^{\textsc{linear}}_1$, for all other datasets, including $\mathcal{D}^{\textsc{linear}}_2$ datasets, we generate two different data realizations. We achieve this by generating two different initial conditions realizations, keeping the values for the remaining parameters fixed. Independent inferences are performed on both data realizations. This helps us gauge the significance of any mis-estimation of the posterior and hence of potential systematic trends in the inferred parameters. Henceforth, we label the different data realizations by the subscript $\{a,b\}$, for example $\mathcal{D}^{\textsc{linear}}_{2,a}$ and $\mathcal{D}^{\textsc{linear}}_{2,b}$ which correspond to the two different realizations of the synthetic dataset listed in the second row of \reftab{synthetic_datasets}.

\subsection{\textsc{2LPT} synthetic data}
\label{sec:2lpt_model_tests}

We generate two types of synthetic datasets for the \textsc{2LPT} forward model described in \refeq{fwd_model_2lpt2d}. They are labelled as $\mathcal{D}^{\textsc{2LPT}}_{1}$ and $\mathcal{D}^{\textsc{2LPT}}_{2}$ and their parameters are listed in the last two rows of \reftab{synthetic_datasets}. As before, we also generate two different realizations of each of these datasets.

The $\mathcal{D}^{\textsc{2LPT}}_{1}$ datasets serve as an input for the internal consistency between the \textsc{2LPT} and \textsc{1LPT} forward models. In particular, these correspond to a noisy, but unbiased tracer of the matter field itself, given that $\bdelta=1$ with all higher-order bias coefficients set to zero. We demonstrate in {\refsec{res_2lpt2d_lpt2d_fwd_models_matter_field}} that we exactly recover the fiducial values of parameters in case of \textsc{2LPT} and that we also recover the expected shifts of parameters in the case of the \textsc{1LPT} forward model. We calculate these shifts analytically in \refapp{relation_between_lpt2d_and_2lpt}.

Finally, the $\mathcal{D}^{\textsc{2LPT}}_{2}$ datasets contain nonzero higher-order bias coefficients.
These are the most realistic datasets considered here, in the sense that they contain complications due to both nonlinear gravity and nonlinear bias. Since the calculation of the running of the bias parameters with cutoff $\Lambda$ is substantially more involved in this case, we instead validate our field-level forward model based on the inference of the $\alpha$ parameter for this case. If the EFT likelihood is able to correctly absorb the effect of modes above the cutoff, then it should lead to an unbiased inference of $\alpha$ for all values of $\Lambda$.

\section{Consistency test results}
\label{sec:analysis}

In this section, we describe the analysis procedure of our MCMC chains. Since the dimensionality of our posterior is exceptionally high, fully characterizing it is challenging {\cite{feng2018exploring}}, even with HMC sampling. To ensure that our samples fairly represent the true posteriors, we strictly adopt the following setup and procedure:

\begin{itemize}

    \item We run MCMC chains with both free initial conditions (\freeIC) and initial conditions fixed to the ground truth (\fixedIC). Since it is expected that the posterior of the {\fixedIC} case is within the typical set of the {\freeIC} posterior, in case of no strong multi-modality, it serves as a good reference point. Indeed, we find a good agreement between {the joint posteriors of} these two cases {(see also the next bullet point)}, suggesting no strong multi-modality is present in the tests we consider in this paper.
    
    \item For {\freeIC} runs, we run at least three chains: two starting from randomized values of initial conditions and sampled parameters, and one more chain starting from the ground-truth. {The latter serves as an additional check on multimodality of the posterior, and of the convergence of our chains.}The remaining parameters differ among the forward models we consider here and we always indicate which parameters are actually sampled.
          
    \item In our analysis, we discard the initial part of each chain, which is typically $5-10$ correlation lengths long (see \refapp{results_convergence_tests} on how we obtain the correlation lengths). Throughout, for each reported inference, if we run more than one MCMC chain as described in the previous point, we combine the chains into one single set of posterior contours. The consistency between different chains being combined is verified with the Gelman-Rubin statistics described in the next point.
    	
    \item We evaluate the Gelman-Rubin statistics \cite{gelman1992inference, gelman1995bayesian, vats2021revisiting} for our MCMC chains as described in \refapp{results_convergence_tests}. In doing so, we also quantify the (combined) effective sample size. The results for both {the Gelman-Rubin statistics and the effective sample size}, for all our chains, are listed in \reftab{gelman_rubin_stats_and_ess_unmarged}, \ref{tab:gelman_rubin_stats_and_ess_unmarged_part2} and \ref{tab:gelman_rubin_stats_and_ess_marged}. We require all of our chains to have $\gtrsim 100$ effective samples. This allows us to have the MCMC sampling error reduced to $\sim 10\%$, which is sufficient for the purposes of this paper.
\end{itemize}
	
It is also important to note that for the {\freeIC} chains, we use different $k$-binned quantities in order to check their statistics and convergence. These are the power spectrum of $\shat$, the mean deviation from $\shat_{\rm{true}}$ and the corresponding power spectrum of this deviation. Additionally, we have verified that the convergence of individual $\shat$ modes is well represented by the $k$-binned quantities for the different bins. We also note that both the $k$-binned $\shat$ quantities and the individual $\shat$ modes converge much faster than other parameters of the model, namely $\alpha$, $\{b_O\}$ and $\sigma_\eps$. 

We will also compare the sampled $\shat$ posterior with analytical predictions. For the latter, we always first calculate the per-$k$-mode prediction, and then compute the $k$-bin average, which is then compared with the corresponding sampled posterior in the same $k$-bin.


\subsection{\textsc{linear} forward models}
\label{sec:res_linear_fwd_model}
In this section, we focus on the forward models described by \refeq{trivial_fwd_model} and \refeq{trivial_fwd_model_2nd_order_bias}. 

\subsubsection{Linear bias}
\label{sec:linear_b1_sigma_tests}
\begin{figure}[htbp!]
	\centering
	\includegraphics[width=.475\linewidth]{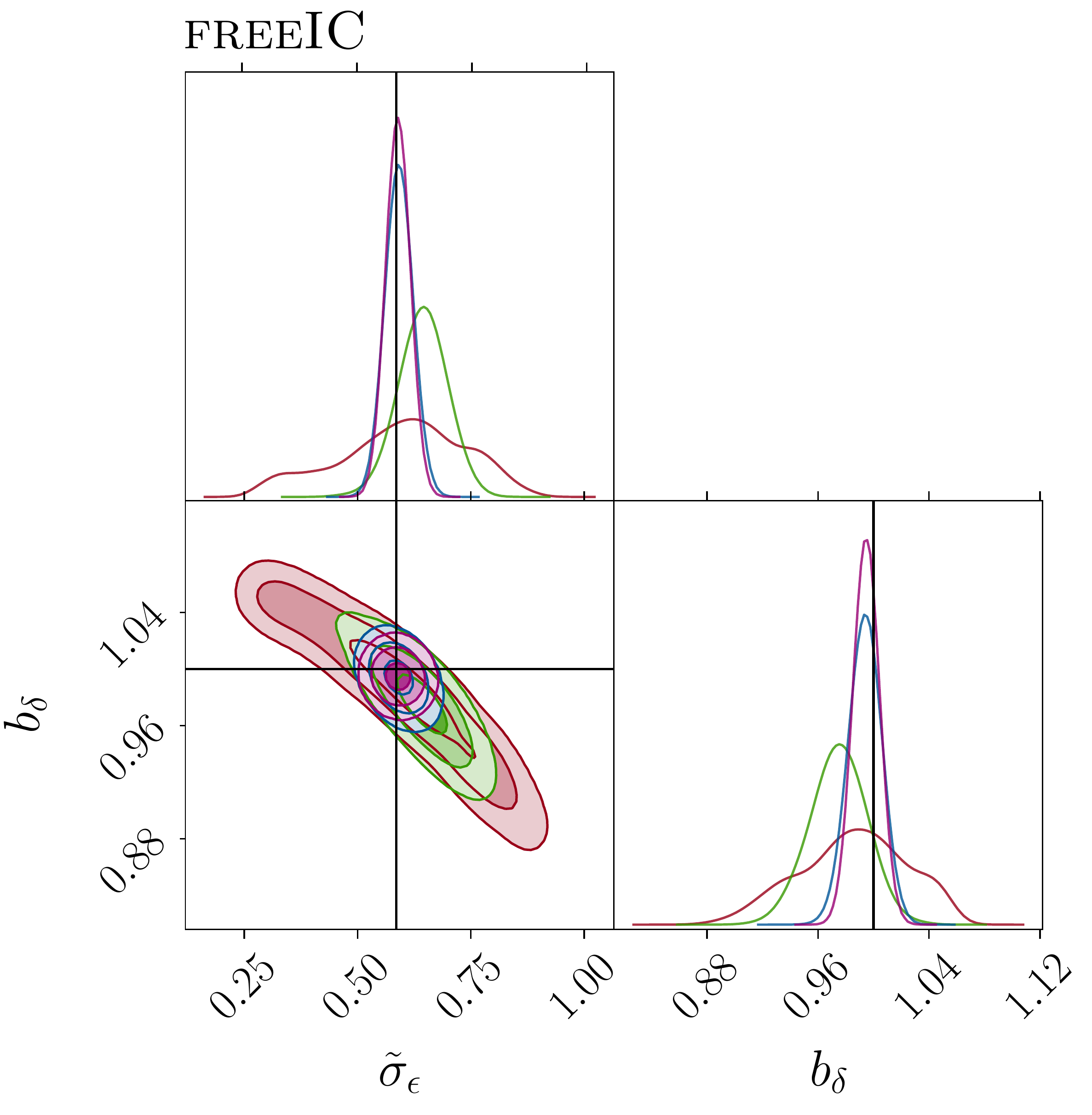}
	\includegraphics[width=.495\linewidth]{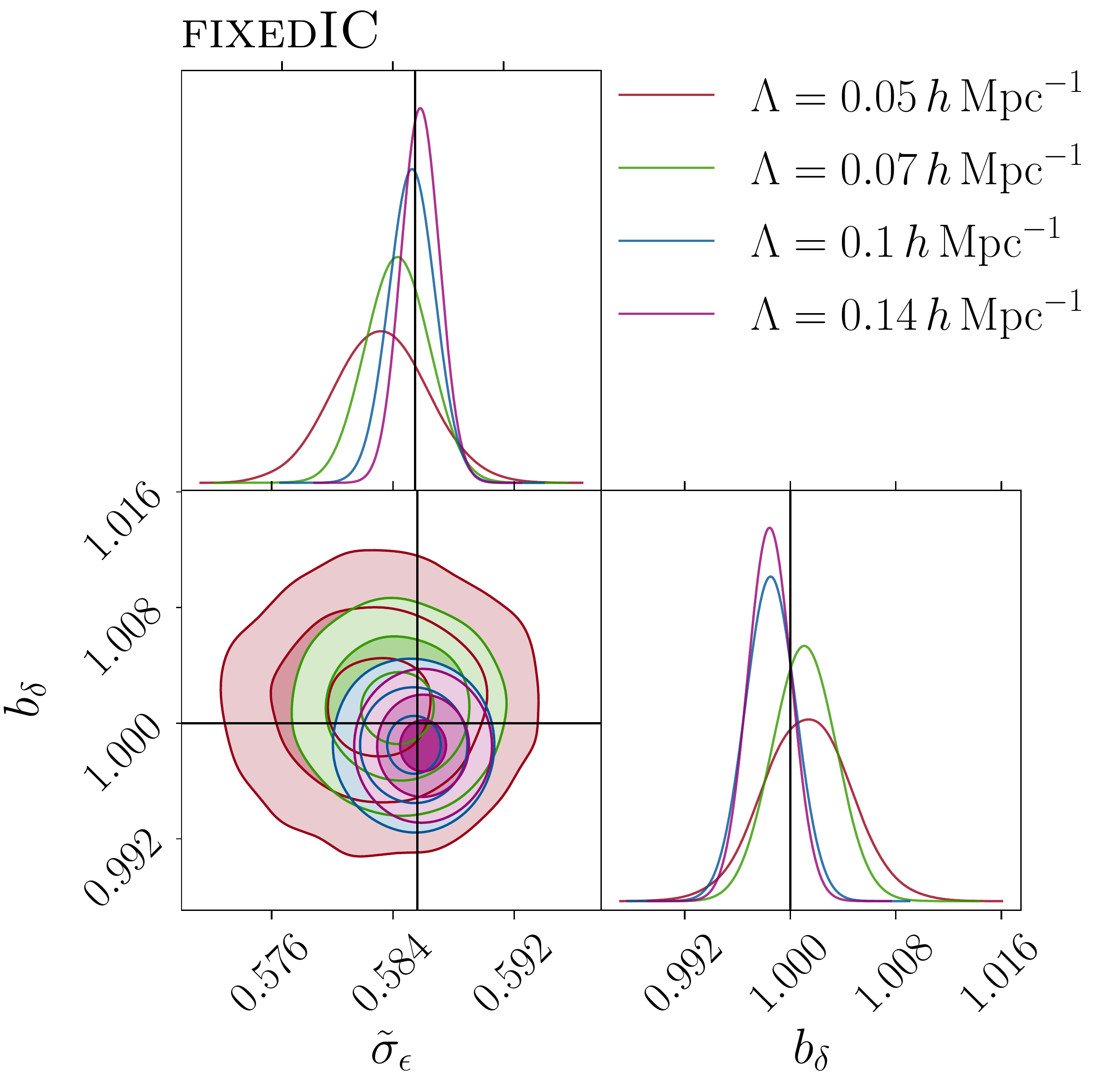}
	
	\caption{Inferred posterior of the parameters of the forward model from \refeq{trivial_fwd_model} on the $\mathcal{D}^{\textsc{linear}}_1$ synthetic dataset. On the left, the posterior projections to the $\tilde{\sigma}_{\epsilon}-\bdelta$ plane for the {\freeIC} case and different cutoffs are shown, while on the right we show the corresponding case of {\fixedIC} posteriors. {Note the difference in the axis ranges between the {\fixedIC} and {\freeIC} posteriors.} The black lines indicate expected values for both parameters. As can be seen, the {\freeIC} and {\fixedIC} posteriors agree with each other. Also, note the difference in the posterior uncertainties between forward models with different cutoffs.}
    \label{fig:res_trivial_fwd_model_b1_sigma_tests_params}
\end{figure}
First, we discuss the results of the forward model with linear bias expansion. In this case, it is possible to calculate the posterior of the initial conditions analytically. The comparison with the sampled posterior then verifies whether our sampling approach indeed fully explores the posterior in this case. As a first result, we focus on \reffig{res_trivial_fwd_model_b1_sigma_tests_params}. In this figure, we show the projection of the posterior to the {$\bdelta \,-\,\tilde{\sigma}_{\eps}$} plane. We distinguish two cases. The case where the posterior contours were obtained by fully marginalizing over the initial conditions ({\freeIC}), shown {in the left panel}, and the case where the initial conditions were fixed to the ground-truth ({\fixedIC}), shown in the right panel. The two panels indicate that the {\freeIC} and {\fixedIC} posterior means are consistent. Note the stark contrast in the posterior widths between the {{\fixedIC} and {\freeIC}} cases. This is explained by the fact that, in the case of {\fixedIC}, only two parameters need to be constrained, while in the case of {\freeIC} the joint posterior simultaneously constrains $\gtrsim 10^5$ degrees of freedom. More specifically, the free initial conditions also allow for an overall change in the amplitude, leading to the wider posteriors in $b_\d$. We also observe that the posterior contours shrink with increasing $\Lambda$, as expected. The degeneracy between the amplitudes of the signal ($\propto \bdelta$) and noise {($\propto \tilde{\sigma}_\eps$)} is harder to break at lower $\Lambda$ due to the shallower slope of the linear power spectrum, hence resulting in posterior uncertainties that grow faster toward smaller $\Lambda$ than expected merely from mode counting arguments; i.e., the error bar on $b_\delta$ grows toward smaller $\Lambda$ more rapidly than $\Lambda^{-3/2}$.
The stronger degeneracy in the {$\bdelta - \tilde{\sigma}_{\eps}$} plane also results in a slower exploration by the samplers, evidenced by a longer correlation length, which we do not explicitly show here for conciseness.

Another point to emphasize is that the correct fiducial noise level has also been recovered at {$39.3 - 86.5 \% \rm{CL}$} in all cases. This means that our inferences clearly disentangle between the actual signal and Gaussian noise contributions. {We emphasize that the contour lines in 2D posteriors always indicate $39.3\% - 86.5\% - 98.9\%\, \rm{CL}$ in that order, which corresponds to the $1-$, $2-$ and $3-\sigma$ levels respectively.} 
\begin{figure}[htbp!]
	\centering
	\includegraphics[width=\linewidth]{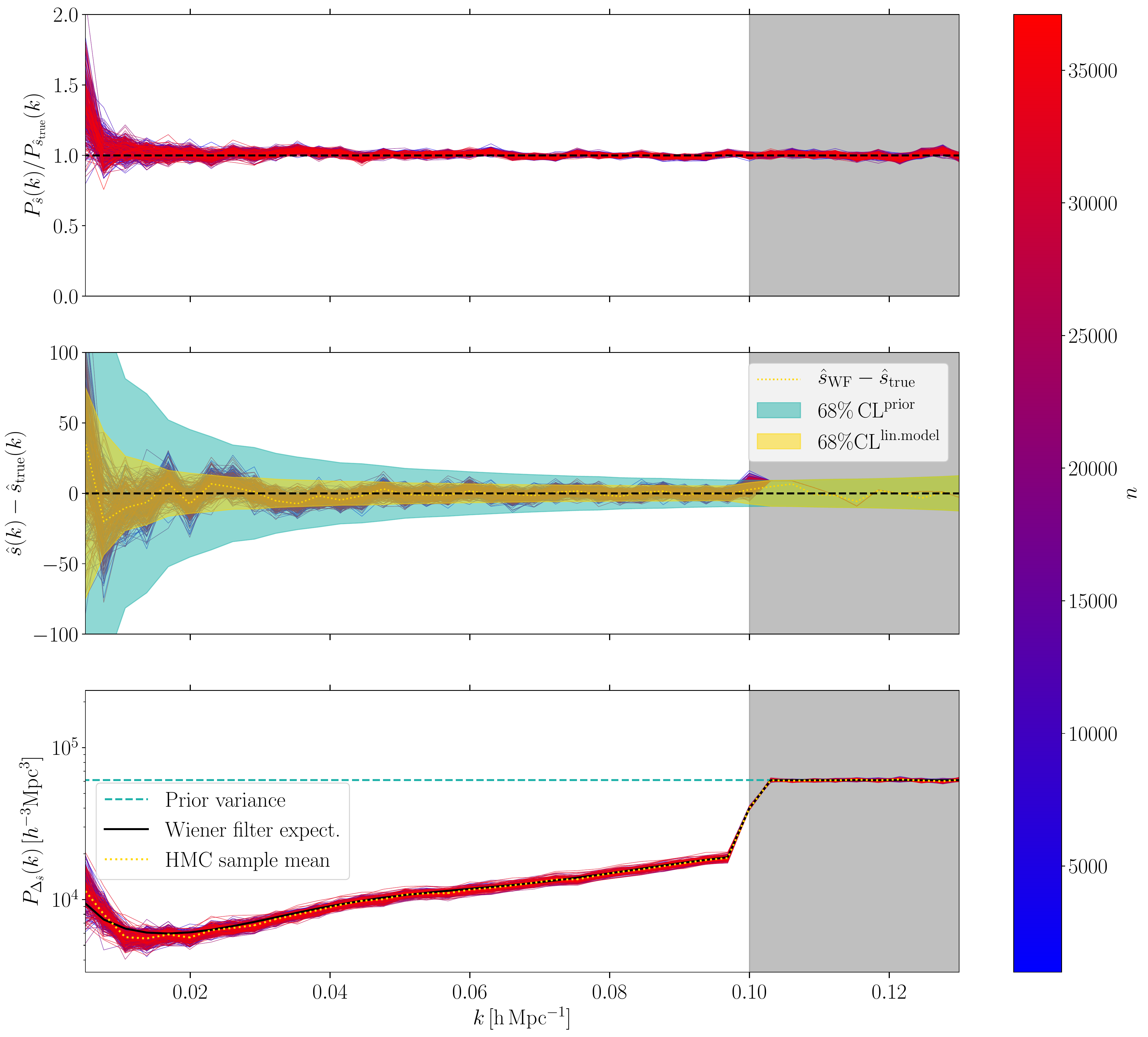}
	\caption{Samples of different binned statistics of the inferred $\shat$ field, using $\Lambda = 0.1\iMpch$, for the linear forward model from \refeq{trivial_fwd_model} on the $\mathcal{D}^{\textsc{linear}}_1$ synthetic dataset. {We keep the other model parameters $\{\bdelta, \tilde{\sigma}_{\epsilon}\}$ free as well, whose posteriors are shown on \reffig{res_trivial_fwd_model_b1_sigma_tests_params} (the $\Lambda=0.1 \iMpch$ contours)}. The gray {band} shows the modes with $k>\Lambda$, {while the color bar traces the sample number $n$}. The top panel shows the ratio of the sampled $\shat$ power spectra and the corresponding ground-truth $\shat_{\rm{true}}$ power spectrum. Clearly, $P_{\shat_{\rm true}}$ is well within the posterior. The middle panel shows the binned residuals, $\Delta_{\shat} = \shat - \shat_{\rm{true}}$, which are found to be centered around 0 as expected. Note that, above the cutoff, we have $\Delta_{\shat} \to - \< \shat_{\rm{true}} \>$, since for these modes, the sampled $\shat$ values follow the prior which is zero-centered. In addition, we also show the $68\% \, \rm{CL}$ intervals estimated from the Wiener-filter solution ($68\% \, \rm{CL}^{\rm{lin. model}}$) and the prior ($68\% \, \rm{CL}^{\rm{prior}}$). The bottom panel shows the power spectrum of the $\Delta_{\shat}$ field. Alongside this, we also show the Wiener filter expectation of the posterior $\shat$ covariance as well as the prior. As can be seen, the modes below the cutoff agree well with the Wiener filter prediction, while above the cutoff of $\Lambda=0.1\, \iMpch$ they follow the prior covariance. As discussed above \refeq{operator_summary}, the Wiener-filter solution is obtained by fixing $b_\delta$ and {$\tilde{\sigma}_{\eps}$} to their fiducial values.
    The corresponding trends are also found for other cutoff values $\Lambda$.}
	\label{fig:res_linear_model_b1_sigma_tests_phases}
\end{figure}

We now turn to investigating the posterior of the initial conditions $\shat$, {focusing in particular on whether the inference is able to recover the true initial conditions used in generating the synthetic data, $\shat_{\rm{true}}$}.
\reffig{res_linear_model_b1_sigma_tests_phases} compares {the bin averaged quantities of the} inferred and true initial conditions as a function of Fourier wave number $k$. The inference was performed on the $\mathcal{D}^{\textsc{linear}}_1$ dataset, at the cutoff of $\Lambda=0.1 \iMpch$, {keeping the other model parameters $\{\bdelta, \tilde{\sigma}_{\epsilon}\}$ free}. The top panel of \reffig{res_linear_model_b1_sigma_tests_phases} shows the ratio between the power spectrum of $\shat$ and that of $\shat_{\rm{true}}$. The ratio is consistent with unity, indicating that the two fields agree in terms of power, or mean amplitude.

The middle panel of \reffig{res_linear_model_b1_sigma_tests_phases} depicts the $k$-bin {averaged} statistics of the residuals $\Delta_{\shat}(\vk) = (\shat-\shat_{\rm{true}})(\vk)$.
Indeed, their distribution is centered around 0, clearly implying that the bulk of the $\shat$ posterior closely {traces} the $\shat_{\rm{true}}$ field.

Note that we do not expect the posterior to be always centered around the $\shat_{\rm{true}}$ field, but that $\shat_{\rm{true}}$ is within the typical set. We can in fact be more precise. As we show in \refapp{shat_posterior}, the $\shat$ posterior mean and covariance for the linear model considered here is, \emph{in the case when the parameters $b_\d$ and $\tilde{\sigma}_\eps$ are fixed to their fiducial values}, given by
\ba
	\shat_{\mathrm{WF}}	&= C_{\rm{WF}} R^{T}C_{\eps}^{-1}\d_{d,\Lambda}, \\
	C_{\rm{WF}} &= (1+SR^\dagger C_{\eps}^{-1}R)^{-1}S,
	\numberthis
	\label{eqn:res_linear_model_b1_sigma_tests_posterior_mean}
\ea
with
\ba
\resp{\vk_1}{\vk_2} 
&= 
{\delta_D}^{\vk_1, \vk_2} \bdelta T(k_1) \\ 
(C_{\eps})^{\vk_1}_{\,\,\, \vk_2} &= {\delta_D}^{\vk_1, \vk_2}{P_{\eps}} \\
{S^{\vk_1}}_{\vk_2} &= {\delta_D}^{\vk_1, \vk_2} \left(N_g^{\Lambda}\right)^3,
\numberthis
\label{eqn:operator_summary}
\ea
and $\d_{d,\Lambda}$ the reduced density field of the $\mathcal{D}^{\textsc{linear}}_1$ synthetic dataset. As seen from \refeq{res_linear_model_b1_sigma_tests_posterior_mean}, the $\shat$ posterior exhibits two limits. First, in the limit of uninformative data, i.e.\ large noise $C_\eps$, the posterior approaches to the prior, and {the posterior mean of $\shat$ approaches zero} while $C_{\rm{WF}}\to S$. Second, in the limit of very informative data, i.e.\ small noise $C_\eps$, the posterior mean and covariance approach $\shat \to R^{-1}\d_{d,\Lambda} \to \shat_{\mathrm{true}}$, following \refeq{trivial_fwd_model}, and $C_{\rm{WF}}\to (S R^\dagger C_\eps^{-1}R)^{-1}S\to 0$, respectively.

The yellow dotted line in the middle panel of \reffig{res_linear_model_b1_sigma_tests_phases} represents the residual between the Wiener-filter solution $\shat_{\mathrm{WF}}$, i.e. the linear model analytical prediction from above, and $\shat_{\rm{true}}$. Clearly, the sampled posterior is precisely centered around $\shat_{\rm WF}$, indicating that it is unbiased {also in the case when $\bdelta$ and $\tilde\sigma_\eps$ are left free.}
Above the cutoff, indicated by the gray band, the sampled modes follow the prior, hence $\Delta_{\shat} \to -\< \shat_{\rm{true}} \> \to 0 $. However, due to the finite number of modes per $k$-bin, the calculated mean will not be zero exactly but vary around it within the prior bounds, which is indeed what we see.
\begin{figure}[htbp!]
	\centering
	\includegraphics[width=.8\linewidth]{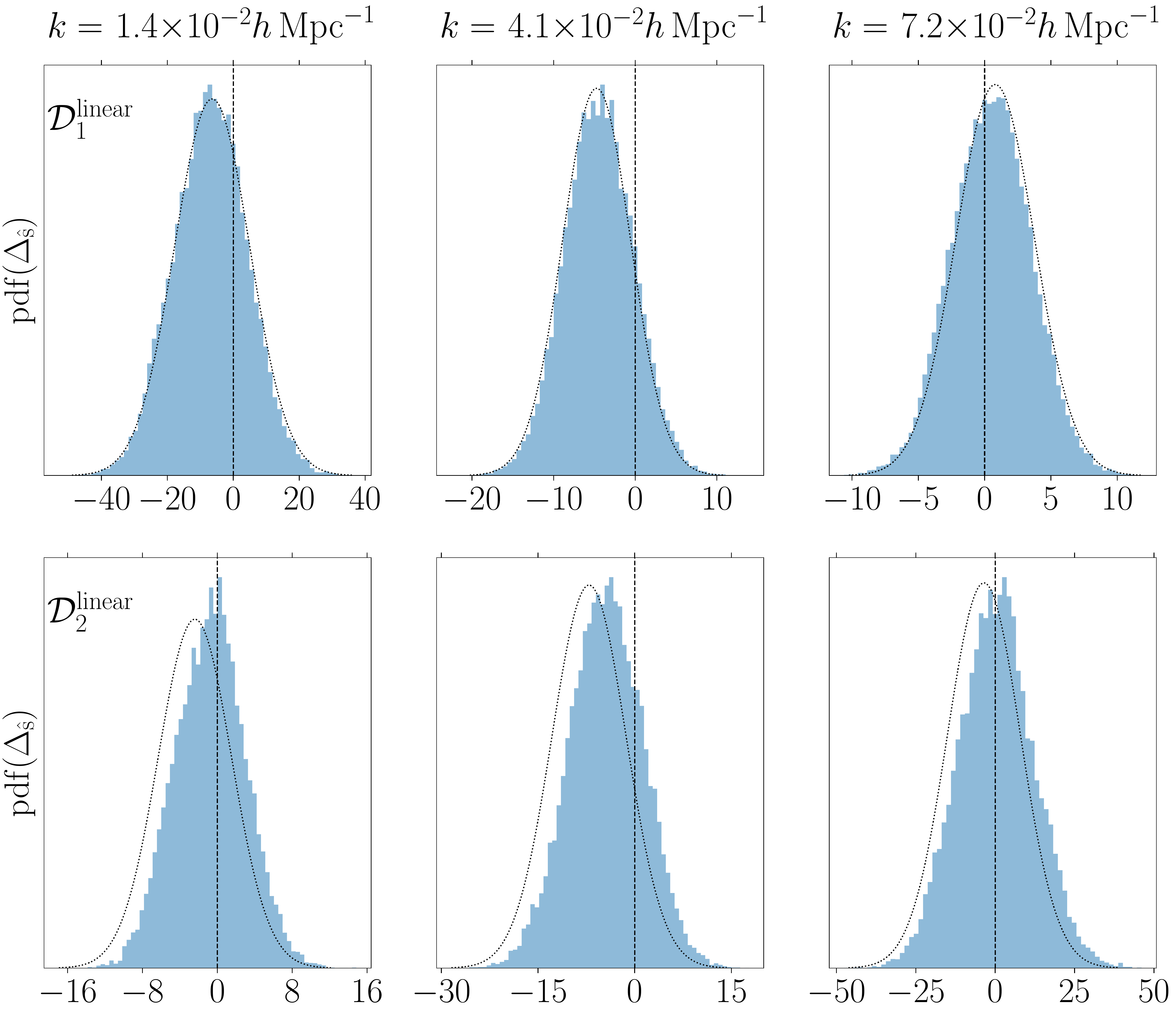}
	\caption{A closer look into the binned statistics of $\Delta_{\shat}$. The units on the $x$-axis correspond to the units of the $\shat(\vk)$ field, {which is} dimensionless based on our discrete Fourier convention. The $k$-bin centers are indicated at the top and are the same for both figures in each column. The dotted line indicates the predicted Wiener filter posterior for the statistics of the initial conditions, while the histogram represents the posterior obtained through sampling. For the top row, the setup is identical to that of \reffig{res_linear_model_b1_sigma_tests_phases}, {i.e. it shows the posterior of the statistics of $\Delta_{\shat}$ for the case of forward model from \refeq{trivial_fwd_model} applied to} the $\mathcal{D}^{\rm linear}_1$ dataset. As can be seen, the sampled posterior follows the Wiener filter prediction closely for all {considered} $k$ bins. The bottom panel shows the corresponding results when applying the forward model with second-order bias, {from \refeq{trivial_fwd_model_2nd_order_bias}}, to $\mathcal{D}^{\rm{linear}}_2$. Here the deviation from the Gaussian posterior of the Wiener filter prediction is prominent, and the latter is generally more biased than the sampled posterior. }
	\label{fig:res_linear_model_b1_sigma_tests_phases_pdf_of_Deltashat}
\end{figure}

Finally, the bottom panel of \reffig{res_linear_model_b1_sigma_tests_phases} shows the power spectrum of the $\Delta_{\shat}$ field. The theoretical expectation is that the power spectrum of $\Delta_{\shat}$ in a given $k$-bin  is $\Gamma$ distributed, $P_{\Delta\shat}\hookleftarrow\Gamma(a,b)$, where the shape parameter is $a=N_{\rm mode}/2$, $N_{\rm mode}$ being the number of modes within the Fourier space shell centered on $k$, and the scale parameter is $b=2 C_{\rm{WF}}(k)$. This conclusion follows from considering the distribution of a sum of squares of Gaussian-distributed variables, in this case, $\shat$. Since the mean of the $\Gamma$ distribution is given by the product of its scale and shape parameter, it follows immediately that the expectation value of $P_{\Delta_{\shat}}$ within a given $k$-bin is $C_{\rm{WF}}(k)$.

This prediction is shown as the black line in the bottom panel of \reffig{res_linear_model_b1_sigma_tests_phases}. We find good agreement with the sampled results, indicating that the posterior is fully explored by the sampler. We expect this to be the case, even though the Wiener filter calculation assumes fixed parameters. {Given that $\bdelta$ {and} $\tilde{\sigma}_\eps$ parameters are very well constrained, the propagated effect of their variance is a subdominant contribution to the $\shat$ posterior variance}.
We also show the $\shat$ prior covariance for comparison, and as we can see, modes above the cutoff indeed follow the prior.

To further investigate the posterior, in \reffig{res_linear_model_b1_sigma_tests_phases_pdf_of_Deltashat}, we plot, for three selected $k$-bins, both the histogram of $\Delta_{\shat}$ and the corresponding Wiener filter prediction, $\shat_{\rm{WF}} - \shat_{\rm{true}}$. The latter is, of course, Gaussian and plotted as the dotted line. To guide the eyes, we also show a vertical line on zero, indicating the ground truth. The top panel of this figure depicts the inferred $\shat$ residual statistics of the linear forward model from \refeq{trivial_fwd_model}.
In this case, the predicted (Wiener-filter) and sampled posteriors fully agree. 
On the other hand, the bottom panel shows the same, but for the forward model including second-order bias (\refeq{trivial_fwd_model_2nd_order_bias}). Here we see clear deviations between the analytical and sampled posteriors. The $\shat$ posterior for this simple but nonlinear forward model is not well approximated by the Wiener-filter solution (see \refsec{linear_b1_b2_sigma_tests} and \refapp{shat_posterior_quadratic} for more details). This highlights the importance of going beyond the Wiener filter approach when trying to extract information from even mildly nonlinear scales.
\begin{figure}[t]
	\centering
	\includegraphics[width=.495\linewidth]{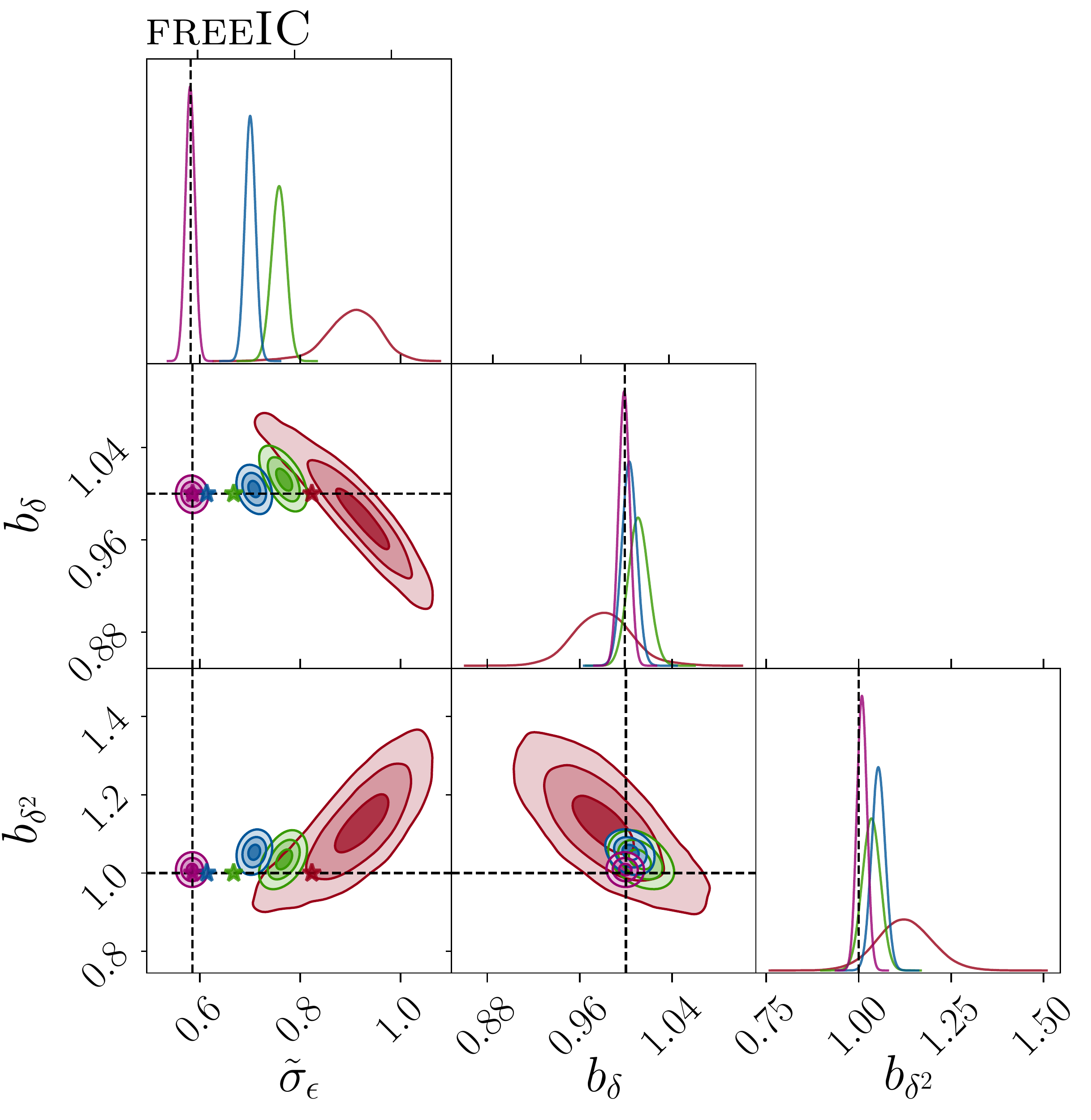}
	\includegraphics[width=.495\linewidth]{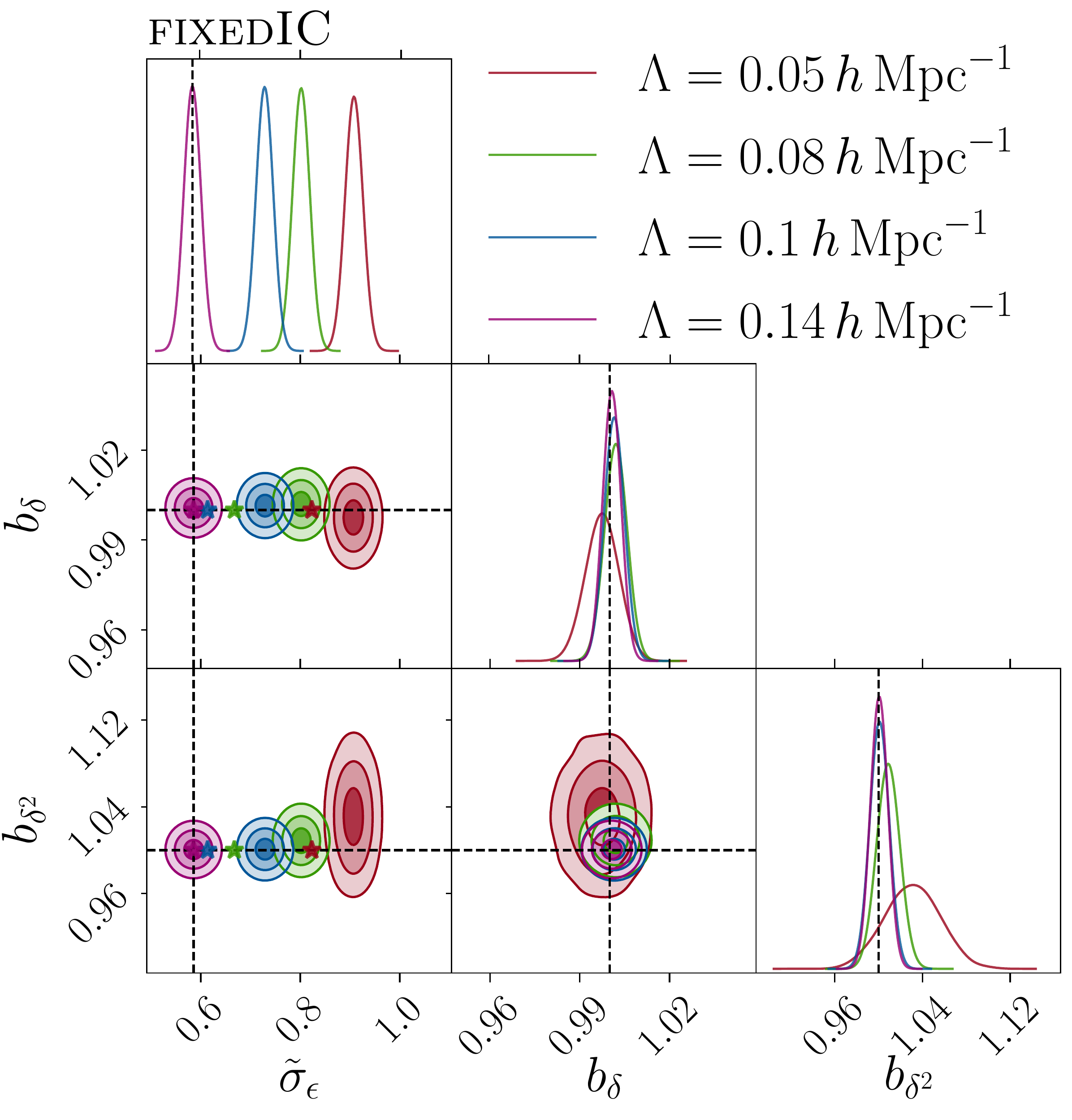}
	
        \caption{Same as \reffig{res_trivial_fwd_model_b1_sigma_tests_params} but for the forward model from \refeq{trivial_fwd_model_2nd_order_bias}. Again, the left panel shows the {\freeIC} case, while the right one shows the {\fixedIC} case. Good agreement is again found between the means of the two posteriors. {Note the difference in the inferred parameter ranges between the {\fixedIC} and {\freeIC} posteriors.} The dashed lines indicate the {fiducial} values for {$\{\bdelta, \bdeltatwo, \tilde{\sigma}_{\eps}\}$} {used for generating the {$\mathcal{D}_2^{\textsc{linear}}$} synthetic data}. Note that the {inferred} noise level increases towards smaller cutoffs for both the {\fixedIC} and {\freeIC} posteriors. {We further elaborate on this in the text and indicate the predicted running of $\tilde{\sigma}_{\eps}$ parameter from \refeq{delta_s_powerspec_correction} by the star symbols.}}
	\label{fig:res_trivial_fwd_model_b1_b2_sigma_tests_params}
\end{figure}

\subsubsection{Second-order bias}
\label{sec:linear_b1_b2_sigma_tests}

Next, we perform consistency tests of the EFT likelihood on the $\mathcal{D}^{\textsc{linear}}_2$ datasets. These synthetic datasets are generated using the forward model in \refeq{trivial_fwd_model_2nd_order_bias}. They additionally include a non-negligible quadratic bias contribution. Note that this contribution however still involves only the linearly evolved density field $\d^{(1)}_{\Lambda_0}$.
This results in a non-Gaussian posterior of the initial conditions, $\shat$, about which nonetheless we are able to make some qualitative analytical statements (see \refapp{shat_posterior_quadratic} and \refapp{b2_running}).

First, we analyze the posterior of the inferred parameters {$\{\bdelta, \bdeltatwo, \tilde{\sigma}_{\eps}\}$} shown in \reffig{res_trivial_fwd_model_b1_b2_sigma_tests_params}. As before, we consider forward models with different cutoffs $\Lambda$. Here, since the synthetic data $\mathcal{D}^{\textsc{linear}}_2$ is generated with a nonzero $\bdeltatwo$, it introduces mode couplings across the whole available range of modes up to the synthetic data cutoff of $\Lambda_0=0.14\, h\, \rm{Mpc}^{-1}$.

We first look at the noise amplitude $\tilde{\sigma}_{\eps}$. \reffig{res_trivial_fwd_model_b1_b2_sigma_tests_params} shows that the inferred value is a function of the cutoff for both {\fixedIC} and {\freeIC} cases; the inferred value is largest for the forward model with $\Lambda=0.05 \iMpch$ and lowest for $\Lambda=\Lambda_0= 0.14 \iMpch$. This can be understood as follows. The synthetic dataset $\mathcal{D}^{\textsc{linear}}_2$ is generated using the forward model from \refeq{trivial_fwd_model_2nd_order_bias}, but with a cutoff of $\Lambda_0 = 0.14 \iMpch$ (see also \reftab{synthetic_datasets}). We can then split the linear density field from which the synthetic dataset is constructed as
\be
\d^{(1)}_{\Lambda_0}(\vx) = \d^{(1)}_{\Lambda}(\vx) + \d^{(1)}_{s}(\vx),
\ee
where $\d^{(1)}_{\Lambda}$ and $\d^{(1)}_{s}$ represent the parts of $\d^{(1)}_{\Lambda_0}$ containing modes up to $\Lambda$, and from $\Lambda$ to $\Lambda_0$, respectively. Thus, $\d_{d,\Lambda}$ contains a contribution $b_{\d^2} (\d^{(1)}_{s})^2(\vx)$ (see \refeq{trivial_fwd_model_2nd_order_bias}). Since $\d^{(1)}_s$ is uncorrelated with $\d^{(1)}_\Lambda$, this contribution to the data corresponds to an additional noise that is absorbed in $P_\eps$ {during} the inference. We thus expect $P_\eps$ to shift by the power spectrum of $(\d^{(1)}_s)^2$, leading to
\ba
P_\eps(k) &= P_\eps^{{\rm no}\  b_{\d^2}} + 2 b_{\d^2} \int_{\vp} P_{\rm L}^{[\Lambda,\Lambda_0]}({p}) P_{\rm L}^{[\Lambda,\Lambda_0]}(|\vk-\vp|) \qquad
{(k < \Lambda)}
\numberthis 
\label{eqn:delta_s_powerspec_correction}\\
\mbox{where}\quad P_{\rm L}^{[\Lambda,\Lambda_0]}(p) &= W_{\Lambda_0}(p) [1 - W_{\Lambda}(p)] P_{\rm L}(p)
\ea
and $P_{\rm L}(p)$ is the linear power spectrum. Notice that only Fourier modes in the shell $[\Lambda,\Lambda_0]$ contribute. Evaluating this $P_\eps(k)$ at $k=\Lambda$ leads to {the results represented with stars in \reffig{res_trivial_fwd_model_b1_b2_sigma_tests_params}. Note that for the case of $\Lambda=\Lambda_0$ (purple), there is no running of $\tilde{\sigma}_{\eps}$ parameter and the corresponding star is right in the center of the posterior contours for this case.}
In general, we find that the analytical result predicts the right trend, although the shift in the sampled posterior mean is generally larger than the prediction, in particular for $\Lambda$ values that approach $\Lambda_0$. The most likely explanation is that the inferred $\tilde\sigma_\eps$ also has to absorb the scale-dependence of the induced noise, since \refeq{delta_s_powerspec_correction} has a significant $k$-dependence in particular if $\Lambda$ is not much smaller than $\Lambda_0$.

\begin{figure}[t]
	\centering
	\includegraphics[width=0.8\linewidth]{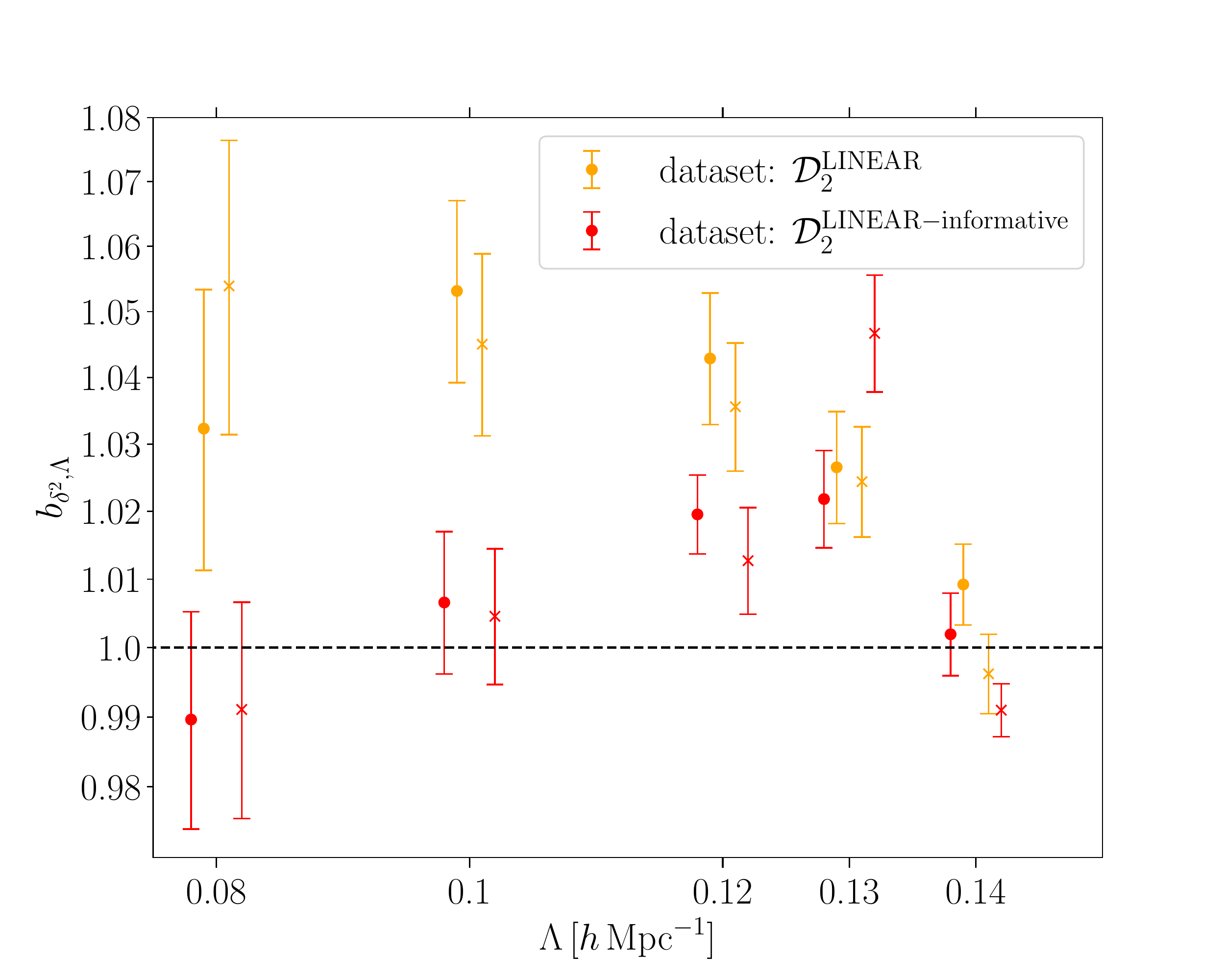}
	\caption{Mean a-posteriori second-order bias parameter {for} the forward model from \refeq{trivial_fwd_model_2nd_order_bias} inferred from different realizations (indicated by circle and {cross}) of the synthetic data $\mathcal{D}^{\textsc{linear}}_2$ (yellow). {The red color represents the inferred values of $\bdeltatwo$ on $\mathcal{D}^{\textsc{linear} - \text{informative}}_2$ dataset}. The error bars indicate $68\% \, \text{CL}$ intervals estimated from the chains.}
	\label{fig:res_linear_model_b1_b2_sigma_tests_b2_running_calcs}
\end{figure}

We now turn to the bias parameters. First, we expect no running of $\bdelta$ with the cutoff $\Lambda$, as the former only multiplies the linear density $\dLambda$ in the forward model \refeq{trivial_fwd_model_2nd_order_bias}. In other words, $b_{\d, \Lambda} = b_{\d,\Lambda_0}$ for all $\Lambda$. In fact, this argument can be made rigorous by examining the maximum likelihood point of the EFT likelihood (see for example Sec. 4 of \cite{paperI}). Note that the maximum likelihood argument assumes the initial conditions $\shat$ fixed to the ground truth $\shat_{\rm{true}}$. Strictly speaking, the argument applies only to the {\fixedIC} case. However, we generally expect the {\freeIC} posterior to overlap the {\fixedIC} one, and \reffig{res_trivial_fwd_model_b1_b2_sigma_tests_params} confirms that this is indeed the case.
\begin{figure}[t!]
	\centering
	\includegraphics[width=\linewidth]{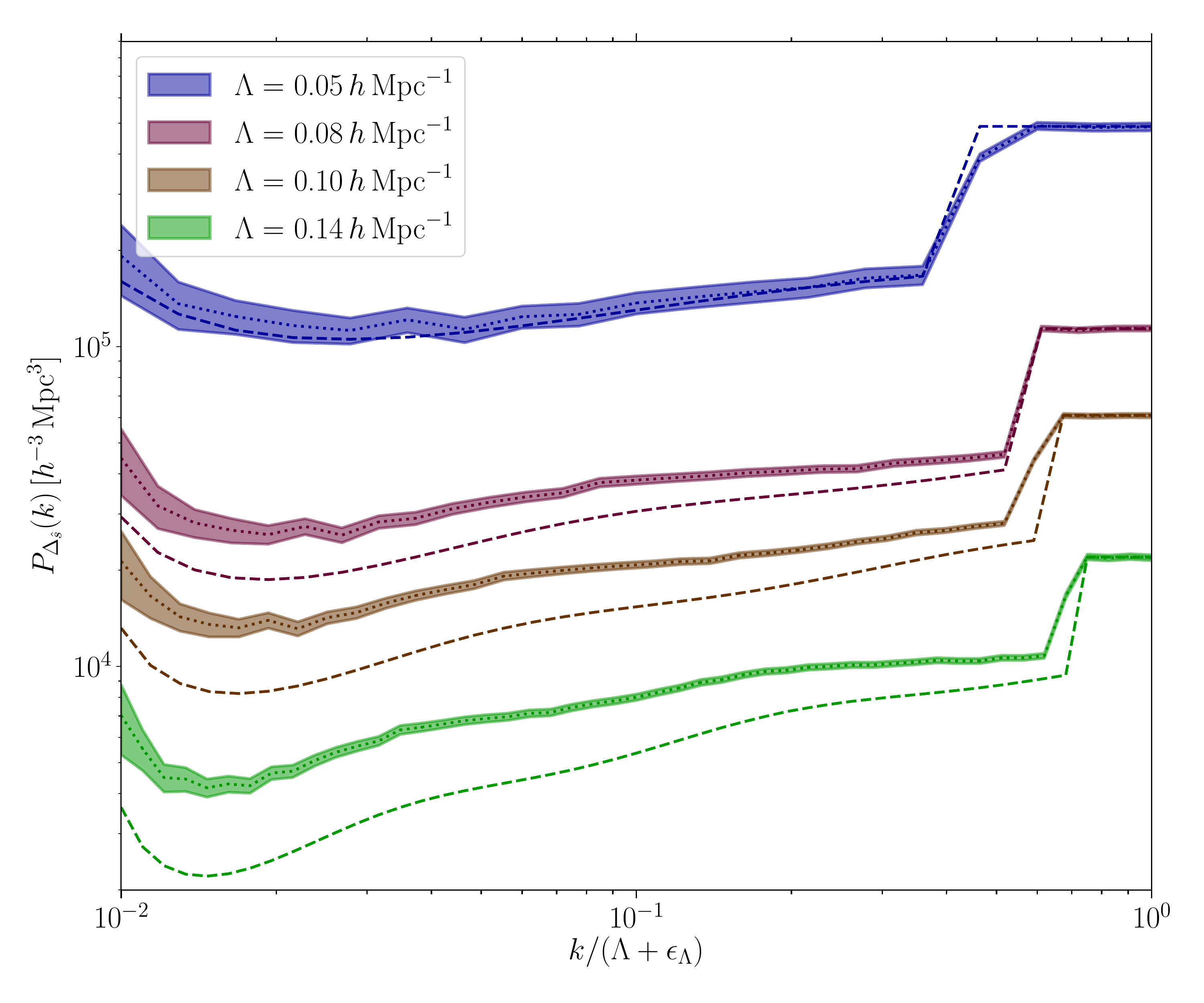}
	\caption{The power spectrum of $\Delta_{\shat}$ for MCMC chains using the forward model from \refeq{trivial_fwd_model_2nd_order_bias} for different cutoffs $\Lambda$, shown as a function of $k/(\Lambda + \eps_{\Lambda})$. We rescale the $x$-axis {for plotting purposes}, such that all the modes can be seen for all forward models with different cutoffs, suitably adjusting the $\eps_{\Lambda}$ parameter {for each cutoff $\Lambda$}. The inference here was performed on $\mathcal{D}^{\textsc{linear}}_{2,b}$ (inferences on the $\mathcal{D}^{\textsc{linear}}_{2,a}$ dataset yield consistent results). The shaded regions represent the mean and $68\% \, \rm{CL}$ of the sampled posterior. In addition, we also show the Wiener filter prediction, assuming a fully linear forward model. This approximation deteriorates toward higher cutoffs at fixed $k/\Lambda$, both because the mode coupling in the forward model becomes more important, and because $\shat$ is better constrained (see also \refapp{shat_posterior_quadratic}). 
	}
	\label{fig:res_linear_model_b1_b2_sigma_tests_phases}
\end{figure}

{We derive the running of the $\bdeltatwo$ parameter with the cutoff $\Lambda$ in \refapp{b2_running}} and show that it vanishes as well; i.e., ${b}_{\d^2, \Lambda} = {b}_{\d^2,\Lambda_0}$. We plot the results for $\bdeltatwo$ for the different {\freeIC} chains in \reffig{res_linear_model_b1_b2_sigma_tests_b2_running_calcs}. There is a residual shift away from the expected value of $\bdeltatwo$ when considering synthetic datasets with {$\tilde{\sigma}_{\eps} = 0.586$}, as shown by the yellow data points. While we expect the model from \refeq{trivial_fwd_model_2nd_order_bias} to be more accurate toward lower $\Lambda$, we in fact observe that the shift in $b_{\d^2,\Lambda}$ with respect to the expected result increases as we lower the cutoff. The most plausible explanation for this is a prior volume effect resulting from the weakening constraint on $\bdeltatwo$ and the growing degeneracy with $\tilde\sigma_\eps$ toward lower $\Lambda$ (see \reffig{res_trivial_fwd_model_b1_b2_sigma_tests_params}). To confirm this, we performed an inference with substantially lower noise ($\tilde{\sigma}_{\eps} = 0.002$), indicated by the red points in \reffig{res_linear_model_b1_b2_sigma_tests_b2_running_calcs}. Indeed, the systematic shift is substantially reduced, showing that more informative datasets help with breaking the degeneracy.

To conclude, the expected values for the bias parameters $\{\bdelta, \bdeltatwo\}$ are precisely recovered at $1-$ to $2-\sigma$ levels by our {\freeIC} posteriors. The only stronger deviation occurs for forward models with cutoffs $0.1 \iMpch <\Lambda < \Lambda_0$, which persists even in the case of highly informative data (see \reffig{res_linear_model_b1_b2_sigma_tests_b2_running_calcs}).
It is possible that including the subleading, $k^2$ contribution to the noise, which is expected to become more important as $\Lambda$ approaches $\Lambda_0$, would help with this residual shift. 

Next, we focus on analyzing the $\shat$ posterior. As before, we look at the first and second moments of the posterior of initial conditions. However, now the Wiener filter prediction for the posterior mean is less accurate than for the case shown in \reffig{res_linear_model_b1_sigma_tests_phases} due to the presence of the quadratic term in the forward model. \reffig{res_linear_model_b1_b2_sigma_tests_phases} compares the estimated $68\% \, \rm{CL}$ from the chain samples (dotted lines, with bands indicating sample variance) with the Wiener-filter expectation for the variance (dashed lines). As expected, the deviation from the Wiener-filter solution is stronger, at fixed $k/\Lambda$, as we go toward higher cutoffs. This is both because the typical amplitude of density fluctuations grows on smaller scales, and the uncertainty on $\shat$ shrinks. 

We can in fact make some qualitative statements about the behavior seen in \reffig{res_linear_model_b1_b2_sigma_tests_phases}. In the case of the quadratic bias forward model, the posterior {of initial conditions} contains terms proportional to $\sim \shat^3$ and $\sim \shat^4$, in addition to the terms $\sim \shat$ and $\sim \shat^2$ present in the purely linear case. The $\shat$ posterior covariance depends on all of these terms, as demonstrated in \refapp{shat_posterior_quadratic}. There, we consider under what conditions the posterior for $\shat$ can be approximated analytically and describe the cause for the discrepancy between the Wiener filter prediction and the sampled joint posterior.

\begin{figure}[ht!]
    \centering
    \begin{minipage}{\textwidth}
       \includegraphics[width=1\linewidth]{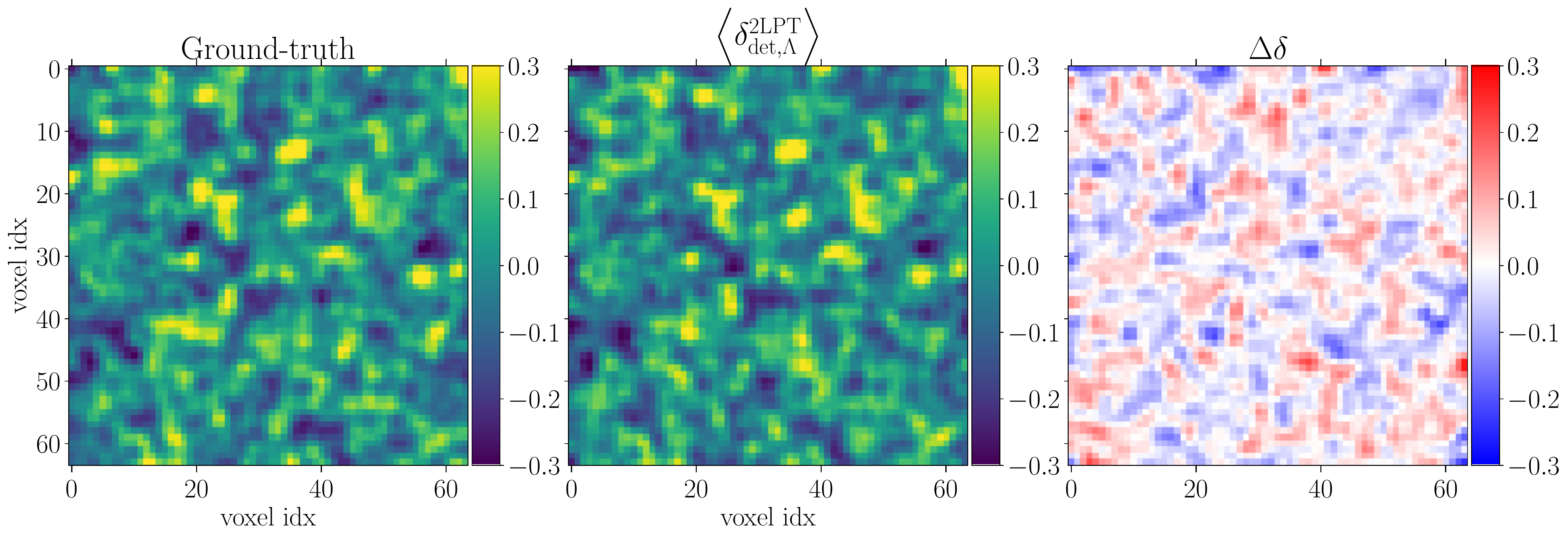}
       \caption*{$\Lambda=0.1\iMpch$}  \label{fig:real_space_slice_Lambda_0.1_linearly_biased_LPT_dataset} 
    \end{minipage}\\
    \begin{minipage}{\textwidth}
       \includegraphics[width=1\linewidth]{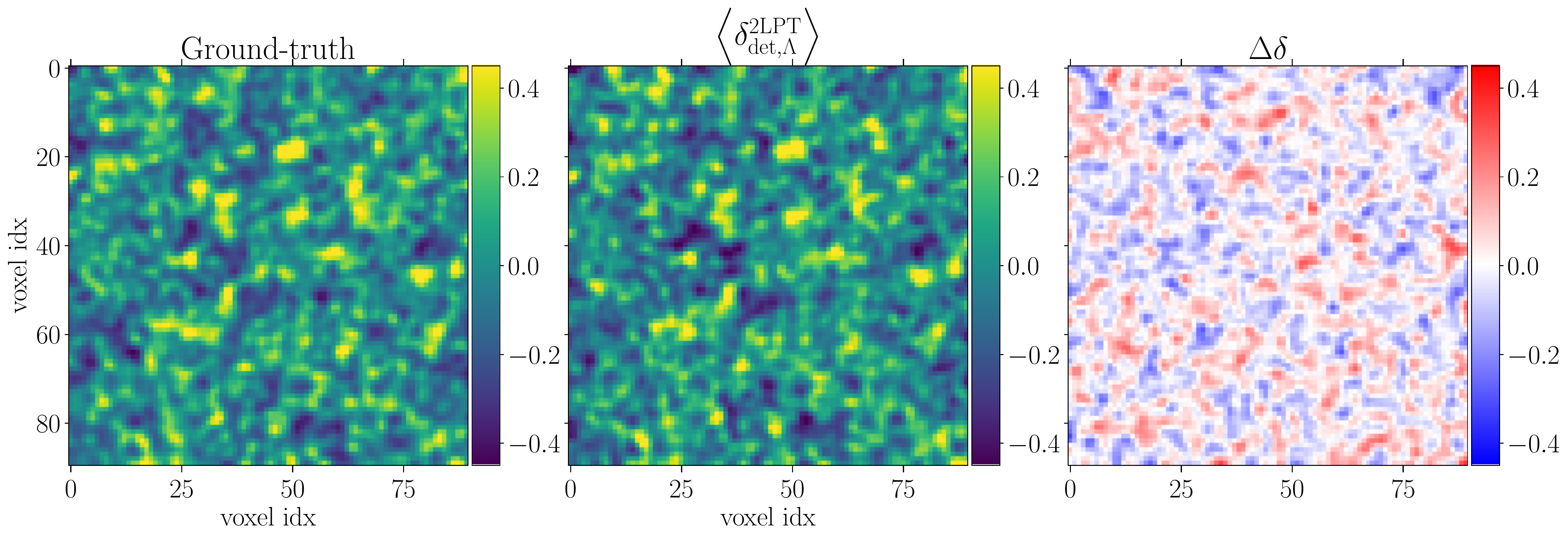}
       \caption*{$\Lambda=0.14\iMpch$}       \label{fig:real_space_slice_Lambda_0.14_linearly_biased_LPT_dataset}
    \end{minipage}
    \caption{\label{fig:real_space_slices_linearly_biased_LPT_dataset}{The comparison of the ground-truth signal within the $\mathcal{D}_{1,a}^{\textsc{2LPT}}$ dataset (left), the posterior mean of the inference performed with the forward model from \refeq{fwd_model_2lpt2d} (middle), and the corresponding residuals (right) for a selected 2D slice through the box. We present the results for $\Lambda=0.1\iMpch$ (upper panel) and $\Lambda=0.14\iMpch$ lower panel. The results were additionally smoothed with a Gaussian kernel of size $5 \mathrm{px}$ along each axis for aesthetic reasons. The posterior mean follows well the structure present in the ground-truth signal, especially in the regions of high overdensities, which is also shown by the residuals being very small at those locations.}}
\end{figure}
\subsection{\textsc{1LPT} and \textsc{2LPT} forward models}
\label{sec:res_2lpt2d_lpt2d_fwd_models}
\begin{figure}[htbp!]
	\centering
	\includegraphics[width=\linewidth]{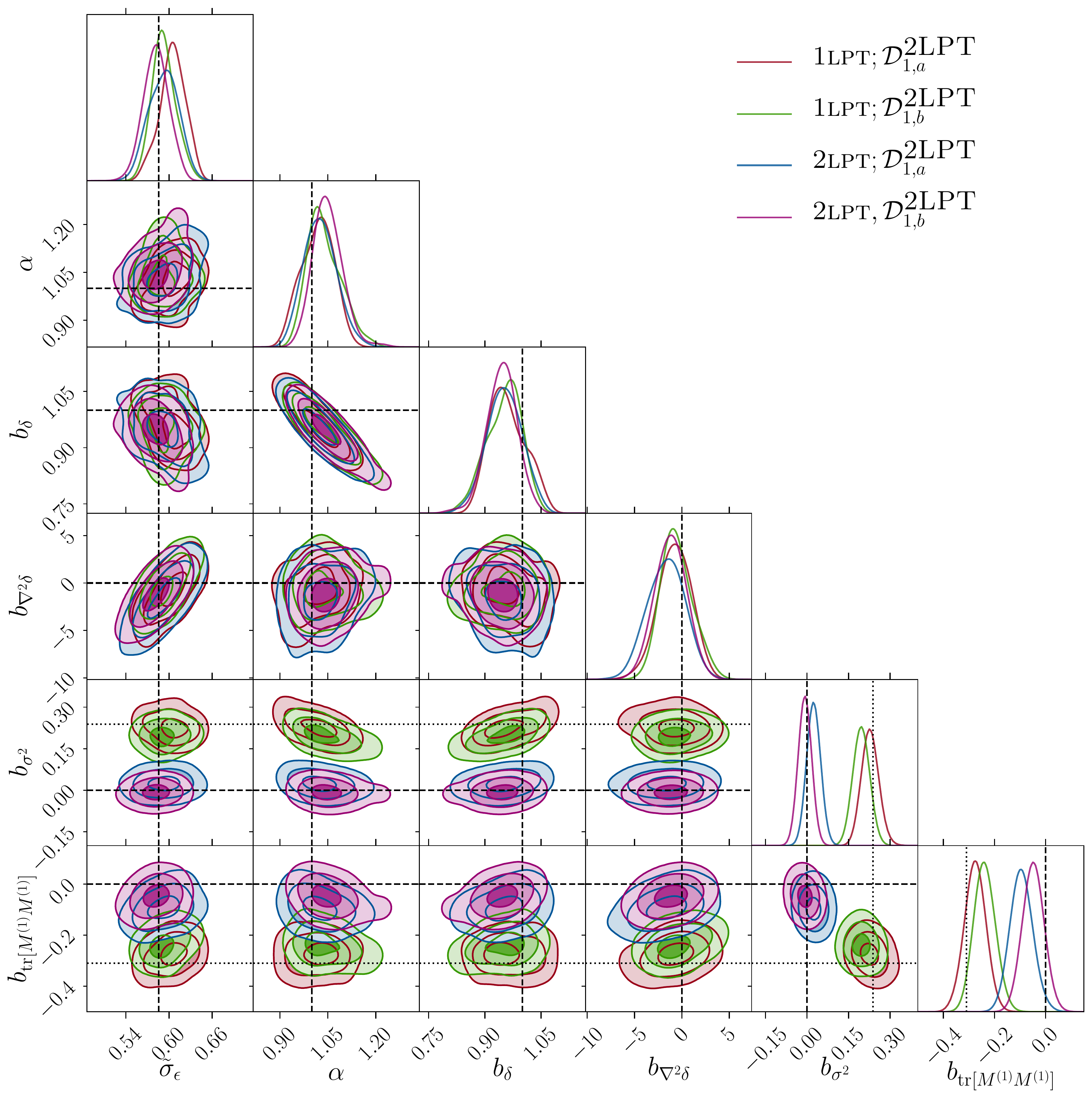}
	\caption{{\freeIC} parameter posteriors for the \textsc{2LPT} and \textsc{1LPT} {inferences}. The synthetic datasets used for this inference are $\mathcal{D}^{\textsc{2LPT}}_{1,a}$ and $\mathcal{D}^{\textsc{2LPT}}_{1,b}$, both generated at $\Lambda_0 = 0.14\iMpch$. The dashed lines indicate the fiducial input parameters of the synthetic datasets. The blue and dark-purple contours show \textsc{2LPT} inferences, while the red and green contours show the same for the \textsc{1LPT} {inferences}, each for the two independent datasets. The cutoff used for all the forward models is $\Lambda=0.1\iMpch$. Note the positive correlation in the $\blapldelta - \tilde{\sigma}_{\eps}$ plane and negative correlation in the $\bdelta - \alpha$ plane, as well as the shifts away from zero in the higher-order bias parameters $\bsigmasigma$ and $\btrMoneMone$ in cases of the \textsc{1LPT} inferences. The dotted lines indicate the expected values of $\bsigmasigma$ and $\btrMoneMone$ from Lagrangian perturbation theory (see {\refeq{res_1lpt_vs_2lpt_relation_bias_coeffs}} and \refapp{relation_between_lpt2d_and_2lpt} for more details).}
	\label{fig:res_lpt2D_vs_2lpt2D_model_2lpt_tests_params}
\end{figure}

We now turn toward the forward models involving nonlinear gravity, i.e.\ \textsc{1LPT} and \textsc{2LPT}. The forward models we consider in this section are the most realistic ones, and allow for reducing the degeneracy between the bias parameters and the scaling parameter, $\alpha$, as we elaborate below. Furthermore, as already hinted in \cite{paperI}, these forward models promise information gains beyond the leading-, next-to-leading order power spectrum, and leading-order bispectrum. We leave it for future work to explicitly demonstrate this. Instead, below we focus on the performance of these forward models on synthetic datasets listed in \refsec{2lpt_model_tests}.

\subsubsection{Linearly biased case}
\label{sec:res_2lpt2d_lpt2d_fwd_models_matter_field}

The synthetic dataset used in this section is $\mathcal{D}^{\textsc{2LPT}}_1$, with different realizations denoted by $\mathcal{D}^{\textsc{2LPT}}_{1,a}$ and $\mathcal{D}^{\textsc{2LPT}}_{1,b}$. These correspond to a linearly biased tracer of the 2LPT-evolved matter density field, and we thus test the consequences of a mismatch in the nonlinear matter forward model. {For the $n\textsc{lpt}$ forward models considered here}, the exact degeneracy between  $\bdelta$ and $\alpha$ that is present for trivial linear evolution is broken, as $\d_{n\rm LPT}$ contains terms scaling as $\propto \alpha$ and $\propto \alpha^2$ both multiplied by the same $\bdelta$ (see \cite{paperII,paperIIb} for more discussion).

{In \reffig{real_space_slices_linearly_biased_LPT_dataset} we show how well the posterior mean of the inferred $\d_{\mathrm{det},\Lambda}$ field in the {\freeIC} case with randomized initial conditions taken as the starting point compares to the ground-truth signal used for generating the $\mathcal{D}_{1,a}^{\textsc{2LPT}}$ dataset. We show results for both cutoffs considered: $\Lambda=0.1\iMpch$ (upper panel of the figure) and $\Lambda=0.14\iMpch$ (lower panel of the figure). For obtaining the posterior mean of $\d_{\mathrm{det},\Lambda}$, the last $2000$ samples of the inference chain were taken. The final results were additionally smoothed with a Gaussian kernel {of size $5 \mathrm{px}$ along each axis}, for aesthetic reasons. Overall, the reconstructed field in the selected slice matches well the ground-truth signal. In the regions with highest density peaks, the residuals are quite small as expected, however due to the stochastic nature of our forward model, the reconstructed and ground-truth field don't match exactly. Once more we emphasize that this result has been obtained by joint sampling of the initial conditions, $\alpha$ and bias parameters as well as the noise parameter $\tilde{\sigma}_{\eps}$, and therefore represents a non-trivial result. We focus on the parameter posteriors next.}

\reffig{res_lpt2D_vs_2lpt2D_model_2lpt_tests_params} shows the parameter posteriors after explicitly marginalizing over the posterior of initial conditions. Here we compare the inferences employing the \textsc{1LPT} (red and green contours) and \textsc{2LPT} (blue and dark-purple contours) forward models. The parameters $\alpha$, $b_\d$, and {$\tilde{\sigma}_{\eps}$}, which are not expected to run, agree well among the two different gravity models. {The running is not present since} $\alpha$ is a cosmological parameter, while $b_\d$ and $\tilde\sigma_\eps$ are protected from running thanks to the absence of nonlinear bias {in the synthetic datasets $\mathcal{D}^{\textsc{2LPT}}_1$}. Note however that $b_{\nabla^2\d}$ is expected to absorb the effect of modes between $\Lambda$ and $\Lambda_0$, hence to be shifted from its fiducial value. We also observe an expected anti-correlation in the $\bdelta - \alpha$ plane, given that these two parameters appear together as a product in the linear bias term in the forward model.
\begin{figure}[t!]
	\centering
	\includegraphics[width=.9\linewidth]{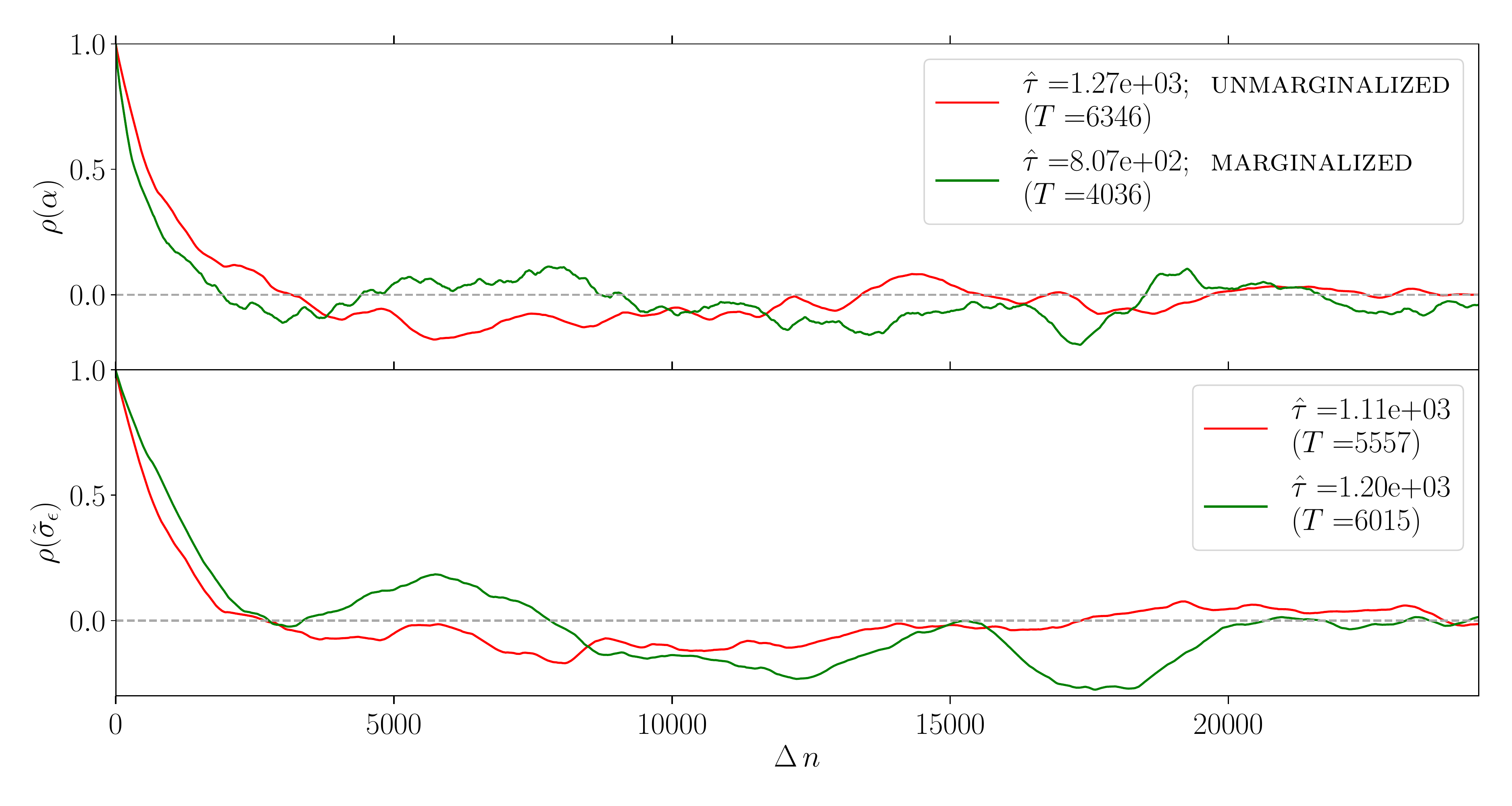}
	\caption{The normalized auto-correlation function (see \refeq{autocorr_rho}) for the \textsc{2LPT} forward model applied to the $\mathcal{D}_1^{\textsc{2LPT}}$ dataset for $\alpha$ (top panel) and $\tilde{\sigma}_{\eps}$ (bottom panel). {The $x$-axis shows the separation between samples in the chain, denoted with $\Delta n$. The labels in the legend show the estimated correlation length $\hat{\tau}$ and $T$, the corresponding maximal sample separation considered for making this estimate (see \refapp{results_convergence_tests} for more details).} In both cases, we compare the chains using the unmarginalized (red) and marginalized (green) likelihoods {(see \refeqs{unmarg_like}{marg_like})}. A faster decay of the auto-correlation function can be seen for the marginalized likelihood in case of $\alpha$, while for $\tilde{\sigma}_\eps$ the correlation lengths are comparable.}
	\label{fig:autocorr_marged_vs_unmarged}
\end{figure}

Another interesting degeneracy is in the $\blapldelta - \tilde{\sigma}_{\eps}$ plane, which shows positive correlation. This can be understood by recalling how these parameters affect the leading order observable, the power spectrum. 
The dominant contribution to the tracer power spectrum that contains $\blapldelta$ is $-2 k^2\blapldelta b_\delta P_L(k) \sim -k^{0.5}$ at $k \approx 0.1 \iMpch$, while the noise contribution scales as $\sim k^0$. Since these two contributions have a similar $k$ dependence, but opposite signs, they result into a positive correlation between the two corresponding parameters. 

{We also notice the difference between the $\textsc{1LPT}$ and $\textsc{2LPT}$ posteriors} in the bottom two rows of \reffig{res_lpt2D_vs_2lpt2D_model_2lpt_tests_params}. Namely, the higher-order bias coefficients inferred using the \textsc{1LPT} forward model are shifted away from their fiducial values of zero. The shifts of these bias coefficients can in fact be predicted using a second-order LPT calculation. Specifically, by solving for the displacement field and then substituting that back into the second-order bias expansion, one can derive the relations between the bias coefficients in the \textsc{1LPT} and \textsc{2LPT} forward model. Following the calculation done in \refapp{relation_between_lpt2d_and_2lpt}, one derives the following relations between the bias coefficients of the two forward models
\ba
\bdelta^{\textsc{1LPT}} &= \bdelta^{\textsc{2LPT}}, \\
\bsigmasigma^{\textsc{1LPT}} 
&= 
\bsigmasigma^{\textsc{2LPT}} + \frac{3}{14}, \\
\btrMoneMone^{\textsc{1LPT}} 
&= 
\btrMoneMone^{\textsc{2LPT}} - \frac{3}{14}.
\numberthis
\label{eqn:res_1lpt_vs_2lpt_relation_bias_coeffs}
\ea
These values are indicated with dotted lines in \reffig{res_lpt2D_vs_2lpt2D_model_2lpt_tests_params}, and are within $68-95\% \rm{CL}$ of the corresponding \textsc{1LPT} posteriors.

The results we have discussed so far were obtained using the likelihood from \refeq{unmarg_like}, i.e. the unmarginalized likelihood. Using the marginalized likelihood from \refeq{marg_like} gives entirely consistent results (see \refapp{unmarged_vs_marged_like}). However, the marginalized likelihood offers the important advantage of a reduced correlation length in the remaining parameters $\alpha, \tilde\sigma_\eps$. Namely, marginalizing over the bias parameters allows for a $\sim 60\%$ reduction in the correlation length of the $\alpha$ parameter (see the top panel of \reffig{autocorr_marged_vs_unmarged}).\footnote{Note that the correlation length was estimated by taking the average over three (two) independent chains for the unmarginalized (marginalized) likelihoods, respectively.}

This in turn means that for the same CPU time, the number of effective samples produced by the marginalized likelihood is correspondingly increased by a factor of 1.6. We expect the performance gain with the marginalized likelihood is more significant as more bias parameters appear in the model.
\begin{figure}[t!]
	\centering
	\includegraphics[width=.8\linewidth]{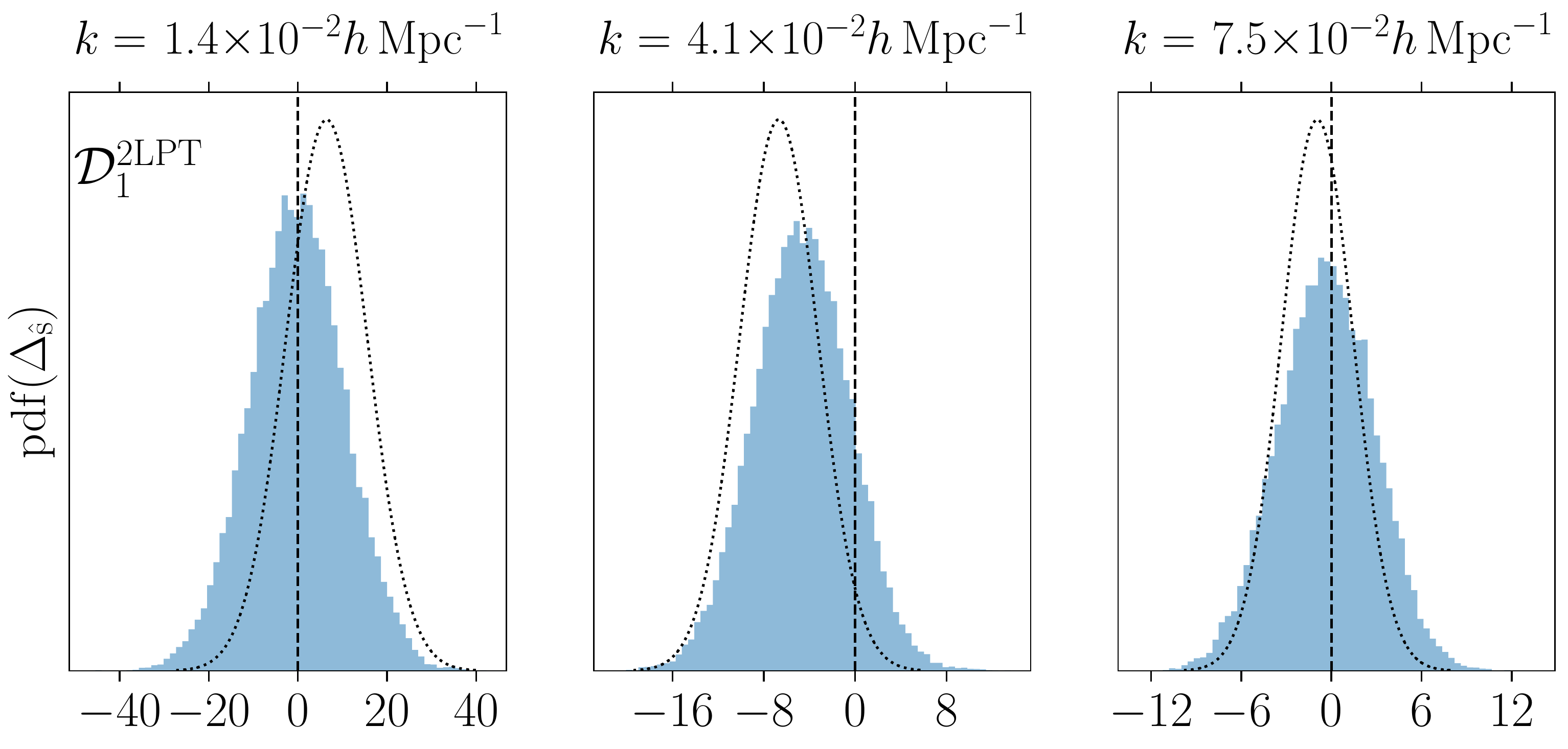}
	\caption{Same as \reffig{res_linear_model_b1_sigma_tests_phases_pdf_of_Deltashat} but for the \textsc{2LPT} forward model from \refeq{fwd_model_2lpt2d} applied to {the} $\mathcal{D}^{\rm{2LPT}}_1$ synthetic data. We can see that {the inferred} posterior {differs significantly} from the predicted Wiener-filter solution, and is closer to the ground truth, $\shat_{\rm{true}}$, as expected. Even on large scales, the deviation is significant. Note also that the distribution of $\shat$ becomes non-Gaussian, and increasingly so toward smaller scales.
    }
	\label{fig:res_2lpt2D_model_phases_pdf_of_Deltashat}
\end{figure}

As a final remark, we also show the posterior of initial conditions within different $k$-bins in \reffig{res_2lpt2D_model_phases_pdf_of_Deltashat}. As anticipated {already} in \reffig{res_linear_model_b1_sigma_tests_phases_pdf_of_Deltashat}, the posterior is indeed non-Gaussian, showing stronger deviations from the Gaussian case as one goes toward smaller scales (reflected in the heavier tails of the distribution). Furthermore, even on the largest scales covered by our simulated volume, the prediction from the Wiener filter is biased with respect to the inferred posterior which is correctly centered around the ground truth (see the left-most panel).
\begin{figure}[t!]
    \centering
    \begin{minipage}{\textwidth}
       \includegraphics[width=1\linewidth]{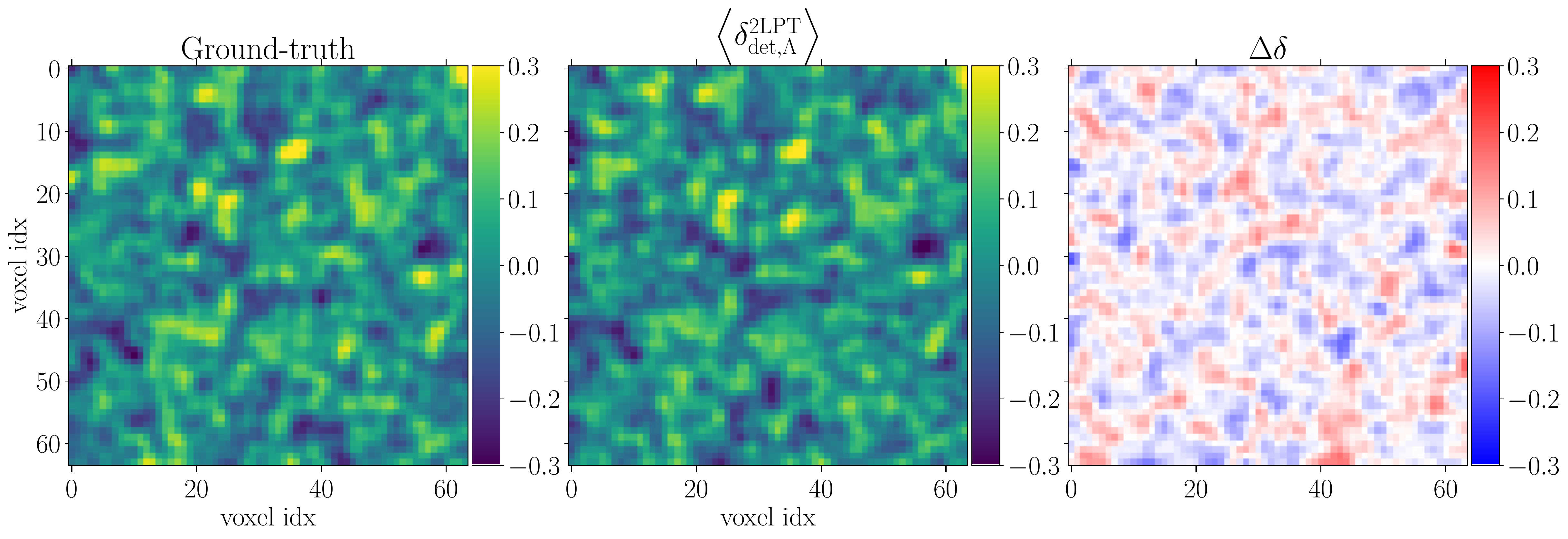}
       \caption*{$\Lambda=0.1\iMpch$}  \label{fig:real_space_slice_Lambda_0.1_higher_bias_nonzero_LPT_dataset} 
    \end{minipage}\\
    \begin{minipage}{\textwidth}
       \includegraphics[width=1\linewidth]{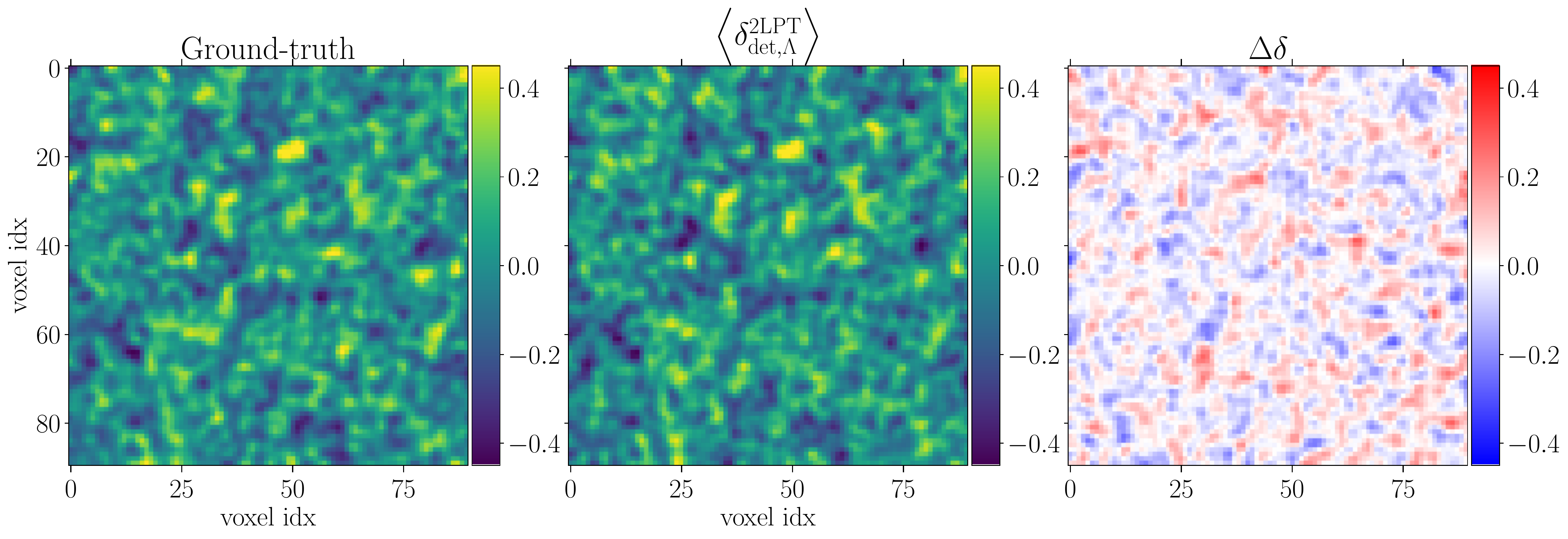}
       \caption*{$\Lambda=0.14\iMpch$}       \label{fig:real_space_slice_Lambda_0.14_higher_bias_nonzero_LPT_dataset}
    \end{minipage}
    \caption{Same as in \reffig{real_space_slices_linearly_biased_LPT_dataset} but for the case of $\mathcal{D}_{2,a}^{\textsc{2LPT}}$ dataset. \label{fig:real_space_slices_higher_bias_nonzero_LPT_dataset}}
\end{figure}

\subsubsection{Biased tracers}
\label{sec:res_2lpt2d_lpt2d_fwd_models_biased_tracers}
\begin{figure}[htbp!]
	\centering
	\includegraphics[width=\linewidth]{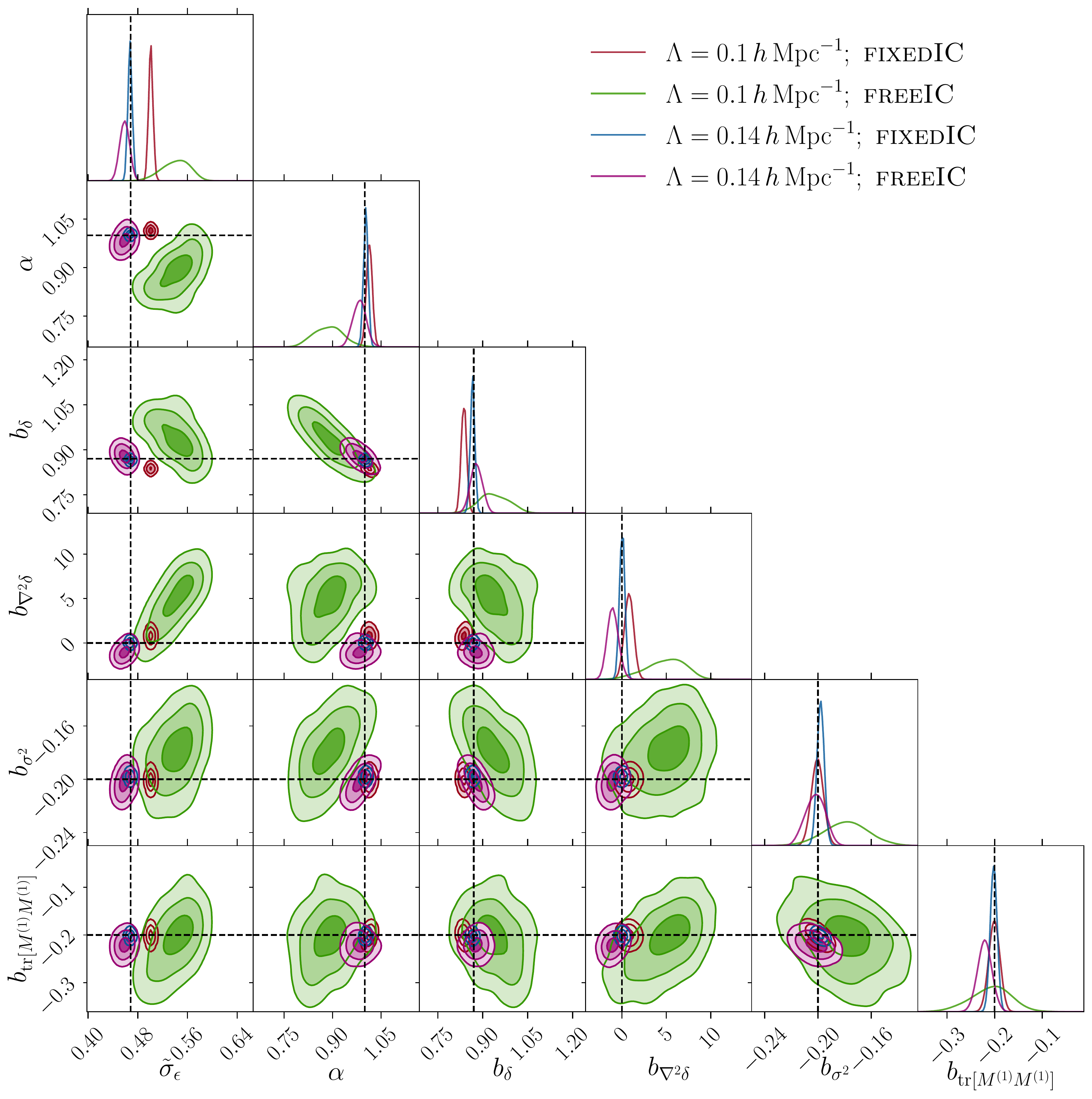}
	\caption{{Parameter posteriors for the} $\mathcal{D}^{\textsc{2LPT}}_2$ synthetic dataset using $\Lambda=0.1h \, \rm{Mpc}^{-1}$ (green and red contours) and $\Lambda=0.14h\,\rm{Mpc}^{-1}$ forward models (blue and purple contours) with \textsc{2LPT} displacement fields. We contrast the {\fixedIC} and the {\freeIC} inference chains for both forward models. As expected, the $\Lambda=0.14h\,\rm{Mpc}^{-1}$ {\freeIC} forward model recovers consistent results as the corresponding {\fixedIC} case. However, the $\Lambda=0.1h \, \rm{Mpc}^{-1}$ {\freeIC} chains, i.e. those with a cutoff mismatch, show a shift with respect to the {\fixedIC} contour. Possible explanations for this are discussed in the text.}
	\label{fig:res_L01_vs_L014_higherbias_nonzero_params_fixed_vs_free_combined}
\end{figure}

The synthetic datasets from \refsec{res_2lpt2d_lpt2d_fwd_models_matter_field} consisted merely of the evolved {\textsc{2LPT}} matter field, rescaled by the linear bias $b_\d$ with added Gaussian noise. Here, we consider synthetic datasets with nonzero higher-order bias coefficients, as well as a cutoff mismatch. This means that we test for the ability of our forward model to extract correct $\alpha$ values from the biased tracers while marginalizing over plausible initial conditions realizations. As before, the $\mathcal{D}_2^{\textsc{2LPT}}$ dataset is generated using a cutoff $\Lambda=0.14 \iMpch$, i.e. restricting to mildly nonlinear scales (recall that all synthetic data sets are at $z=0$). A more realistic test case would adopt dark matter halo or simulated galaxy fields identified in cosmological N-body or hydrodynamic simulations as input. We leave this for future work.

{We show in \reffig{real_space_slices_higher_bias_nonzero_LPT_dataset} a comparison of the real-space posterior mean of our inference and the underlying ground-truth signal of the $\mathcal{D}_{2,a}^{\mathrm{2LPT}}$. As in \reffig{real_space_slices_linearly_biased_LPT_dataset}, we also smooth with the same Gaussian kernel for aesthetic reasons. The overall trend of underdensities and overdensities is well captured by the posterior mean {in this case as well}. For a more quantitative analysis we present parameter posteriors next.}

{Since} we currently do not have analytical predictions for the cutoff dependence of the bias coefficients, we {focus instead on} the value of $\alpha$ as a measure of the model performance. {Being} a cosmological parameter, {it} should be consistent across different cutoffs $\Lambda$.

We {present} here the results using the unmarginalized likelihood from \refeq{unmarg_like}, shown in \reffig{res_L01_vs_L014_higherbias_nonzero_params_fixed_vs_free_combined}. We find a difference between the fiducial noise level and the one inferred at a lower cutoff, similar to the case in \refsec{linear_b1_b2_sigma_tests}. The reason is the same as there, namely the presence of second-order bias terms in the $\mathcal{D}_2^{\textsc{2LPT}}$ dataset and the fact that $\Lambda_0 > \Lambda$. Estimating this shift could be done similarly as in \refeq{delta_s_powerspec_correction}, but taking into account the presence of higher-order nonlinear terms. We leave this calculation for future work.

Turning to the $\alpha$ parameter, both the $\Lambda=0.14 \iMpch$ (blue) and $\Lambda=0.1 \iMpch$ (red) {\fixedIC} posteriors consistently infer the correct value as expected. For the {\freeIC} case, the $\Lambda=0.14 \iMpch$ forward model posterior (purple contours) is able to recover the fiducial $\alpha$ value and is furthermore consistent with the corresponding {\fixedIC} posterior (blue contours). However, the $\Lambda=0.1 \iMpch$ {\freeIC} posterior (green contours) shows a preference for smaller $\alpha$, and excludes the fiducial $\alpha$ value at $95 \% \rm{CL}$ contour. Similar systematic shifts can also be observed in the bias parameters.

{Given that we have already found evidence for prior-volume effects in the case of a linear gravity model with quadratic bias expansion in \refsec{linear_b1_b2_sigma_tests} (see \reffig{res_linear_model_b1_b2_sigma_tests_b2_running_calcs}) it is natural to suspect a similar cause here. {We leave further investigation for the upcoming work}.

\section{Conclusions and Summary}
\label{sec:conclusions_and_summary}

In this paper, we have investigated the robustness of field-level inference based on the EFT {framework} with respect to a mismatch between theory and data, as well as the ability to constrain the amplitude of initial conditions ($\sigma_8$) when marginalizing over the initial conditions. 
Such tests do not only probe the robustness of the chosen forward model and likelihood but also allow for a better understanding regarding what types of forward model physics are necessary in order to capture all relevant effects and to obtain unbiased inference of the cosmological parameters.

We have focused on several different types of forward models described in \refsec{forward_data_models}, each probing different limits of our modeling framework from \refeq{data_model}, as well as different types of likelihood, one with explicit marginalization over bias parameters (\refeq{marg_like}) and the other without (\refeq{unmarg_like}). We perform all the tests on a suite of synthetic datasets with realistic noise levels, described in \refsec{synthetic_datasets}.

We have demonstrated in \refsec{linear_b1_sigma_tests} that, in the case of the purely linear forward model (see \refeq{trivial_fwd_model}), our sampling approach coupled with the EFT likelihood attains a full exploration of the high-dimensional posterior, which in this case can be derived analytically (see \reffigs{res_trivial_fwd_model_b1_sigma_tests_params}{res_linear_model_b1_sigma_tests_phases} as well as the top panel of \reffig{res_linear_model_b1_sigma_tests_phases_pdf_of_Deltashat}). This is a nontrivial result given the high-dimensional ($N_\text{dim} \sim 10^5 - 10^6$) posterior surface involved.

In \refsec{linear_b1_b2_sigma_tests} we have considered a simple but nontrivial extension by adding the $\bdeltatwo$ term in the bias expansion as given in \refeq{trivial_fwd_model_2nd_order_bias}. This term leads to a mode coupling between two linear density fields, which in turn yields a non-Gaussian posterior of the initial conditions, as can be seen from the bottom panel of \reffig{res_linear_model_b1_sigma_tests_phases_pdf_of_Deltashat} as well as \reffig{res_linear_model_b1_b2_sigma_tests_phases}.
Nevertheless, the inferred parameters show the expected behavior also in this non-Gaussian case (\reffig{res_trivial_fwd_model_b1_b2_sigma_tests_params}). We further find good agreements between the {\fixedIC}, where the initial conditions are fixed to their ground-truth values, and {\freeIC} posteriors. This illustrates that our forward modeling and sampling approaches explore the posterior around the correct solution. 

We have also examined cases that include a model mismatch in the \textsc{2LPT} gravity model, by way of choosing a lower cutoff in the inference than the one used to generate the synthetic data. As synthetic datasets, we consider matter fields in \refsec{res_2lpt2d_lpt2d_fwd_models_matter_field} and nonlinearly biased tracer fields in \refsec{res_2lpt2d_lpt2d_fwd_models_biased_tracers}, both including white Gaussian noise. For these cases, we find that, even in the presence of model mismatch, the EFT likelihood is still able to obtain unbiased estimates of cosmological parameters, specifically $\alpha$ {which is our proxy for $\sigma_8$}. 
We have also demonstrated the advantage of using the marginalized over the unmarginalized likelihood from \refeqs{marg_like}{unmarg_like} respectively, by showing a significantly reduced correlation length, in particular for $\alpha$.

We do find signs of a mild discrepancy in the inferred $\alpha$ value in the case for the synthetic data set including nonlinear bias when also allowing for a $\Lambda$ mismatch.
We leave the exploration of possible causes to upcoming work; while prior volume effects could be responsible, higher-order bias terms and density-dependent noise could also be relevant for this particular data set. The flexibility in generating different synthetic data sets will allow for a disentangling of the possible causes.
Such future tests should also include the generalization to synthetic data involving nontrivial noise, such as scale- and density-dependent or non-Gaussian (for example, Poisson) noise.

{Even though we focused on real-space realizations of the synthetic data and our forward model, we note that recent developments within the \codefont{LEFTfield} framework allow for a consistent treatment of the redshift-space distortions at any order within the EFT framework \cite{stadler2023leftyrsd}, although this has only been tested the {\fixedIC} scenario for now. We leave the development of the corresponding {\freeIC} sampling scheme, including the redshift-space distortions for future work.}

{Finally, while we are refraining from making a precise quantitative statement here, it is worth noting that the posteriors on $\alpha$ obtained on our biased \textsc{2LPT} mock data sets (\reffig{res_lpt2D_vs_2lpt2D_model_2lpt_tests_params} and \reffig{res_L01_vs_L014_higherbias_nonzero_params_fixed_vs_free_combined}; with average 68\% CL errors on $\alpha$ of $0.044$ and $0.017$ for $\Lambda = 0.1\iMpch$ and $0.14\iMpch$, respectively) are significantly tighter than what a Fisher forecast for joint power spectrum and bispectrum analysis yields on a similar configuration (cf. Sec. 4.1.3 of \cite{biasreview}, for example). This will likewise be explored in more detail in upcoming work.}

\newpage

\section{Acknowledgements}
\label{sec:acknowledgements}

We thank Andrej Obuljen, Marko Simonović, Uroš Seljak, Matias Zaldariagga, Henrique Rubira for useful discussions during the Šmartno 2022 conference. AK thanks Philipp Frank for the discussion on the results of \refapp{shat_posterior_quadratic} and Florent Leclercq for the discussion on higher-order symplectic integrator schemes for HMC. MN thanks Nickolas Kokron, Emmanuel Schaan, and Chirag Modi for useful discussions on model mis-specification, Wiener-filter solution, and HMC performance, respectively. The authors also thank Deaglan Bartlett, Eiichiro Komatsu, Henrique Rubira, and Julia Stadler for useful feedback on the initial manuscript, which significantly improved the quality of the text. {The authors also thank the anonymous editor and referee for providing useful comments on the initial draft of the manuscript}. AK and FS acknowledge support from the Starting Grant (ERC-2015-STG 678652) ``GrInflaGal'' of the European Research Council. MN acknowledges support from the Leinweber Foundation and the NASA grant under contract 19-ATP19-0058.
All MCMC chains in this paper were produced on the HPC cluster \codefont{FREYA}, maintained by the Max Planck Computing \& Data Facility.
All 1D and 2D posterior plots shown in this paper were obtained through the use of a modified version of \codefont{corner.py}\footnote{\url{https://corner.readthedocs.io/}} \cite{corner}. This work has been done within the Aquila Consortium\footnote{\url{https://www.aquila-consortium.org}}.

\clearpage
\appendix

\section{Fourier space convention}
\label{app:fourier_hartley}

Below, we summarize the Fourier convention and notation adopted throughout the paper. {We} give the explicit relation between the Fourier- and Hartley-representation of $\shat$. The latter is of relevance for understanding our prior choices in \refsec{full_posterior} and following the calculations done in \refapp{shat_posterior} {and \refapp{shat_posterior_quadratic}}.

First, we define the forward Fourier transform $f(\vk)\equiv\{\mathbb{F}f(\vx)\}$ and its inverse transform $f(\vx)\equiv\{\mathbb{F}^{-1}f(\vk)\}$ as
\ba
 f(\vk) &\equiv  \int \dd^3 \vx f(\vx) e^{-i \vk\cdot\vx} \equiv \int_{\vx} f(\vx) e^{-i \vk \cdot \vx}, \\
 f(\vx) &\equiv  \int \frac{\dd^3 \vk}{(2\pi)^3} f(\vk) e^{i \vk\cdot\vx} \equiv \int_{\vk} f(\vk) e^{i \vk \cdot \vx}. 
\ea
In practice, we operate on finite grids, {thus we use} the discrete Fourier transforms given by
\ba
\d(\vk) &= \sum_{i}^{N_g^3}\d(\vx_i)e^{-i \vk \cdot \vx_i},\\
\d(\vx) &= \frac{1}{N_g^3}\sum_{\vk_i} \d(\vk_i) e^{i \vk_i \cdot \vx},
\ea
where $\vk \in (n_x,n_y,n_z)k_F$ with $k_F = 2\pi/L$, and $n_i \in \{-N_g/2, \cdots, N_g/2\}$. The Nyquist frequency is given by $k_{\rm{Ny}} \equiv N_g k_F / 2$.
$L$ stands for the box size. With this, the two-point {function of the $\shat$} becomes
\be
\< \shat(\bm{n}k_F) \shat(\bm{n}'k_F) \> = \frac{1}{L^3} \d_D^{\bm{n},-\bm{n}'} P_{\shat,\shat}(\bm{n}k_F),
\ee
where $\d_D^{\bm{n},-\bm{n}'} = \d_D^{n_x,-n'_x}\d_D^{n_y,-n'_y}\d_D^{n_z,-n'_z}$, with $\d_D^{n_i,n_j}$ representing the Kronecker delta. For clarity, we also write this Kronecker delta in wavenumber space as $\d_D^{\vk,\vk'}$. For a field $\shat(\vx)$ drawn from a unit normal distribution in real space, it follows that $P_{\shat,\shat} = L^3 N_g^3$, and hence $\< \shat(\bm{n}k_F) \shat(\bm{n}'k_F) \> = N_g^3 \d_D^{\bm{n},\bm{-n'}}$. 

In order to implement the cutoff $\Lambda$, we use the isotropic sharp-$k$ filter $W_\Lambda$ defined as
\be
W_\Lambda(\vk) = \Theta_{\rm{H}}(k - \Lambda),
\ee
with $\Theta_{\rm{H}}$ being the Heaviside function.

{Finally, we note that} any field $\shat(\vk)$ can be represented either in the Fourier or Hartley convention, with our \codefont{LEFTfield} code utilizing the latter. The two representations are related through (see \cite{olejniczak2000hartley} and Sec. 3.4 in \cite{jones2010regularized} for more details)
{
\be
\shat(k)\equiv\{\mathbb{H}\shat\}(\vk)=\Re\left[\{\mathbb{F}\shat\}(\vk)\right] - \Im\left[\{\mathbb{F}\shat\}(\vk)\right],
\label{eqn:Fourier-Hartley_relation}
\ee
}
with $\mathbb{F}$ and $\mathbb{H}$ denoting Fourier and Hartley transforms respectively. This is the field whose mean, residual, and variance are shown in the figures in \refsec{analysis}.

\section{Gaussian expectation for $\shat$ posterior}

In this section, we present efforts towards analytical understanding of the shapes of $\shat$ posteriors for the forward models represented by \refeq{trivial_fwd_model} (\refapp{shat_posterior}) and \refeq{trivial_fwd_model_2nd_order_bias} (\refapp{shat_posterior_quadratic}), {restricting to the case where the bias parameters and noise amplitude are fixed to the ground truth. It is much more difficult to obtain an analytical expression for the posterior when also varying the latter.}

In the linear case, an analytical form of the posterior exists, whose mean and variance coincide with the Wiener-filter solution (see, for example, \cite{Kitaura:2008, ensslin2009information} and references therein). In the nonlinear case, however, only a perturbative approach is possible {and we elaborate on this in \refapp{shat_posterior_quadratic}}. 

\subsection{Linear model}
\label{app:shat_posterior}

As discussed around \refeq{trivial_fwd_model}, the covariance structure of cosmological initial conditions is diagonal in Fourier space. Specifically, using the Fourier-space representation of our prior covariance (see our Fourier convention from \refapp{fourier_hartley}), one obtains
\ba
S^{\vk}_{\,\,\, \vk'} = N_g^3 {\d_D}^{\vk,\vk'},
\ea 
where ${\d_D}^{\vk,\vk'}$ represents the Kronecker delta. The noise is likewise assumed to be Gaussian with diagonal covariance $(C_{\eps})^{\vk}_{\,\,\,\vk'}$ related to $P_{\eps}$ (see \refeq{sigma_eps}) as
\ba
(C_{\eps})^{\vk}_{\,\,\, \vk'} &= {\delta_D}^{\vk,\vk'}P_{\eps}.
\ea
These two assumptions allow us to derive the expected posterior on $\shat$. In order to more easily see this, we can rephrase \refeq{trivial_fwd_model} as follows
\ba
	\d_d^{\vk} &= \resp{\vk}{\vk'} \shat^{\vk'} + \eps^{\vk},\\
	\resp{\vk}{\vk'} 
	&= 
	{\delta_D}^{\vk,\vk'} \bdelta T(\vk),
\ea
where repeated indices are summed over; in the following, we will drop the repeated indices. We also drop the explicit $\alpha$ dependence, since here we are only interested in the posterior of initial conditions with $\alpha$ fixed to the ground truth.
In the following, we will further fix the parameters $\bdelta$ and $\sigma_\eps$; only for this case can we derive the posterior for $\shat$ analytically.

The likelihood for $\d_d$ can be derived by marginalizing over the noise distribution, which yields
\ba	
	\mathcal{P}(\d_d|\shat,\bdelta,\sigma_{\eps}) = \G(\d_d ; R\shat, C_{\eps}),
\ea
with $R\shat$ denoting the mean and $C_{\eps}$ denoting the covariance of this Gaussian. Going forward, we consider log probabilities for convenience. This leads to (suppressing the conditional on $\{\bdelta, \sigma_{\eps}\}$ parameters for clarity)
\ba
	- \ln \mathcal{P} (\d_d,\shat) 
	&= 
	- \ln \mathcal{P}(\d_d|\shat) - \ln \mathcal{P} (\shat) \\
	&=
	\frac{1}{2}
	\left(
	\d_d - R\shat
	\right)^\dagger
	C_{\eps}^{-1}
	\left(
	\d_d - R\shat
	\right)
	+
	\frac{1}{2}
	\ln 
	\lvert
	2 \pi C_{\eps}
	\rvert
	+
	\frac{1}{2}
	\shat^\dagger S^{-1} \shat
	+
	\frac{1}{2}
	\ln
	\lvert
	2 \pi S
	\rvert \\
	&=
	\frac{1}{2}
	\left(
	\shat^\dagger(R^\dagger C_{\eps}^{-1} R + S^{-1})\shat
	-
	\d_d^\dagger C_{\eps}^{-1} R\shat
	-
	\shat^\dagger R^{T}C_{\eps}^{-1}\d_d
	\right)\\
	&
	+
	\frac{1}{2}
	\left(
	\d_d^\dagger
	C_{\eps}^{-1}
	\d_d
	+
	\tr\ln C_{\eps}
	+ 
	\tr \ln S
	+
	N_{\d_d} \ln 2\pi
	+
	N_{\shat} \ln 2 \pi
	\right),
	\numberthis
	\label{eqn:WF_posterior_eq1}
\ea    
where $N_{\d_d}$ and $N_{\shat}$ represent the total number of modes in the $\d_d$ and $\shat$ fields, respectively. In our applications, these are always the same. 

Defining $j = R^{T}C_{\eps}^{-1}\d_d$ and $(C_{\mathrm{WF}})^{-1}=(S^{-1} + R^\dagger C_{\eps}^{-1}R)$ {we can rewrite} \refeq{WF_posterior_eq1} as
\be
	\ln \mathcal{P}(\d_d,\shat)
	=
	\frac{1}{2}
	\left(
	\shat - C_{\rm{WF}}\,j
	\right)^\dagger
	(C_{\rm{WF}})^{-1}
	\left(
	\shat - C_{\rm{WF}}\,j
	\right)
	+
	\rm{const.}
\label{eqn:linear_Gaussian_posterior}
\ee
where we have accumulated all the $\shat$-independent terms inside $\rm{const.}$, i.e.\ 
\be
\mathrm{const.}
\equiv
\frac{1}{2}
\left(
\d_d^\dagger
C_{\eps}^{-1}
\d_d
+
\tr\ln C_{\eps}
+ 
\tr \ln S
+
N_{\d_d} \ln 2\pi
+
N_{\shat} \ln 2 \pi
-
j^\dagger
C_{\rm{WF}} \,
j
\right).
\label{eqn:posterior_normalizing_const}
\ee
It is now clear that the posterior of $\shat$ is Gaussian:
\be
\mathcal{P}(\shat|\d_d) 
= 
\G(\shat ; \shat_{\rm{WF}}, C_{\rm{WF}}),
\label{eqn:wf_posterior}
\ee
with mean $\shat_{\rm{WF}}$ and covariance $C_{\rm{WF}}$ given as
\ba
\shat_{\rm{WF}}
&=
C_{\rm{WF}} \, j \\
C_{\rm{WF}}
&= 
(1+S R^\dagger C_{\eps}^{-1}R)^{-1}S.
\numberthis 
\label{eqn:linear_model_WF_mean_cov}
\ea
Substituting into the second line of \refeq{linear_model_WF_mean_cov} the expression for response $R$, noise covariance $C_{\eps}(k)$ and the prior $S$ yields the {following} expression for the $\shat$ posterior covariance
\be
(C_{\rm{WF}})^{\vk}_{\,\,\, \vk'}
= 
{\d_D}^{\vk,\vk'} 
\left( 
1 
+ 
N_g^3\frac{\bdelta^2 \Plin(k)}{P_{\eps}} 
\right)^{-1} 
N_g^3.
\label{eqn:Dmatrix_final_expression}
\ee
For the results shown in the main text, we calculate \refeq{Dmatrix_final_expression} for every mode. We note that when comparing our analytical expression from \refeq{Dmatrix_final_expression} to the results for the $\Delta_{\shat}$ power spectrum obtained from sampling shown in the bottom panel of \reffig{res_linear_model_b1_sigma_tests_phases} and in \reffig{res_linear_model_b1_b2_sigma_tests_phases}, we account for the fact that $P_{\Delta_{\shat}}$ is in fact $\Gamma$ distributed within each $k$-bin. The shape parameter is given by $N_{\rm mode}/2$, with $N_{\rm mode}$ being the number of modes within the Fourier space shell centered on $k$, while the scale parameter is $2 C_{\rm{WF}}(k)$ (see \refsec{linear_b1_sigma_tests}).
We {again emphasize} that the posterior mean and variance \refeq{linear_model_WF_mean_cov} coincide with the Wiener filter result only for a linear forward model,
Gaussian prior and likelihood, and fixed parameters $\alpha, \bdelta, \tilde{\sigma}_\eps$. In fact, \reffig{res_linear_model_b1_b2_sigma_tests_phases} indicates that the Wiener filter estimate of the residual variance is biased low, i.e. $C_{\rm{WF}}$ is relatively lower than the actual variance $P_{\Delta_{\shat}}$, for nonlinear forward models.

\subsection{Quadratic model}
\label{app:shat_posterior_quadratic}

We now consider the quadratic bias model with linearized gravity (see \refeq{trivial_fwd_model_2nd_order_bias}). {Readers are referred to} \refapp{fourier_hartley} for our discrete Fourier convention. We start with writing out the full likelihood expression
\ba
\ln\mathcal{L}(\d_{d, \Lambda} | \ddetLambda[\{O,b_O\}], \sigma_{\eps}) \,
&=
-\frac{1}{2}
\sum_{\vk \neq 0}^{\kmax}
\left[
\ln{2\pi\sigma^2_{\eps}} 
+\frac{1}{\sigma^2_{\eps}}
\lvert
\d_{d, \Lambda}(\vk) - \ddetLambda[\{O,b_O\}](\vk)
\rvert^2
\right], 
\ea
{dropping the explicit $\alpha$ dependence since this parameter is held fixed for our forward model from \refeq{trivial_fwd_model_2nd_order_bias}. Writing explicitly the $\d_{\mathrm{det}}$ from} \refeq{trivial_fwd_model_2nd_order_bias} reads
\ba
\ln\mathcal{L}(\d_{d, \Lambda} | \ddetLambda[\{O,b_O\}], \sigma_{\eps}) \,
=
&-\frac{1}{2}
\sum_{\vk \neq 0}^{\kmax}
\left[
\ln{2\pi\sigma^2_{\eps}} + \frac{1}{\sigma^2_{\eps}}|\d_{d,\Lambda}|^2(\vk) 
\right. \\
&
-\frac{1}{\sigma^2_{\eps}}
\d_{d,\Lambda}^*(\vk)
\Bigl(
(R\shat)(\vk)
+
(R_2\shat\shat)(\vk)
\Bigr) \\
&-
\frac{1}{\sigma^2_{\eps}}
\d_{d,\Lambda}(\vk)
\Bigl(
(R\shat)^*(\vk)
+
(R_2\shat\shat)^*(\vk)
\Bigr) \\
&\left. 
+ \frac{1}{\sigma^2_{\eps}}
\Bigl|
(R\shat)(\vk)
+
(R_2\shat\shat)(\vk)
\Bigr|^2
\right],
\numberthis
\label{eqn:likelihood_trivial_fwd_2nd_order} 
\ea
with $R$ and $R_2$ operations defined as
\ba
\resp{\vk}{\vk_1} [\, . \,] ^{\vk_1}
&\equiv 
\d_D^{\vk, \vk_1} \bdelta W_\Lambda(\vk_1)T(k_1)[\, .\, ]^{\vk_1} \\
\resptwo{\vk}{\vk_1}{\vk_2} [\, .\, , \, . \,]^{\vk_1, \vk_2}
&\equiv 
\bdeltatwo 
\frac{1}{N_g^3} 
\sum_{\vk_1, \vk_2} 
\d_{D}^{\vk, \vk_1 + \vk_2}
W_\Lambda(\vk_1)W_\Lambda(\vk_2)
T(k_1)T(k_2)
[\, . \, , \, . \, ]^{\vk_1,\vk_2} .
\numberthis
\label{eqn:r1_and_r2_ops}
\ea
Note that the $R$ operator is the same as that in the linear forward model described in \refapp{shat_posterior}. The $R_2$ operator implements the second-order bias via a convolution in Fourier space, with the kernel represented by the product of the two transfer functions and the sharp-$k$ cutoffs.

We now add the Gaussian log-prior on $\shat$ to \refeq{likelihood_trivial_fwd_2nd_order} and expand in powers of $\shat$. This results in the following ordering of terms (repeated indices are summed over)
\begin{equation}
\begin{aligned}
o(\shat)&: \hspace{1cm} 
\frac{2}{\sigma_{\eps}^2} 
\d_{d,\Lambda}^{\vk}\resp{\vk}{{\vk}_1}\shat^{\vk_1}  \\
o(\shat^2)&: \hspace{1cm} 
\Bigl(
\frac{1}{\sigma_{\eps}^2}
\resp{\vk}{\vk_1}\resp{\vk}{\vk_2}
-
\frac{2}{\sigma_{\eps}^2}
\d_{d,\Lambda}^{\vk}
\resptwo{\vk}{\vk_1}{\vk_2}
+
\d_{D}^{\vk_1,\vk_2} N_g^{-3}
\Bigr) \shat^{\vk_1} \shat^{\vk_2} 
\\
o(\shat^3)&: \hspace{1cm} 
\frac{2}{\sigma_{\eps}^2}
\resp{\vk}{\vk_1}\resptwo{\vk}{\vk_2}{\vk_3} \shat^{\vk_1}\shat^{\vk_2}\shat^{\vk_3} \\
o(\shat^4)&: \hspace{1cm} 
\frac{1}{\sigma_{\eps}^2}
\resptwo{\vk}{\vk_1}{\vk_2}
\resptwo{\vk}{\vk_3}{\vk_4}
\shat^{\vk_1}\shat^{\vk_2}\shat^{\vk_3}\shat^{\vk_4}.
\end{aligned}
\label{eqn:shat_order_quadr_posterior}
\end{equation}
In other words, the final log-posterior is given by (in matrix notation)
\begin{align*}
\mathcal{H}(\shat | \d_{d,\Lambda}) 
\equiv -\ln \mathcal{P}(\shat|\d_{d,\Lambda}) 
&=
\underbrace{
2R^\dagger C_{\eps}^{-1}\d_{d,\Lambda}
}_{j^\dagger} \shat
+
\frac{1}{2}
\shat^\dagger
\underbrace{
\left(
R^\dagger C_\eps^{-1} R
-
2 R_2 C_{\eps}^{-1} \d_{d,\Lambda}
+
\UnitM \, N_g^{-3}
\right)}_{(D')^{-1}}
\shat \\
&+
\frac{1}{2}
\shat^\dagger
\underbrace{
R^\dagger
C_\eps^{-1}
R_2^\dagger}_{\M^{(3)}}
\shat\shat
+\frac{1}{2}
\shat^\dagger\shat^\dagger
\underbrace{R_2^\dagger
C_\eps^{-1}
R_2}_{\M^{(4)}}
\shat\shat,
\numberthis
\label{eqn:log_pshat_quadratic}
\end{align*}
where we have introduced the third and fourth order coupling kernels {with $\M^{(3)}$ and $\M^{(4)}$} respectively. Also, we have relabeled the operators in the quadratic and linear term with $(D')^{-1}$ and $j$ respectively. In the absence of the $\bdeltatwo$ term, the posterior covariance is given exactly by the Wiener-filter solution {for posterior covariance}, i.e.\ second line of \refeq{linear_model_WF_mean_cov}.
For the posterior given in \refeq{log_pshat_quadratic}, it is not straightforward to compute the corresponding first and second moments. Instead, we expand the posterior around the Wiener-filter solution. While this expansion is strictly only valid if the correction due to $\bdeltatwo$ is small, this expansion nevertheless offers some interesting insights.

We thus define $\shat' \equiv \shat - \shat_{\rm{WF}}$, where $\shat_{\rm{WF}}$ represents the Wiener filter prediction of the initial conditions given by \refeq{linear_model_WF_mean_cov}. In this case, the formalism of Information Field Theory, as presented in \cite{ensslin2009information}, suggests the following diagrammatic representation of the solution (see also Sec. V.C of \cite{ensslin2009information} and, for the Feynman rules, Sec. IV.A.2 of the same paper)
\ba
	\< \shat' (\shat')^{\dagger} \> 
	&= 
	\includegraphics[width=\diagramwidth, valign=c]{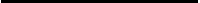}
	+ 
	\includegraphics[width=\diagramwidth, valign=c]{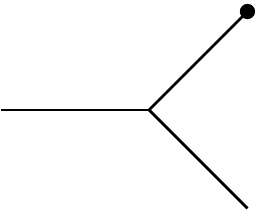}
	+ \text{2 perm.}
	+ 
	\includegraphics[width=\diagramwidth, valign=c]{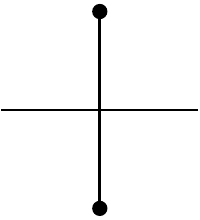} 
	+ \text{5 perm.} \\
	&+ 
	\includegraphics[width=\diagramwidth, valign=c]{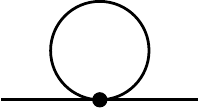}
	\,+ \includegraphics[width=\diagramwidth, valign=c]{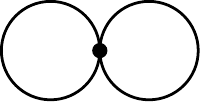},
\ea
where the diagrams correspond to the following expressions
\ba
	{}^{\vk_1}
	\includegraphics[width=\diagramwidth, valign=c]{ift_diagrams/D.pdf}
	{}^{\vk_2}
	&= D'_{\vk_1,\vk_2}	,\\
	\includegraphics[width=\diagramwidth, valign=c]{ift_diagrams/DLDjD.pdf}
	&\sim
	(D')^{\vk_1 \vk'} 
	(\M^{(3)})^{\vk}_{\,\,\, \bm{l} \bm{m}}
	(D')^{\bm{l}\bm{l}_1}j_{\bm{l}_1}
	(D')^{\bm{m}\vk_2}	,\\
	\includegraphics[width=\diagramwidth, valign=c]{ift_diagrams/DLDjDjD.pdf}
	&\sim
	(D')^{\vk_1 \vk} 
	(\M^{(4)})^{\vk}_{\,\,\, \bm{l} \bm{m} \bm{n}}
	(D')^{\bm{l}\bm{l}_1}j_{\bm{l}_1}
	(D')^{\bm{m}\bm{m}_1}j_{\bm{m}_1}		
	(D')^{\bm{n}\vk_2} ,\\
	\includegraphics[width=\diagramwidth, valign=c]{ift_diagrams/loop1.pdf}
	&\sim
	(D')^{\vk_1 \vk} 
	(\M^{(4)})^{\vk}_{\,\,\, \bm{l} \bm{m} \bm{n}}
	(D')^{\bm{l}\bm{m}}
	(D')^{\bm{n}\vk_2} ,\\
	\includegraphics[width=\diagramwidth, valign=c]{ift_diagrams/loop2.pdf}
	&\sim
	(D')^{\vk \bm{l}} 
	(\M^{(4)})^{\vk}_{\,\,\, \bm{l} \bm{m} \bm{n}}
	(D')^{\bm{m}\bm{n}},
	\numberthis
	\label{eqn:shat_covariance_IFT_prediction}
\ea
where we assume that the external lines (without dots) are fixed at $\vk_1$ and $\vk_2$ as indicated in the first line.

We can see that at leading order the posterior covariance is given exactly by $D'$, while the corrections to it depend on the exact form of the coupling kernels $\M^{(3)}$ and $\M^{(4)}$. The evaluation of $D'$ requires an explicit matrix inversion.

To avoid this, and keeping in mind that this expansion is only valid for small corrections to the Wiener-filter posterior, we expand $D'$ to obtain the leading correction to the posterior covariance
\be
D' = 
\left(C_{\rm{WF}}^{-1} - 2 R_2 C_{\eps}^{-1} \d_{d,\Lambda}\right)^{-1}
\approx
C_{\rm{WF}} 
+ 
2 C_{\rm{WF}}
\left(R_2 C_{\eps}^{-1} \d_{d,\Lambda}\right)
C_{\rm{WF}}
+ \cdots
\label{eqn:perturbed_WF}
\ee
{Writing out the leading correction term, one obtains
\be
\left[C_{\rm{WF}}
\left(R_2 C_{\eps}^{-1} \d_{d,\Lambda}\right)
C_{\rm{WF}}\right]_{\vk}{}^{\vk'} \propto b_{\delta^2} (W_\Lambda T \d_{d,\Lambda})(\vk-\vk').
\ee
}
{
This implies that the leading correction to the covariance around the Wiener-filter solution only contributes to the off-diagonal elements. This is an expected result, given the structure of the mode coupling introduced by the $\sim (\dLambda )^2$ term from \refeq{trivial_fwd_model_2nd_order_bias}. In order to compute the correction to the diagonal part of the posterior covariance shown in \reffig{res_linear_model_b1_b2_sigma_tests_phases}, one would need to compute the next-to-leading correction to the posterior covariance. At this order, one also has to include the shift of the maximum of the posterior from the Wiener-filter solution, which is also of order $b_{\delta^2}$. This is a more involved calculation which we leave for future work.}

{The considerations above indicate that obtaining even approximate analytical posteriors for $\shat$ is very difficult already for the simplest nonlinear models. These difficulties are correspondingly exacerbated for more nonlinear models, such as those involving $n$LPT forward models. Thus, the explicit sampling approach appears to be the only path toward obtaining trustable posteriors for initial conditions inference using nonlinear forward models.}

\section{Running of $\bdeltatwo$ with the cutoff}
\label{app:b2_running}

In this section, we describe the 1-loop calculation of the expectation value of the $\bdeltatwo$ parameter as a function of the forward model cutoff $\Lambda$. In order to derive this relation, it is sufficient to look at the maximum likelihood point of the unmarginalized likelihood (see \refeq{unmarg_like})
\be
\frac{\partial}{\partial \bdeltatwo} \ln \mathcal{L}(\d_{d, \Lambda} | \ddetLambda[\{O,b_O\}, \shat], \sigma_{\eps}) = 
\sum_{\vk \neq 0}^{\Lambda} 
\left[
\frac{1}{\sigma^2_{\epsilon}}
\left(\dLambda\right)^2(\vk)
\left(
\d_{d,\Lambda}(\vk) - \ddetLambda(\vk)
\right)^{*} 
\right]
= 0.
\label{eqn:app_mle_point_for_bdeltatwo}
\ee
We have again suppressed the explicit $\alpha$ dependence within $\ddetLambda$, since for the forward model from \refeq{trivial_fwd_model_2nd_order_bias} we keep it fixed. Since $\sigma_{\epsilon}$ is a constant, we can factor it out and using the fact that the forward model here is given by \refeq{trivial_fwd_model_2nd_order_bias} one can rearrange the above \refeq{app_mle_point_for_bdeltatwo} to obtain
\ba
\sum_{\vk \neq 0}^{\Lambda} 
\left[
\left(\dLambda\right)^2(\vk)\d_{d,\Lambda}(-\vk)
\right]
=
\sum_{\vk \neq 0}^{\Lambda} 
\left[
b_{\d,\Lambda}\left(\dLambda\right)^2(\vk)\dLambda(-\vk)
+
b_{\d^2,\Lambda}\left(\dLambda\right)^2(\vk)\left(\dLambda\right)^2(-\vk) 
\right],
\ea
where we have explicitly stated the cutoff dependence of the bias coefficients, which holds in general and used the fact that the density fluctuation field is hermitian. 

Before proceeding, we note that the above equation holds for a given realization of initial conditions, as explicitly stated in \refeq{app_mle_point_for_bdeltatwo}.

In what follows, {we evaluate the MAP relation for $b_{\d^2}$ at the ground-truth initial conditions $\shat_{\rm true}$. This is obviously the correct choice when comparing to {\fixedIC} chains. However, since $\shat_{\rm true}$ is expected to be in the typical set of the {\freeIC} posterior, the result can also be translated to {\freeIC} chains. After evaluating on the ground truth, we then take the ensemble average over data realizations. This allows us to compute the result analytically and gives the following (see also section 3 of \cite{paperI})
\be
b_{\d^2,\Lambda_0}
\sum_{\vk\neq 0}^{\Lambda}
\left\<
\left(\dLambda\right)^2(\vk)
\left(W_\Lambda \left(\dLambdazero\right)^2\right)(-\vk)
\right\>
= 
b_{\d^2,\Lambda}
\sum_{\vk \neq 0}^{\Lambda}
\left\<
\left(\dLambda\right)^2(\vk)
\left(\dLambda\right)^2(-\vk)
\right\>,
\label{eqn:MLE_point_bdeltatwo}
\ee
keeping only the non-zero correlators. Given that the procedure for evaluating all correlators is essentially the same, we focus only on the correlator from the left-hand side of \refeq{MLE_point_bdeltatwo}. {Hence,} we get
\ba
\left\<
\left(\dLambda\right)^2(\vk)
\left(W_{\Lambda} \left(\dLambdazero\right)^2\right)(-\vk)
\right\>
=
\int_{\vk_1}
\int_{\vk_2}
&W_{\Lambda}(\vk - \vk_1)
W_{\Lambda}(\vk_1)
W_{\Lambda_0}(\vk_2 + \vk)
W_{\Lambda_0}(\vk_2) \\
&\left\<
\d^{(1)}(\vk - \vk_1) \d^{(1)}(\vk_1)
\d^{(1)}(\vk - \vk_2) \d^{(1)}(\vk_2)
\right\>.
\ea
Using repeated Wick contractions one gets 
\begin{alignat*}{2}
\left\<
\left(\dLambda\right)^2(\vk)
\left(W_{\Lambda} \left(\dLambdazero\right)^2\right)(-\vk)
\right\>
\,
&=
\,
2\int_{\vk_1}
\begin{aligned}[t]
	&W_{\Lambda}(\vk - \vk_1)
	W_{\Lambda}(\vk_1)
	W_{\Lambda_0}(\vk_1-\vk)
	W_{\Lambda_0}(-\vk_1) \\
	&P_L(\lvert \vk - \vk_1 \rvert)P_L(k_1) 
\end{aligned} \\
&+ 
\d_{D}(\vk)\d_{D}(-\vk)
\int_{\vk_1}\int_{\vk_2}
\begin{aligned}[t]
	&W_{\Lambda}(\vk - \vk_1) W_{\Lambda}(\vk_1) \\
	&W_{\Lambda_0}(\vk - \vk_2) W_{\Lambda_0}(\vk_2) \\
	&P_L(k_1) P_L(k_2),
\end{aligned}
\numberthis
\label{eqn:full_correlator_expression_powerspectra}
\end{alignat*}
with $P_L(k)$ representing the linear power spectrum. From the above equation, it is clear that the correlator is a diagonal matrix in Fourier space. Furthermore, using the fact that the sharp-$k$ cutoff $W_{\Lambda}$ cares only about the magnitude of the given $k$-mode, we can rewrite \refeq{full_correlator_expression_powerspectra} as
\ba
\left\<
\left(\dLambda\right)^2
\left(W_{\Lambda} \left(\dLambdazero\right)^2\right)
\right\>(\vk)
&\,
=
\,
2 
\int_{\vk'}
W_{\Lambda}(\lvert \vk - \vk' \rvert)
W_{\Lambda}(k')
P_L(\lvert \vk - \vk' \rvert)P_L(k') \\
&\quad \stackrel{\vk = 0}{+} 
\int_{\vk_1}
W_{\Lambda}(k_1) P_L(k_1)
\int_{\vk_2}
W_{\Lambda_0}(k_2) P_L(k_2),
\numberthis
\label{eqn:full_correlator_final}
\ea
indicating that the second line only contributes to the $\vk=0$ mode, which is not included in the likelihood evaluation, so this line can be dropped from further {consideration}. Therefore, the only relevant piece is the loop integral on the first line, which in fact matches the correlator on the right-hand side of \refeq{MLE_point_bdeltatwo}. This shows that $b_{\d^2,\Lambda_0} = b_{\d^2,\Lambda}$, and hence no running of $\bdeltatwo$ is expected for the forward model represented by \refeq{trivial_fwd_model_2nd_order_bias}. This behavior is confirmed within our inference chains in \reffig{res_linear_model_b1_b2_sigma_tests_b2_running_calcs}. Note that this result is specific to this simple forward model, and does not apply to the $n$LPT forward models.

\begin{table}[htbp!]
\centering
\begin{tabular}{ | c | c | c | c | c |} 
  \hline
 Forward model, $\Lambda \, [\iMpch]$ & Dataset & $(\hat{R}_C - 1) \times 10^{3}$ & $(\hat{R}_r - 1, \mathcal{T}_\eps - 1) \times 10^{4}$ & $\hat{E}_s$ \\ 
 \hline
 \textsc{linear} - \refeq{trivial_fwd_model}, $\Lambda=0.05$ & $\mathcal{D}^{\textsc{linear}}_1$ & {$5.3$} & {$(129.8, 7.97)$} & {113} \\
 \textsc{linear} - \refeq{trivial_fwd_model}, $\Lambda=0.07$ & $\mathcal{D}^{\textsc{linear}}_1$ & $29.6$ & $(15.65, 7.97)$ & 320 \\
 \textsc{linear} - \refeq{trivial_fwd_model}, $\Lambda=0.1$ & $\mathcal{D}^{\textsc{linear}}_1$ & $1.79$ & $(21.79, 7.97)$ & 230 \\
 \textsc{linear} - \refeq{trivial_fwd_model}, $\Lambda=0.14$ & $\mathcal{D}^{\textsc{linear}}_1$ & $2.31$ & $(48.79, 7.97)$ & 102 \\ 

 \textsc{linear} - \refeq{trivial_fwd_model_2nd_order_bias}, $\Lambda=0.05$  & $\mathcal{D}^{\textsc{linear}}_{2,a}$ & $10.2$ & $(4.29, 7.38)$ & 1138 \\ 
 \textsc{linear} - \refeq{trivial_fwd_model_2nd_order_bias}, $\Lambda=0.05$  & $\mathcal{D}^{\textsc{linear}}_{2,b}$ & $15.2$ & $(5.26, 7.38)$ & 933 \\  
 \textsc{linear} - \refeq{trivial_fwd_model_2nd_order_bias}, $\Lambda=0.08$  & $\mathcal{D}^{\textsc{linear}}_{2,a}$ & $2.66$ & $(4.64, 7.38)$ &  1051 \\ 
 \textsc{linear} - \refeq{trivial_fwd_model_2nd_order_bias}, $\Lambda=0.08$  & $\mathcal{D}^{\textsc{linear}}_{2,b}$ & $4.14$ & $(4.54, 7.38)$ & 1075 \\  
 \textsc{linear} - \refeq{trivial_fwd_model_2nd_order_bias}, $\Lambda=0.1$  & $\mathcal{D}^{\textsc{linear}}_{2,a}$ & $1.62$ & $(7.69, 7.38)$ & 635 \\ 
 \textsc{linear} - \refeq{trivial_fwd_model_2nd_order_bias}, $\Lambda=0.1$  & 
 $\mathcal{D}^{\textsc{linear}}_{2,b}$ & $4.26$ & $(7.86, 7.38)$  & 622 \\
 
\textsc{linear} - \refeq{trivial_fwd_model_2nd_order_bias}, $\Lambda=0.12$  & $\mathcal{D}^{\textsc{linear}}_{2,a}$ & $8.71$ & $(30.18, 7.38)$ & 330 \\ 
\textsc{linear} - \refeq{trivial_fwd_model_2nd_order_bias}, $\Lambda=0.12$  & 
$\mathcal{D}^{\textsc{linear}}_{2,b}$ & $11.21$ & $(35.96, 7.38)$  & 277 \\
 
\textsc{linear} - \refeq{trivial_fwd_model_2nd_order_bias}, $\Lambda=0.13$  & $\mathcal{D}^{\textsc{linear}}_{2,a}$ & $13.84$ & $(42.91, 7.38)$ & 232 \\ 
\textsc{linear} - \refeq{trivial_fwd_model_2nd_order_bias}, $\Lambda=0.13$  & 
$\mathcal{D}^{\textsc{linear}}_{2,b}$ & $10.74$ & $(34.56, 7.38)$  & 288 \\
 
 \textsc{linear} - \refeq{trivial_fwd_model_2nd_order_bias}, $\Lambda=0.14$  & $\mathcal{D}^{\textsc{linear}}_{2,a}$ & $24.9$ & $(33.69, 7.38)$  & 147\\  
 \textsc{linear} - \refeq{trivial_fwd_model_2nd_order_bias}, $\Lambda=0.14$  & $\mathcal{D}^{\textsc{linear}}_{2,b}$ & $14.8$ & $(25.31, 7.38)$  & 195 \\  

 \textsc{1LPT} - \refeq{fwd_model_2lpt2d}, $\Lambda=0.1$  & $\mathcal{D}^{\textsc{2LPT}}_{1,a}$ & $3.25$ & $(38.38, 6.89)$ & 256 \\
 \textsc{1LPT} - \refeq{fwd_model_2lpt2d}, $\Lambda=0.1$  & $\mathcal{D}^{\textsc{2LPT}}_{1,b}$ & $5.37$ & $(73.68, 6.89)$  & 135 \\
 
 \textsc{1LPT} - \refeq{fwd_model_2lpt2d}, $\Lambda=0.14$  & $\mathcal{D}^{\textsc{2LPT}}_{1,a}$ & $41.8$ & $(28.39, 6.89)$ & 171 \\
 \textsc{1LPT} - \refeq{fwd_model_2lpt2d}, $\Lambda=0.14$  & $\mathcal{D}^{\textsc{2LPT}}_{1,b}$ & $23.2$ & $(42.95, 6.89)$  & 114 \\

 \textsc{2LPT} - \refeq{fwd_model_2lpt2d}, $\Lambda=0.1$  & $\mathcal{D}^{\textsc{2LPT}}_{1,a}$ & $2.46$ & $(36.32, 6.89)$ & 273 \\
 \textsc{2LPT} - \refeq{fwd_model_2lpt2d}, $\Lambda=0.1$  & $\mathcal{D}^{\textsc{2LPT}}_{1,b}$ & $23.5$ & $(40.71, 6.89)$ & 243 \\
 \textsc{2LPT} - \refeq{fwd_model_2lpt2d}, $\Lambda=0.14$  & $\mathcal{D}^{\textsc{2LPT}}_{1,a}$ & $17.5$ & $(31.66, 6.89)$  & 311 \\
 \textsc{2LPT} - \refeq{fwd_model_2lpt2d}, $\Lambda=0.14$  & $\mathcal{D}^{\textsc{2LPT}}_{1,b}$ & $95.1$ & $(25.56, 6.89)$  & 385 \\

 \textsc{2LPT} - \refeq{fwd_model_2lpt2d}, $\Lambda=0.1$  & $\mathcal{D}^{\textsc{2LPT}}_{2,a}$ & $2.78$ & $(53.73, 6.89)$  & 184 \\  
 \textsc{2LPT} - \refeq{fwd_model_2lpt2d}, $\Lambda=0.1$  & $\mathcal{D}^{\textsc{2LPT}}_{2,b}$ & $0.17$ & $(63.17, 6.89)$  & 157 \\ 
 \textsc{2LPT} - \refeq{fwd_model_2lpt2d}, $\Lambda=0.14$  & $\mathcal{D}^{\textsc{2LPT}}_{2,a}$ & $94.3$ & $(51.18, 6.89)$  & 125 \\  
 \textsc{2LPT} - \refeq{fwd_model_2lpt2d}, $\Lambda=0.14$  & $\mathcal{D}^{\textsc{2LPT}}_{2,b}$ & $99.6$ & $(47.42, 6.89)$  & 103 \\
 
 \hline 
\end{tabular}
\caption{Gelman-Rubin test statistics, {showing} both $\hat{R}_C$ (see \refeq{classical_GR_stat}) and $\hat{R}_r$ (see \refeq{revised_GR_stat}) for our MCMC chains using the unmarginalized likelihood {(see \refeq{unmarg_like})}. We also show the idealized convergence threshold value, $\mathcal{T}_\eps$, corresponding to having the same number of chains, but instead requiring that $95\%$ of samples lie within $<10\%$ of the posterior volume around the posterior mean which is reported by the \codefont{target.psrf} method of \codefont{stableGR} package (see \cite{vats2021revisiting}). The last column indicates the estimated effective sample size, $\hat{E}_s$, calculated using the \codefont{n.eff} method of \codefont{stable.GR}. The results in each line are obtained from at least 2 chains. For calculating the $\hat{R}_C$ value, we considered $\bdelta$ parameter in case of forward models from \refsec{res_linear_fwd_model}, and $\alpha$ parameter for the chains from \refsec{res_2lpt2d_lpt2d_fwd_models} (see text for more details about this choice). }
\label{tab:gelman_rubin_stats_and_ess_unmarged}
\end{table}
\begin{table}[htbp!]
\centering
\begin{tabular}{ | c | c | c | c | c |} 
  \hline
 Forward model, $\Lambda\,[\iMpch]$ & Dataset & $(\hat{R}_C - 1) \times 10^{3}$ & $(\hat{R}_r - 1) \times 10^{4}$ & $\hat{E}_s$ \\ 
 \hline
\textsc{linear} - \refeq{trivial_fwd_model_2nd_order_bias}, $\Lambda=0.08$  & $\mathcal{D}^{\textsc{linear}-\text{inform.}}_{2,a}$ & $0.21$ & $3.24$ & 1516 \\ 
\textsc{linear} - \refeq{trivial_fwd_model_2nd_order_bias}, $\Lambda=0.08$  & $\mathcal{D}^{\textsc{linear}-\text{inform.}}_{2,b}$ & $0.22$ & $3.49$ & 1408 \\  
\textsc{linear} - \refeq{trivial_fwd_model_2nd_order_bias}, $\Lambda=0.1$  & $\mathcal{D}^{\textsc{linear}-\text{inform.}}_{2,a}$ & $31.4$ & $10.58$ & 469 \\ 
\textsc{linear} - \refeq{trivial_fwd_model_2nd_order_bias}, $\Lambda=0.1$  & $\mathcal{D}^{\textsc{linear}-\text{inform.}}_{2,b}$ & $3.54$ & $9.49$ & 523 \\    
\textsc{linear} - \refeq{trivial_fwd_model_2nd_order_bias}, $\Lambda=0.12$  & $\mathcal{D}^{\textsc{linear}-\text{inform.}}_{2,a}$ & $9.45$ & $22.26$ & 224 \\ 
\textsc{linear} - \refeq{trivial_fwd_model_2nd_order_bias}, $\Lambda=0.12$  & $\mathcal{D}^{\textsc{linear}-\text{inform.}}_{2,b}$ & $0.89$ & $16.81$ & 296 \\  
\textsc{linear} - \refeq{trivial_fwd_model_2nd_order_bias}, $\Lambda=0.13$  & $\mathcal{D}^{\textsc{linear}-\text{inform.}}_{2,a}$ & $31.81$ & $37.81$ & 132 \\ 
\textsc{linear} - \refeq{trivial_fwd_model_2nd_order_bias}, $\Lambda=0.13$  & $\mathcal{D}^{\textsc{linear}-\text{inform.}}_{2,b}$ & $36.64$ & $41.13$ & 121 \\
\textsc{linear} - \refeq{trivial_fwd_model_2nd_order_bias}, $\Lambda=0.14$  & $\mathcal{D}^{\textsc{linear}-\text{inform.}}_{2,a}$ & $9.95$ & $24.46$ & 205 \\
\textsc{linear} - \refeq{trivial_fwd_model_2nd_order_bias}, $\Lambda=0.14$  & $\mathcal{D}^{\textsc{linear}-\text{inform.}}_{2,b}$ & $9.64$ & $23.57$ & 211 \\
 \hline 
\end{tabular}
\caption{Supplement table for \reftab{gelman_rubin_stats_and_ess_unmarged} containing Gelman-Rubin test statistics for chains obtained from applying the forward model from \refeq{trivial_fwd_model_2nd_order_bias} to the $\mathcal{D}^{\textsc{linear}-\text{informative}}_2$ datasets appearing in \reffig{res_linear_model_b1_b2_sigma_tests_b2_running_calcs}. {The convergence threshold, $\mathcal{T}_{\eps}$, is the same for all the chains, given that all chains are using the same forward model. The threshold is $(\mathcal{T}_\eps - 1)\times 10^4 = 7.38$.}}
\label{tab:gelman_rubin_stats_and_ess_unmarged_part2}
\end{table}

\begin{table}[htbp!]
\centering
\begin{tabular}{ | c | c | c | c | c | } 
 \hline
 Forward model, $\Lambda\, [\iMpch]$ & Dataset & $(\hat{R}_C - 1) \times 10^{3}$ & $(\hat{R}_r - 1) \times 10^{4}$ & $\hat{E}_s$ \\ 
 \hline
\textsc{1LPT} - \refeq{fwd_model_2lpt2d}, $\Lambda=0.1$  & $\mathcal{D}^{\textsc{2LPT}}_{1,a}$ & $6.75$ & $116.51$ & 128 \\
\textsc{1LPT} - \refeq{fwd_model_2lpt2d}, $\Lambda=0.1$  & $\mathcal{D}^{\textsc{2LPT}}_{1,b}$ & $19.5$ & $137.9$ & 108 \\ 

\textsc{1LPT} - \refeq{fwd_model_2lpt2d}, $\Lambda=0.14$  & $\mathcal{D}^{\textsc{2LPT}}_{1,a}$ & $22.5$ & $96.3$ & 155 \\
\textsc{1LPT} - \refeq{fwd_model_2lpt2d}, $\Lambda=0.14$  & $\mathcal{D}^{\textsc{2LPT}}_{1,b}$ & $44.9$ & $102.9$ & 145 \\ 

 \textsc{2LPT} - \refeq{fwd_model_2lpt2d}, $\Lambda=0.1$  & $\mathcal{D}^{\textsc{2LPT}}_{1,a}$ & $4.78$ & $60.61$ & 166 \\
 \textsc{2LPT} - \refeq{fwd_model_2lpt2d}, $\Lambda=0.1$  & $\mathcal{D}^{\textsc{2LPT}}_{1,b}$ & $4.54$ & $63.89$ & 156 \\
 \textsc{2LPT} - \refeq{fwd_model_2lpt2d}, $\Lambda=0.14$  & $\mathcal{D}^{\textsc{2LPT}}_{1,a}$ & $99.5$ & $80.76$ & 185  \\
 \textsc{2LPT} - \refeq{fwd_model_2lpt2d}, $\Lambda=0.14$  & $\mathcal{D}^{\textsc{2LPT}}_{1,b}$ & $28.3$ & $85.84$ & 174 \\

 \textsc{2LPT} - \refeq{fwd_model_2lpt2d}, $\Lambda=0.1$  & $\mathcal{D}^{\textsc{2LPT}}_{2,a}$ & $98.4$ & $108.9$ & 137 \\ 
 \textsc{2LPT} - \refeq{fwd_model_2lpt2d}, $\Lambda=0.1$  & $\mathcal{D}^{\textsc{2LPT}}_{2,b}$ & $94.2$ & $115.6$ & 129 \\
 \textsc{2LPT} - \refeq{fwd_model_2lpt2d}, $\Lambda=0.14$  & $\mathcal{D}^{\textsc{2LPT}}_{2,a}$ & $6.47$ & $113.86$ & 131  \\
 \textsc{2LPT} - \refeq{fwd_model_2lpt2d}, $\Lambda=0.14$  & $\mathcal{D}^{\textsc{2LPT}}_{2,b}$ & $78.8$ & $51.95$ & 192 \\

 \hline
\end{tabular}
\caption{Same as \reftab{gelman_rubin_stats_and_ess_unmarged}, but for chains using the marginalized likelihood from \refeq{marg_like}. Note that these chains were not run as long as the chains from \reftab{gelman_rubin_stats_and_ess_unmarged} and hence have a smaller number of effective samples overall. {The convergence threshold, $\mathcal{T}_{\eps}$, for these chains is $(\mathcal{T}_\eps - 1)\times 10^4 = 7.97$.}}
\label{tab:gelman_rubin_stats_and_ess_marged}
\end{table}

\section{Chain convergence and sample correlation analysis}
\label{app:results_convergence_tests}

MCMC samples are not entirely independent. In practice, this correlation between neighboring samples introduces further uncertainty in any estimate based on averaging over those samples, such as the posterior mean, variance, and all higher moments. The correlation is often measured by the integrated autocorrelation time, while the resulting uncertainty in the posterior is quantified by the effective sample size. We refer readers to \cite{gelman1995bayesian} for more details.

We define the normalized autocorrelation function, $\rho(t)$, as
\ba
	\rho(t) &\equiv \frac{\mathcal{A}(t)}{\mathcal{A}(0)}, \\
	\mathcal{A}(t) &= \< \gamma_s \gamma_{s + t} \>_s - \< \gamma_s \>_s^2,
	\numberthis
\label{eqn:autocorr_rho}
\ea
where $\{\gamma_s\}_{s = 1 \cdots N}$ is the set of chain samples, and the brackets indicate the average over samples, i.e.\ $\< \gamma_s \>_s {\equiv \bar{\gamma} = \frac{1}{N}\sum_{s=1}^N \gamma_s}$, while $t$ indicates the sample separation. \refeq{autocorr_rho} highlights the significance of having a sufficient number of MCMC samples, i.e.\ running sufficiently long MCMC chains, since $\bar{\gamma}$ and $\mathcal{A}(0)$ are noisy estimates of the true mean and variance, whose noise propagates nonlinearly into $\rho(t)$. For the autocorrelation function, $\mathcal{A}(t)$, we use the estimator presented in \cite{Sokal1996MonteCM} which shows better asymptotic behavior than the one given in \refeq{autocorr_rho}. The normalized autocorrelation function $\rho(t)$ is exactly what is shown in \reffig{autocorr_marged_vs_unmarged}. We then use the {$\rho(t)$} to estimate the correlation length of the chain as
\be
\hat{\tau}(T) = \sum_{t=-T}^T \rho(t) = 1 + 2\sum_{t=1}^T\rho(t),
\label{eqn:Sokal_tau_estimator}
\ee
with $T$ representing the maximal separation between the samples considered. This estimator has a vanishing variance in the limit of large chain lengths, i.e.\ number of samples. We have adopted the approach of \cite{madras1988pivot} for choosing $T$. In short, $T$ is chosen such that it corresponds to the smallest integer satisfying $T \geq C \hat{\tau}(T)$ with a constant $C$ chosen such that the variance of the estimator is minimized, at the cost of introducing a negative bias in the estimate of $\hat{\tau}$. This is typically achieved for $C \in [5,10]$. We report the $\hat{\tau}$ estimates, as well as the used window $T$ in \reffig{autocorr_marged_vs_unmarged}, when comparing the sampling performance of the marginalized and unmarginalized likelihood.

We now describe the two tests of convergence we perform for all chains analyzed in this paper, namely the classical and revised Gelman-Rubin (G-R) diagnostic. The revised G-R statistics \cite{vats2021revisiting} makes a clear connection to the chain effective sample size (see Eq. (12) in \cite{vats2021revisiting}). We exploit this connection to link the (revised) G-R value and our target effective sample size.

For the classical G-R statistics \cite{gelman1992inference,brooks1998general,gelman1995bayesian}, we adopt the following procedure. First we calculate the inter- and intra-chain variances
\ba
B &= \frac{N}{M-1}\sum_{j=1}^{M}(\bar{\gamma}_{.j} - \bar{\gamma}_{..})^2, \\
W &= \frac{1}{M}\sum_{j=1}^{M} s_{j}^2, 
\numberthis
\ea
with
\ba
\bar{\gamma}_{.j} &= \frac{1}{N}\sum_{i=1}^{N}\gamma_{ij} 
\,\, , \,\,\, 
\bar{\gamma}_{..} = \frac{1}{M} \sum_{j=1}^{M}\bar{\gamma}_{.j}
\,\, , \,\,\, 
s_j^2 = \frac{1}{N-1}\sum_{i=1}^N (\gamma_{ij} - \bar{\gamma}_{.j})^2 \,\, ,
\ea
and $N$,$M$ being the chain length and number of chains considered respectively. From the above expression, one can see that $B$ represents an estimate of the variance between the chains while $W$ is the mean of the variance within individual chains. These two quantities then can be combined into an estimate of the true underlying target distribution variance 
\be
\hat{\sigma}^2 = \frac{N-1}{N} W + \frac{1}{N}B.
\ee 
The authors of \cite{gelman1992inference, brooks1998general} argue that for a properly dispersed set of chains, the $\hat{\sigma}^2$ estimate is typically over-estimating the underlying variance, while the mean of the within-the-chain variances, $W$, under-estimates it. Hence, they propose the following quantity as a measure of chain convergence, which is known as the classical G-R test statistics
\be
\hat{R}_C = \sqrt{\frac{\hat{\sigma}^2}{W}}.
\label{eqn:classical_GR_stat}
\ee
Specifically, we apply this univariate G-R test to the $\alpha$ parameter (for chains from \refsec{2lpt_model_tests}) and $\bdelta$ parameter (for chains from \refsec{res_linear_fwd_model}), since these parameters typically have the longest correlation lengths and largest $\hat{R}_C$ values. The resulting values are reported in \reftab{gelman_rubin_stats_and_ess_unmarged}, \ref{tab:gelman_rubin_stats_and_ess_unmarged_part2} and \ref{tab:gelman_rubin_stats_and_ess_marged} as $\hat{R}_C$.

The revised G-R test statistics can be estimated from the following expression
\be
\hat{R}_r \approx \sqrt{1 + \frac{M}{\hat{E}_s}} \leq \mathcal{T}_{\eps}.
\label{eqn:revised_GR_stat}
\ee
The above \refeq{revised_GR_stat} provides a clear connection between the number of chains $M$, effective sample size $\hat{E}_s$ and the convergence threshold $\mathcal{T}_{\eps}$. It is possible to determine the convergence threshold a priori and hence the corresponding effective size necessary for reaching it. Specifically, in this paper, we set a target of $\hat{E}_s\geq100$ for all MCMC chains. In practice, we use the \codefont{n.eff} multivariate method of the \codefont{stable.GR} package\footnote{\url{https://github.com/knudson1/stableGR}} to estimate $\hat{E}_s$. The \codefont{stable.GR} package is provided by the authors of \cite{vats2021revisiting}. For more details on how $\hat{E}_s$ is calculated, we refer readers to Sec. 5 of \cite{vats2021revisiting}.
This number is reported in the last column of \reftab{gelman_rubin_stats_and_ess_unmarged}, \ref{tab:gelman_rubin_stats_and_ess_unmarged_part2} and \ref{tab:gelman_rubin_stats_and_ess_marged}.
We note that the convergence threshold value $\mathcal{T}_{\eps}$ we report represents an ideal case, which corresponds to having the same number of chains as we do, but with $95\%$ of samples lying within $<10\%$ of the posterior volume around the posterior mean. Such chains will then have $\hat{R}_r \sim \mathcal{T}_{\eps}$.

\clearpage
\section{Relation between \textsc{1LPT} and \textsc{2LPT} second-order bias coefficients}
\label{app:relation_between_lpt2d_and_2lpt}

As described in \refsec{forward_data_models}, the \textsc{1LPT} and \textsc{2LPT} forward models differ only in the order of the LPT displacement field. This difference has an impact on the inference as shown in \reffig{res_lpt2D_vs_2lpt2D_model_2lpt_tests_params}. The goal of this section is to understand whether the observed discrepancy between the {\textsc{1LPT} and \textsc{2LPT} posteriors is expected}. In order to derive this, we first go back to the general setup of both of these models. \par
First, the same order of Lagrangian bias expansion is employed in both, allowing for the following set of Lagrangian bias operators
\ba
O_L &\in \left\{ 1,
\left(\tr{M^{(1)}}\right)^2, \tr\left(M^{(1)} M^{(1)}\right) \right\},
\numberthis
\label{eqn:app_biasops_and_biasparams_2lpt}
\ea
and the corresponding bias coefficients. Recall that we displace a unit field to obtain the Eulerian matter density. The transformation of bias coefficients derived in the following only involves operators at leading order in derivatives, therefore we do not need to consider $\lapl\d_{n-{\rm LPT}}$ in the following. The relation between Eulerian and Lagrangian frames is given by \cite{biasreview}
\ba
1 + \ddetLambda^{n\textsc{LPT}} (\vx) &= \left(\J^{(n)}\right)^{-1}\left(1 + \ddetLambda^{n\textsc{LPT}}(\vq)\right), \\
\J^{(n)} &= \det\left(\UnitM + \partial_{\vq} \bm{{\psi}}^{(n)} \right),
\numberthis
\label{eqn:app_ddet_and_jac}
\ea
with $\J$ being the Jacobian of the transformation from Lagrangian to Eulerian coordinates, and $n$ denoting the order up to which the forward model is to be evaluated. Note that in our field-level forward model, $\J^{-1}$ is computed non-perturbatively, by displacing and depositing pseudo particles within the simulated box. Here, we instead expand $\J^{-1}$ perturbatively up to second order, to obtain the mapping of bias operators between the different LPT orders. Specifically, for \textsc{1LPT} and \textsc{2LPT} the corresponding inverse Jacobians are given by 
\ba
\left(\J^{(1)}\right)^{-1}
&= 
1 - \partial_{q_i} {\psi}_i^{(1)}
+
\frac{1}{2}
\left[
\left(
\partial_{q_i}{\psi}_i^{(1)}
\right)^2 
+ 
\partial_{q_i}{\psi}_j^{(1)}\partial_{q_j}{\psi}_i^{(1)} 
\right]
+ o\left(({\psi}^{(1)})^3\right),  \\
\left(\J^{(2)}\right)^{-1} 
&= 
1 - \partial_{q_i} {\psi}_i^{(1)} - \partial_{q_j}{\psi}_j^{(2)} 
+
\frac{1}{2}
\left[ 
\left(\partial_{q_i}{\psi}_i^{(1)}
\right)^2 
+
\partial_{q_i}{\psi}_j^{(1)}\partial_{q_j}{\psi}_i^{(1)} 
\right]
+ o\left(({\psi}^{(1)})^3\right),
\numberthis
\label{eqn:J_inverse}
\ea
where we kept only second-order terms. We then use the following bias expansion
\be
\ddetLambda(\vq) = \bdelta^L + \bsigmasigma^L \left(\tr M^{(1)}\right)^2 + \btrMoneMone^L \tr\left(M^{(1)}M^{(1)}\right),
\label{eqn:q_bias_expansion}
\ee
where we note that $\bdelta^L$ is the coefficient of the uniform field $O_L=1$.  Plugging the results of \refeqs{J_inverse}{q_bias_expansion} directly into the first line of \refeq{app_ddet_and_jac} we obtain
\ba
\d^{1\textsc{LPT}}_{\rm{det},\Lambda}(\vx) 
&=
(\bdelta^L + 1)(1 - \tr M^{(1)})
\begin{aligned}[t]
	&+ \left(\bsigmasigma^L + \frac{1}{2}(\bdelta^L + 1)\right) \left(\tr M^{(1)}\right)^2 \\
	&+ \left(\btrMoneMone^L + \frac{1}{2}(\bdelta^L + 1)\right)\tr\left(M^{(1)} M^{(1)}\right),
\end{aligned} \\
\d^{\rm{2LPT}}_{\rm{det},\Lambda}(\vx) 
&=
(\bdelta^L + 1)(1-\tr M^{(1)})
\begin{aligned}[t]
	&+\left(\bsigmasigma^L + \frac{1}{2}(\bdelta^L + 1) - \frac{3}{14}\right)\left(\tr M^{(1)} \right)^2 \\
	&+\left(\btrMoneMone^L + \frac{1}{2}(\bdelta^L + 1) + \frac{3}{14} \right)\tr\left(M^{(1)} M^{(1)}\right),
\end{aligned}
\numberthis
\label{eqn:delta_lpt2D_delta_2lpt2D_perturbative}
\ea
again keeping only second-order terms. In \refeq{delta_lpt2D_delta_2lpt2D_perturbative} we also used the solution for the second order displacement field from the equations of motion assuming an Einstein-de Sitter universe (see \cite{bouchet1994perturbative}, as well as Sec. 2.7 in \cite{bernardeau2002large}, Sec. 2.5.2 in \cite{biasreview})
\begin{equation*}
\tr M^{(2)} = 
-\frac{3}{14}
\left( \left(\tr M^{(1)}\right)^2 - \tr\left(M^{(1)} M^{(1)}\right)\right).
\end{equation*}
This then produces the following relationship between the \textsc{1LPT} and \textsc{2LPT} forward model bias coefficients
\ba
\bdelta^{\textsc{2LPT}} &= \bdelta^{\textsc{1LPT}} = \bdelta^L + 1, \\
\bsigmasigma^{\textsc{1LPT}} &= \bsigmasigma^{\textsc{2LPT}} + \frac{3}{14} \\
\btrMoneMone^{\textsc{1LPT}} &= \btrMoneMone^{\textsc{2LPT}} - \frac{3}{14} .
\numberthis
\label{eqn:app_res_1lpt_vs_2lpt_relation_bias_coeffs}
\ea
We can now compare with the results from the \textsc{1LPT} and \textsc{2LPT} {posteriors obtained} on the $\mathcal{D}^{\rm{2LPT}}_1$ {datasets}, shown in \reffig{res_lpt2D_vs_2lpt2D_model_2lpt_tests_params}.
Taking the mean of the \textsc{2LPT} posterior for $\bsigmasigma^{\textsc{2LPT}}\approx 0.0236$ and $\btrMoneMone^{\textsc{2LPT}} \approx -0.099$, leads to $\bsigmasigma^{\textsc{1LPT}} \approx 0.238$ and $\btrMoneMone^{\textsc{1LPT}} \approx -0.31$, which are indicated with dotted lines in \reffig{res_lpt2D_vs_2lpt2D_model_2lpt_tests_params}, which agree within $39.3-86.5\%$ confidence level with the obtained \textsc{1LPT} posterior.
\begin{figure}[t!]
	\centering
	\includegraphics[width=.495\linewidth]{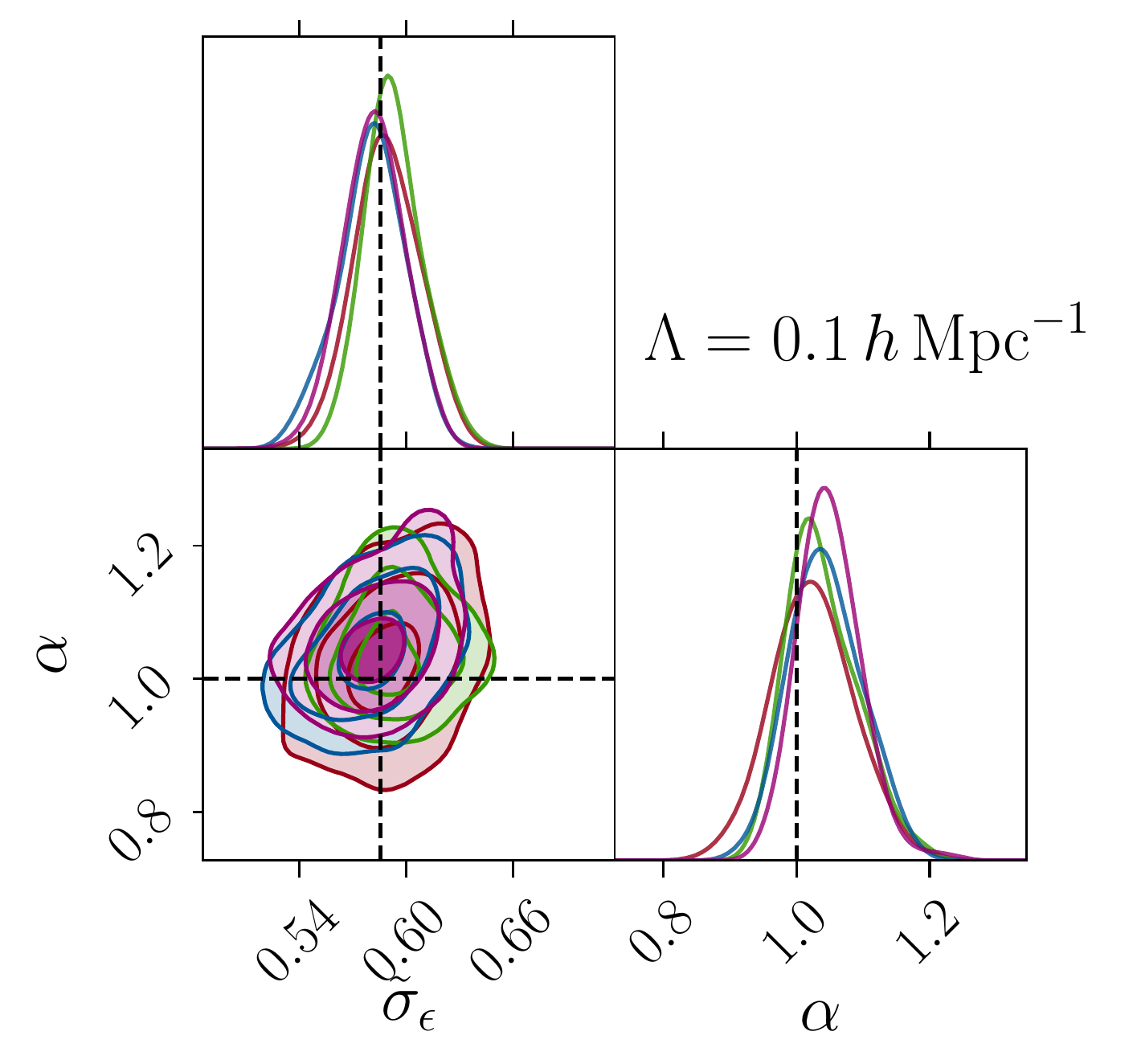}
	\includegraphics[width=.495\linewidth]{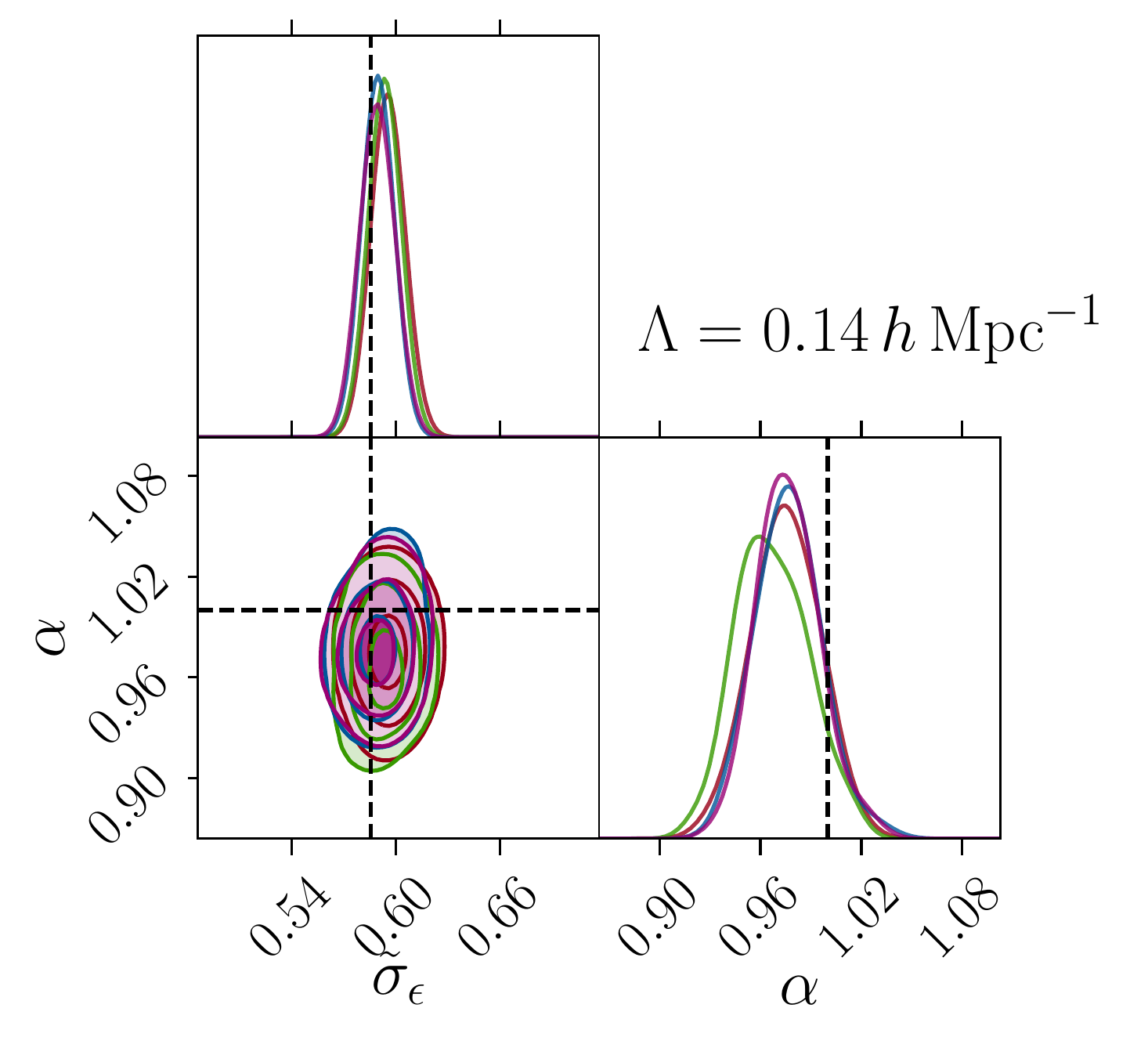}	
	\caption{Parameter posteriors obtained using the marginalized (\refeq{marg_like}) and unmarginalized likelihoods (\refeq{unmarg_like}) in the $\alpha - \tilde{\sigma}_{\eps}$ plane, both inferred from the $\mathcal{D}_{1,b}^{\textsc{2LPT}}$ dataset. We show the results at different cutoffs, on the left, $\Lambda=0.1\iMpch$ and on the right $\Lambda=0.14\iMpch$. Forward models considered are \textsc{1LPT} with unmarginalized likelihood (green), \textsc{1LPT} with marginalized likelihood (red), \textsc{2LPT} with unmarginalized likelihood (purple) and \textsc{2LPT} with marginalized likelihood (blue). {There is also an} excellent agreement among the two likelihoods, while the marginalized likelihood yields a smaller correlation length for the $\alpha$ parameter (see \reffig{autocorr_marged_vs_unmarged}).}
	\label{fig:res_2lpt_vs_lpt2d_marg_vs_unmarg}
\end{figure}
\begin{figure}[htbp!]
	\centering
	\includegraphics[width=.5\linewidth]{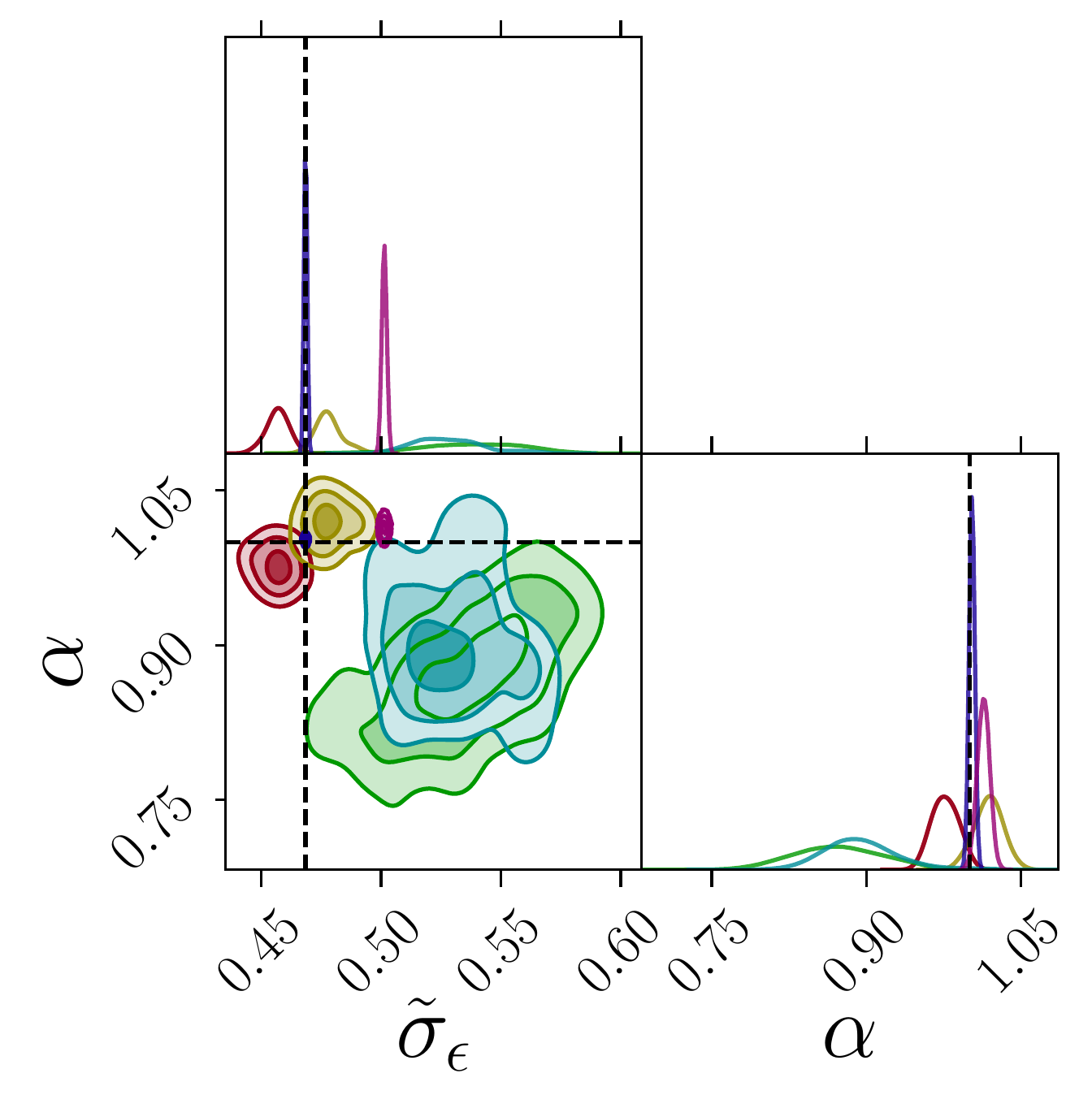}
	\caption{{Parameter posteriors obtained using the marginalized (\refeq{marg_like}) likelihood on the $\mathcal{D}_{2,a}^{\textsc{2LPT}}$ and $\mathcal{D}_{2,b}^{\textsc{2LPT}}$ datasets (see \reftab{synthetic_datasets}). The red and dark-yellow contours represent the {\freeIC} posteriors of the forward model from \refeq{fwd_model_2lpt2d} obtained from $\mathcal{D}_{2,a}^{\textsc{2LPT}}$ and $\mathcal{D}_{2,b}^{\textsc{2LPT}}$ datasets respectively, with a cutoff $\Lambda=0.14 \iMpch$, and similarly for light-green and turquoise but with a cutoff of $\Lambda=0.1 \iMpch$. The blue and pirple posteriors represent the {\fixedIC} posteriors for $\Lambda=0.1 \iMpch$ and $\Lambda=0.14 \iMpch$ respectively.}}
	\label{fig:res_2lpt_higher_order_bias_nonzero_marged}
\end{figure}

\section{Marginalized and unmarginalized likelihood chains}
\label{app:unmarged_vs_marged_like}

{Here, we provide the posteriors obtained with the marginalized likelihood (\refeq{marg_like}), using the priors given in \refeq{priors_marg}.} 

{\reffig{res_2lpt_vs_lpt2d_marg_vs_unmarg} shows the posteriors of the forward model from \refeq{fwd_model_2lpt2d} obtained on the $\mathcal{D}_{1,b}^{\textsc{2LPT}}$ dataset. We find the same {results} for the inference done on $\mathcal{D}_{1,a}^{\textsc{2LPT}}$ dataset. We contrast the marginalized likelihood posteriors (red and blue) to the unmarginalized (\refeq{unmarg_like}) likelihood posteriors (green and purple), using the priors from \refsec{full_posterior}. The posterior contours are entirely consistent with each other for both \textsc{1LPT} and \textsc{2LPT} forward models, and across the different cutoffs ($\Lambda=0.1\iMpch$ and $\Lambda=0.14\iMpch$). However, recall that the marginalized likelihood leads to better efficiency in terms of sampling the $\alpha$ parameter, as depicted in \reffig{autocorr_marged_vs_unmarged} (top panel)}.

{The posterior of the same forward model obtained on the $\mathcal{D}_2^{\textsc{2LPT}}$ dataset, using the marginalized likelihood, is shown in  \reffig{res_2lpt_higher_order_bias_nonzero_marged}. The results for the forward model with cutoff $\Lambda=0.1 \iMpch$, turquoise and light-green for the {\freeIC} case and the purple for the {\fixedIC} posterior, all show higher levels of noise than the corresponding cases for $\Lambda=0.14\iMpch$ posteriors. As discussed in \refsec{res_2lpt2d_lpt2d_fwd_models_biased_tracers}, this is a consequence of the mode couplings present at the higher cutoff at which the synthetic datasets were generated (see also \refsec{linear_b1_b2_sigma_tests}). It is as well worth noting that the posteriors obtained with the marginalized likelihood from \reffig{res_2lpt_higher_order_bias_nonzero_marged} are consistent with the corresponding unmarginalized case shown in \reffig{res_L01_vs_L014_higherbias_nonzero_params_fixed_vs_free_combined}}.

\clearpage
\bibliographystyle{JHEP}
\bibliography{paper}

\end{document}